\documentclass[a4paper,reqno]{amsart}    

\usepackage[utf8]{inputenc}
\usepackage[greek,english]{babel}
\usepackage{textgreek}

\usepackage{etoolbox}

\usepackage{cite}

\usepackage[hang]{footmisc}
\usepackage{graphicx}
\usepackage{subfigure}
\usepackage{float}
\usepackage{xcolor}

\usepackage{amsmath}
\usepackage{amssymb}
\usepackage{amsthm}
\usepackage{mathtools}

\graphicspath{
  {figures/}
  {figures/geom/}
  {figures/geom/1d_alternative/}
  {figures/geom/2d_axisymmetric/}
  {figures/geom/2d_axisymmetric/wormhole/}
}

\title[Numerical initial data deformation exploiting a gluing construction I.]{Numerical initial data deformation exploiting a gluing construction: I. Exterior asymptotic Schwarzschild}

\author[B. Daszuta]{Boris Daszuta}
\address{Department of Mathematics and Statistics, University of Otago, PO Box 56, Dunedin 9010, New Zealand}
\email{bdaszuta@maths.otago.ac.nz}
\author[J. Frauendiener]{J\"org Frauendiener}
\address{Department of Mathematics and Statistics, University of Otago, PO Box 56, Dunedin 9010, New Zealand}
\email{joergf@maths.otago.ac.nz}
\newcommand{\paperabstract}{In this work a new numerical technique
to prepare Cauchy data for the initial value problem (IVP) formulation
of Einstein's field equations is presented.
Directly inspired by the exterior asymptotic gluing
(EAG) result of Corvino \cite{scalarcurvature2000corvino}
our (pseudo)-spectral scheme is demonstrated under the assumption of axisymmetry 
so as to fashion composite Hamiltonian constraint
satisfying initial data featuring internal binary black holes (BBH) as glued to
exterior Schwarzschild initial data in isotropic form. The generality of the 
method is illustrated in a comparison of the ADM mass of EAG initial data 
sets featuring internal BBHs as modelled by Brill-Lindquist and Misner data.
In contrast to the recent work of 
Doulis and Rinne \cite{:ref:numericalconstruction2016doulis}, and
Pook-Kolb and Giulini \cite{:ref:numerical2018pook_kolb} we do not make 
use of the York-Lichnerowicz 
conformal framework to reformulate the constraints.}

\usepackage{enumerate}

\usepackage[colorlinks = true,
            linkcolor = blue,
            urlcolor  = blue,
            citecolor = red,
            anchorcolor = green,
            hyperindex = true,
            unicode = true,
            pdfauthor={.},
            pdfsubject={.},
            pdfcreator={.}
            ]{hyperref}

\usepackage{bookmark}

\usepackage[scaled]{helvet}        
\usepackage{courier}               
\usepackage{euler}                 
\normalfont
\usepackage[T1]{fontenc}

\usepackage[a4paper,left=25mm, right=25mm, top=20mm, bottom=20mm,%
            includefoot]{geometry}
\usepackage[capitalize]{cleveref}          

\newtheoremstyle{theoremsty}
{0pt}        
{-2pt}       
{\upshape}   
{}           
{\bfseries}  
{:}          
{.5em}       
{}           
\theoremstyle{theoremsty}

\theoremstyle{theoremsty}

\newtheorem*{theorem*}{Theorem}

\theoremstyle{definition}

\renewcommand{\d}[1]{\ensuremath{
  \mathrm{d}{#1}
}}

\DeclareMathOperator*{\argmax}{arg\,max}
\newcommand{\dimd}[1]{\ensuremath{{}^{[3]}{#1}}}
\newcommand{\psbrc}[3]{\ensuremath{\prescript{ #1}{#2\hphantom{)}}{#3}}}
\newcommand{\sw}[2]{\ensuremath{{}_{#1}#2}}
\newcommand{\edth}[1]{\ensuremath{\eth\left[#1\right]}}
\newcommand{\edtb}[1]{\ensuremath{\overline{\eth}\left[#1\right]}}
\newcommand{\Lpop}[0]{\ensuremath{\mathrm{L}{}_{\overline{g}}}}
\newcommand{\Lspop}[0]{\ensuremath{\mathrm{L}{}^*_{\overline{g}}}}
\newcommand{\wei}{\ensuremath{\omega}}
\newcommand{\deform}{\delta}
\newcommand{\rmax}{\ensuremath{\rho_{\mathrm{max}}}}
\newcommand{\rmin}{\ensuremath{\rho_{\mathrm{min}}}}

\newcommand{\nabo}{\overset{\circ}{\mathfrak{D}}}
\newcommand{\nab}{{\mathfrak{D}}}
\let\oldsqrt\sqrt                            
\def\sqrt{\mathpalette\DHLhksqrt}            

\def\DHLhksqrt#1#2{%
\setbox0=\hbox{$#1\oldsqrt{#2\,}$}\dimen0=\ht0
\advance\dimen0-0.2\ht0
\setbox2=\hbox{\vrule height\ht0 depth -\dimen0}
{\box0\lower0.4pt\box2}}

\numberwithin{equation}{section}
\begin{document}

\begin{abstract}
  \paperabstract  
\end{abstract}
\maketitle

\pagenumbering{arabic}

\section{Introduction}\label{sec:intro}
Gluing techniques provide for a powerful method of geometric analysis (GA)
which may be exploited to combine multiple, distinct, solutions to a 
(system of) PDE of interest through their gradual deformation over 
some open set $\Omega$ so as to furnish a new, composite solution that
approximately coincides with the original solutions away from $\Omega$
\cite{:ref:mathematicalsampler2010chrusciel,%
  :ref:scalar2011corvino,:ref:antigravity2016chrusciel}. 

In this work we focus on the vacuum Einstein constraint equations
$\mathcal{C}^\mathrm{ADM}[g,\,K]=0$
\cite{:ref:initial2000cook,:ref:constraint2004bartnik,:ref:inival2005pfeiffer,%
  :ref:construction2007gourgoulhon,:ref:initial2017tichy}.
For concreteness, recall that the constraints split into the scalar Hamiltonian
constraint $\mathcal{H}^\mathrm{ADM}[g,\,K]=0$ and the vectorial momentum
constraint $\mathcal{M}{}_j^{\mathrm{ADM}}[g,\,K]=0$.  In general, these are to
be satisfied on a Riemannian manifold $\Sigma$ by a spatial metric $g_{ij}$,
together with extrinsic curvature $K_{ij}$.  From the perspective of the
(numerical) evolution problem, a triplet $(\Sigma,\,g_{ij},\,K_{ij})$
constitutes an initial data set.

Consider a moment-in-time (MIT) symmetry (where $K_{ij}=0$), such that the
constraints reduce to the single, scalar-flat condition $\mathcal{R}[g]=0$.  In
this setting, by exploiting a GA based technique of scalar curvature
deformation, Corvino~\cite{scalarcurvature2000corvino} has shown the following:
let $g$ be an arbitrary asymptotically flat metric on $\mathbb{R}^3$, satisfying
the Hamiltonian constraint $\mathcal{R}[g]=0$, with positive ADM-mass
$m_0$. Then there exists another asymptotically flat metric $\hat{g}$ satisfying
the constraint which agrees with $g$ on a compact set $K_1$ and is identical to
a Schwarzschild solution (with, in general, a different ADM-mass and a shifted
centre of mass) outside another compact set $K_2$ with $K_1 \Subset K_2$. Thus,
the new metric may be regarded as a composite metric obtained by gluing the
original metric to the Schwarzschild metric in the transition region
$\Omega:=K_2\backslash K_1$. The truly novel feature of this exterior asymptotic
gluing (EAG) construction is that the new composite initial data set exactly
coincides with its respective constituents outside the ``transition region''
$\Omega$ (which, as an example, may be imagined to be a spherical shell of
finite thickness). This possibility of local gluing, where the region over which
two initial data sets are spliced together is of compact support, is entirely
due to the underdeterminedness of the constraint equations. The assumption of
asymptotic flatness for the interior metric seems to have been made for
technical reasons in order to guarantee that the non-linear operator in question
is surjective. In the present work we stick with this assumption and glue only
metrics which are asymptotically flat. It would be interesting, however, to see
whether Corvino's approach would also allow us to glue arbitrary scalar-flat
metrics.

By relaxing the MIT condition and applying a similar gluing strategy
it has been shown that the Corvino-Schoen technique 
may also be used to glue to exact Kerr exteriors~\cite{asymptotics2006Corvino}.
Thus, quite general interior gravitational configurations may be glued 
to exterior Schwarzschild or Kerr regions forming a
composite solution with precise asymptotics\footnote{Sacrificing local control
on solution character whilst engineering asymptopia without gluing
has also been investigated~\cite{AsymptoticStaticity2011Avila}.}.

A striking variant of the above, where $\Omega$ is replaced by a conical region
of infinite extent, is Carlotto-Schoen gluing~\cite{:ref:localizing2016carlotto}
(see also~\cite{:ref:antigravity2016chrusciel}). In principle,
effective screening is allowed for by manipulation of vacuum initial data alone.
Furthermore, the Corvino-Schoen and Carlotto-Schoen gluings may
be utilised so as to construct $N$-body initial data 
sets~\cite{:ref:construction2011chrusciel}.

Related to the above is the method of connected sum 
or IMP gluing~\cite{:ref:gluing2002isenberg,:ref:gluing2005isenberg,%
  :ref:topology2003isenberg}.
Here the conformal (Lichnerowicz-York) framework is adopted and consequently 
a determined elliptic system results.
Given any two solutions of the constraints:
$(\check{\Sigma}_0,\,\check{g}_{ij},\,\check{K}_{ij})$ and 
$(\hat{\Sigma}_0,\,\hat{g}_{ij},\,\hat{K}_{ij})$, say,
a new solution may be produced by first removing small neighbourhoods 
$\check{N}$ and $\hat{N}$ about the points
$\check{p}\in \check{\Sigma}_0$ and $\hat{p}\in\hat{\Sigma}_0$ respectively.
Then, new data $(\Sigma_0,\,g_{ij},\,K_{ij})$ is found by
connecting $\partial\check{N}$ along an interpolating tube to $\partial\hat{N}$
with $\Sigma_0$ resulting in a connected sum manifold with the topology of 
$\check{\Sigma}_0\#\hat{\Sigma}_0$. By suitable interpolation of the pairs 
$(\check{g}_{ij},\,\check{K}_{ij})$ and $(\hat{g}_{ij},\,\hat{K}_{ij})$ a 
composite constraint satisfying solution $(\Sigma_0,\,g_{ij},\,K_{ij})$ may be 
found.

Alternatively, identifying $\check{\Sigma}_0$ and 
$\hat{\Sigma}_0$ allows for a handle (wormhole) to be
introduced to a given initial data set. On account of the determinedness
this leads to a global deformation of the initial data set which is small
away from the gluing site. By combining the results of
\cite{scalarcurvature2000corvino,:ref:existence2002chrusciel} it was shown in 
\cite{:reg:gluing2004chrusciel,:ref:initial2005chrusciel} 
how the deformation may be localised. To date there do not appear to have
been attempts made to prepare numerical initial data based on the IMP approach.

Aside from the ability to control asymptopia of initial data sets, 
engineering of exotic properties is interesting in its 
own right and, indeed, a robust scheme for fashioning numerical solutions 
could prove useful in endowing the associated space-time with a particular, 
desired phenomenology. For example, inspired by the result of
\cite{scalarcurvature2000corvino}, a potential path towards
minimisation of so-called ``spurious'' gravitational radiation content 
was provided in~\cite{corvino2005Giulini}. We refer the reader there for further
details.

While the above gluing results have intriguing properties, an unfortunate
aspect is that the GA flavour of proof technique
is quite technical in nature. It is not entirely clear how to proceed 
if direct numerical preparation of an initial data set based on such
methods is desired. This is evidenced by the fact that there exists only a 
single attempt~\cite{corvino2005Giulini} based on formal
perturbation theory to ``embed'' within the conformal framework
a problem that seeks to mimic the setup of Corvino's
result~\cite{scalarcurvature2000corvino}.

Briefly, the idea in~\cite{corvino2005Giulini} was to work at an MIT symmetry, 
assume axisymmetry and fix internal
Brill-Lindquist (BL) data. Then, over an annular $\Omega$
a conformally transformed Brill-Wave~\cite{:ref:positive1959brill} 
ansatz on the form of the conformal factor~$\psi$ is made. 
It was claimed that composite solutions exist to this problem when the 
exterior is a suitably chosen Schwarzschild initial data set.
This approach requires a further, \emph{ad hoc} treatment of
the decay rates of $\psi$ as $\partial \Omega$ is approached.
Following this programme, it appears that a numerical solution may be 
constructed~\cite{:ref:numericalconstruction2016doulis} (see however, the 
modified, Newton-Krylov based approach of~\cite{:ref:numerical2018pook_kolb}).

An additional insight is offered in~\cite{:ref:numericalconstruction2016doulis} 
as to how consistent selection of
exterior data (or parameters on valid internal data) may be made by 
exploitation of an integrability condition. Such arguments are not 
required in the proof of~\cite{scalarcurvature2000corvino}.
It does not appear that numerical evolution has been performed based on
the results of~\cite{:ref:numericalconstruction2016doulis,%
  :ref:numerical2018pook_kolb}.
Indeed, we are not aware of any numerical evolution
of initial data sets which have been prepared based on gluing techniques.

More broadly, the technique of scalar curvature deformation may be of potential
interest in studies involving geometric curvature flow.  Such flows were
introduced to general relativity in~\cite{:ref:energy1973geroch} and the idea
built upon in~\cite{:ref:positive1977jang} to rule out a class of
counterexamples to the cosmic censorship hypothesis proposed by Penrose
in~\cite{:ref:naked1973penrose} as encapsulated by an inequality relating black
hole (ADM) mass and the area of its apparent horizon. The veracity of this
inequality in a special case was first established rigorously in
\cite{:ref:inversemeancurvature2001huisken} by exploiting the inverse mean
curvature flow for MIT data sets. The more general problem without this
restriction remains open and numerical investigation utilising the weak
formulation approach of~\cite{:ref:inversemeancurvature2001huisken} may help
shed light on the matter.  The related Ricci
flow~\cite{:ref:threemanifolds1982hamilton} has also been studied numerically
\cite{:ref:critical2003garfinkle,:ref:visualizing2005rubinstein} where
in~\cite{:ref:critical2003garfinkle} preliminary evidence for critical behaviour
along the flow was presented.  Another potentially novel scenario to consider
may be whether scalar curvature deformation can be employed as a mechanism to
control the appearance of such critical behaviour.

Our goal in this work is to provide some insight as to how the proof
in~\cite{scalarcurvature2000corvino} may be more directly adapted to a numerical
technique itself without making use of the conformal programme for reformulation
of the constraint equations as
in~\cite{corvino2005Giulini,:ref:numericalconstruction2016doulis,%
  :ref:numerical2018pook_kolb}.
In so doing, we shall numerically construct initial data as composite solutions
with an MIT symmetry.  To begin, we elaborate upon Corvino's method at a formal
level in \S\ref{sec:CorvMeth}. Scalar curvature deformation over
$\Omega\Subset\Sigma$ and construction of a solution metric describing it, is
effected iteratively, through solution of a sequence of linear sub-problems. The
basic ingredient of this is described in \S\ref{sec:linweak}. How the iteration
is to proceed, together with an obstruction that occurs in the particular case
of solving the constraints themselves and our proposed remedy is detailed in
\S\ref{sec:nonlinloc}.

Having outlined the problem at the abstract level we next turn our attention to
providing a robust description of geometric quantities required for the problem
that is suitable for numerical work. To this end, a frame based approach is
introduced in \S\ref{sec:GeomPre}. In particular, $\Sigma$ is viewed as foliated
by topological $2$-spheres. With a view towards efficient numerical
implementation, the intrinsic geometry is described through the
$\eth$-formalism, which is briefly recounted in \S\ref{sec:intrinsicsw}.
Details on how it may be adapted to treat topological $2$-spheres are provided 
in \S\ref{sec:topotwosph}. For convenience, the relation between intrinsic and
ambient quantities adapted to our discussion is touched upon in
\S\ref{sec:sigdecomp}.

The success and versatility of pseudo-spectral
methods~\cite{:ref:spectral2009grandclement} motivates our numerical approach in
\S\ref{sec:NumPre}.  In particular, function approximation of intrinsic
quantities cast in the $\eth$-formalism is discussed in \S\ref{sec:fcnapprsig}.
Approximation of more general quantities over $\Sigma$ is described in
\S\ref{sec:funcSigAppr}. For the deformation problem at hand, we supplement the
discussion with some complex analytic considerations that can assist in
improving numerical solution quality in \S\ref{sec:cplxAnSec}.

With particulars of the physical problem and numerical technique fixed we
subsequently investigate prototype problems in \S\ref{sec:ProPro}. As an initial
test, the case of scalar curvature deformation in the context of spherical
symmetry is initially investigated in \S\ref{sec:sphsymred}, and self-consistent
convergence tests performed in \S\ref{sec:sphsymSCCT}.  Following this, a
relaxation to the class of axisymmetric problems is set up in
\S\ref{sec:axisymdeform} and explored in \S\ref{sec:axiToyProblem}.  In
\S\ref{sec:numgluebbh} all previously introduced material is brought together
and we perform gluing of internal BBH data (for Brill-Lindquist and Misner
initial data) to exterior Schwarzschild initial data. Numerical performance of
the approach together with properties of the physical construction are
investigated.  Finally \S\ref{sec:disc} concludes.

\section{Corvino's method}\label{sec:CorvMeth}
The argument for solving the Einstein constraints
$\mathcal{C}^{\mathrm{ADM}}[g,\,K]=0$ presented
in~\cite{scalarcurvature2000corvino} by virtue of exterior asymptotic gluing
(EAG) is quite technical in nature and consequently how one should proceed in
order to fashion a numerical technique is somewhat opaque.  Our goal here is to
provide a sketch of the idea adapted to the aforementioned context (cf. the
general discussions
of~\cite{generalrelativity2015Ashtekar,:ref:antigravity2016chrusciel,%
  :ref:mathematicalsampler2010chrusciel,:ref:scalar2011corvino}).
The physical setting is vacuum with vanishing cosmological constant at an MIT
symmetry $(K_{ij}=0)$ and in what follows $\Sigma$ is to be understood as an
initial Cauchy slice\footnote{Here particularised to
  $\mathrm{dim}(\Sigma)=3$.}. Under these assumptions
$\mathcal{C}^{\mathrm{ADM}}[g,\,K]=0$ reduces to the single, non-trivial,
scalar-flat condition $\mathcal{R}[g]=0$ and $(\Sigma,\,g)$ is sought.

To explain EAG and fix the desired behaviour of $g$ recall that asymptotically
flat (AF) data are characterised by the existence of a diffeomorphism between
the ``end'' of $\Sigma$ and $\mathbb{R}^3$ with a ball $\mathbb{B}$ removed.
Let $\delta_{\mathrm{Euc}}$ be the Euclidean metric. For AF $(\Sigma,\,g_E)$,
end coordinates $\{x^i\}_{i=1:3}$ may be introduced such that decay of $g_E$
(and derivatives thereof) to $\delta_{\mathrm{Euc}}$ is controlled by negative
powers of $|x|$
(see~\cite{generalrelativity2015Ashtekar,:ref:antigravity2016chrusciel}).

The result of~\cite{scalarcurvature2000corvino} concerns an equivalence
class of Schwarzschild initial data where a representative in isotropic form
is provided by:
\begin{equation}
  \begin{aligned}
  g_S =
  \left(1 + \frac{M_{\mathrm{ADM}}}{2|\mathbf{x}- \mathbf{C}|}
  \right)^4 \delta_{\mathrm{Euc}},
  \end{aligned}
\end{equation}
and the $(1+3)$-parameter tuple $(\mathrm{M}_{\mathrm{ADM}},\,C^i)$ describes
the ADM mass and centre of mass. We identify $\Sigma$ with its image in
$\mathbb{R}^3$. Let $\mathbb{B}_\rho\subset \Sigma$ be the ball of radius
$\rho>0$. Introduce the compactly contained domain
$\Sigma\Supset\Omega_\rho:=\mathbb{B}_{2\rho}\setminus \mathbb{B}_\rho$ the
closure of which is a spherical shell of thickness $\rho$ and serves as a
``transition region''.  A selection of sufficiently large $\rho$ allows one to
smoothly combine any AF $(\mathbb{B}_\rho,\,{g}_E)$ satisfying the scalar-flat
condition ${\mathcal{R}}[g_E]=0$ with
$(\Sigma \setminus \mathbb{B}_{2\rho},\,{g}_S)$ over $\Omega_\rho$ via judicious
selection of $g_S$. This latter is accomplished through tuning of the parameters
$(\mathrm{M}_{\mathrm{ADM}},\,C^i)$ and iterative correction of a smooth,
interpolating ``background metric'' $\overline{g}_{\Omega_{\rho}}$.  To 
understand the procedure, introduce the smooth cut-off function $\chi$ equal 
to $1$ on
$\mathbb{B}_\rho$ and $0$ outside $\mathbb{B}_{2\rho}$, and on $\Omega_\rho$
set:
\begin{equation}\label{eq:cutnpaste}
    \overline{g}_{\Omega_\rho}:= \chi g_E + (1-\chi)g_S.
\end{equation}
Clearly, ${\mathcal{R}}[\overline{g}_{\Omega_\rho}]=0$ on
$\Sigma\setminus \Omega_\rho$ whereas on $\Omega_\rho$ we have
${\mathcal{R}}[\overline{g}_{\Omega_\rho}] = \delta$ where $\delta$ 
is a compactly supported function. Furthermore, we shall assume that
${\mathcal{R}}[\overline{g}_{\Omega_\rho}]$ is non-constant on $\Omega_\rho$ to
avoid a technical issue outlined in~\S\ref{sec:appendix}.

To proceed further, the problem is now viewed as a local (i.e., compactly
supported) deformation of the scalar curvature.  Consider the change
$\delta\rightarrow
\check{\delta}:={\mathcal{R}}[\overline{g}_{\Omega_\rho}]+\tilde{\delta}$ for
$\tilde{\delta}$ sufficiently small.  The idea is to seek a suitable correction
$h$ to the background metric such that
${\mathcal{R}}[\overline{g}_{\Omega_\rho} + h] = \check{\delta}$ is satisfied.
The approach of~\cite{scalarcurvature2000corvino} is to linearise about the
background metric:
\begin{equation}
    {\mathcal{R}}[\overline{g}_{\Omega_\rho} + h]
    \simeq
    {\mathcal{R}}[\overline{g}_{\Omega_\rho}]
    + {L}_{\overline{g}_{\Omega_\rho}}[h],
\end{equation}
where the linear problem ${L}_{\overline{g}_{\Omega_\rho}}[h] = \tilde{\delta}$
is investigated so as to characterise properties of the underdetermined elliptic
operator $L_{\overline{g}_{\Omega_\rho}}$. Unfortunately,
${L}_{\overline{g}_{\Omega_\rho}}$ fails to be injective and the question of
surjectivity of ${L}_{\overline{g}_{\Omega_\rho}}$ is addressed with a
demonstration of injectivity of the formal adjoint
${L}^*_{\overline{g}_{\Omega_\rho}}$ which is overdetermined.  This
latter is then utilised, working within weighted function spaces
yielding a so-called ``basic estimate'' over $\Omega_\rho$ where a certain
growth (decay) rate of functions near $\partial\Omega_\rho$ is permitted.  Thus,
boundary behaviour of functions is implicitly controlled by a weight-function
$\omega$ allowing for a variational based solution to the above linear problem.
The control on properties of the linear solution turns out to be sufficiently
strong to also allow for a Picard iteration scheme to obtain the solution $h$ to
the nonlinear local deformation problem.

In principle, the problem of finding $h$ with
${\mathcal{R}}[\overline{g}_{\Omega_\rho} + h]=0$ may also be pursued in this
way. However, an obstruction exists in that an approach to the scalar-flat
condition, as for instance when
${\overline{g}}_{\Omega_\rho} \rightarrow {g}_S$, induces an approximate,
non-trivial kernel 
$\mathscr{K}[{\overline{g}_{\Omega_\rho}}]:=\ker({L}_{\overline{g}_{\Omega_\rho}}^*)$
of the formal linear adjoint.
This may be ameliorated through judicious selection of $(M_{\mathrm{ADM}}, C^i)$
so as to work in a space transverse to $\ker ({L}_{\delta_{\mathrm{Euc}}}^*)$
when solving the previously described variational problem at the linearised
level. Unfortunately, while~\cite{scalarcurvature2000corvino} demonstrates that
such a selection exists a method for \emph{a priori} specification of the 
parameters is not provided and hence we instead adopt a direct, numerical 
linear-algebraic strategy.

We now proceed to provide further details of the variational approach to 
solving the linearised problem in \S\ref{sec:linweak} with sufficient detail 
for our numerical scheme. The nonlinear deformation shall be addressed in 
\S\ref{sec:nonlinloc} together with our method for approaching the issue of
non-trivial kernel.

\subsection{Linear corrections via weak-formulation}\label{sec:linweak}

For the sake of exposition we shall assume geometric quantities to be defined
with respect to $\Omega:=\Omega_{\rho}$. Background quantities will be denoted
by an over-bar. Furthermore, we shall assume that $\overline{\mathcal{R}}$ is
non-constant.  Suppose that a sufficiently small, smooth, local deformation
(i.e. of compact support) of the scalar curvature is made
$\mathcal{R}[\overline{g}]\rightarrow
\mathcal{R}[g]:=\mathcal{R}[\overline{g}]+\tilde{\delta}$ and a metric
correction $h$ satisfying $g=\overline{g} + h$ is sought.  The problem may be
investigated perturbatively by noting that formally
$\overline{g}_{ij}\rightarrow g_{ij}= \overline{g}_{ij} + \varepsilon h_{ij}$
induces a corresponding linear-order correction to the scalar curvature
$\mathcal{R}[g+\varepsilon h]=\mathcal{R}[\overline{g}]+\varepsilon
L_{\overline{g}}[h]$ where standard methods
yield~\cite{:ref:generalrelativity1984wald}:
\begin{equation}\label{eq:scRpert}
    L_{\overline{g}}[h]
    = -h^{i j}\overline{\mathrm{Ric}}_{i j}
    -\overline{\nabla}^2\left[h^k{}_k \right] +  
    \overline{\nabla}_i \overline{\nabla}_j h^{i j}.
\end{equation}
Thus, solution of $L_{\overline{g}}[h]=\tilde{\delta}$ is required and hence
properties of the linear operator $L_{\overline{g}}$ are investigated
in~\cite{scalarcurvature2000corvino}.  While it turns out that
$L_{\overline{g}}$ is underdetermined elliptic one may instead work with the
formal $\mathrm{L}^2$ adjoint $\Lspop[\cdot]_{ij}$ as identified from the inner
product
$\langle \Lpop[h],\,f\rangle_{\mathrm{L}^2(\Omega)} =\langle
h,\,\Lspop[f]\rangle_{\mathrm{L}^2(\Omega)}$:
\begin{equation}\label{eq:LstformalAdjoint}
  \left(L_{\overline{g}}^*[f]\right)_{ij}
  =
  -\overline{\mathrm{Ric}}_{ij}f - \overline{g}_{ij}\overline{\nabla}^2\left[
    f\right] +
  \overline{\nabla}_{(i}\overline{\nabla}_{j)} [f],
\end{equation}
which is injective~\cite{scalarcurvature2000corvino} (see also
\S\ref{sec:appendix}).  Introduce the weighted, Sobolev space functional
$\mathcal{V}: H^2_\wei(\Omega)\rightarrow\mathbb{R}$ defined by:
\begin{equation}\label{eq:minimserProblem}
\mathcal{V}[u] = \int_\Omega \left(\frac{1}{2}\Vert L_{\overline{g}}^* u \Vert^2
-\tilde{\deform}u \right)\wei\,\d{\mu_{\overline{g}}},
\end{equation}
where $\tilde{\delta}\in \mathrm{L}^2_\wei(\Omega)$, $\wei$ is a weight function
(to be defined) and $\d{\mu}_{\overline{g}}$ is the integration measure induced
by $\overline{g}$.  To find the unique $u$
satisfying~\eqref{eq:minimserProblem}, we introduce the test-function
$\eta\in C^\infty_c(\Omega)$ and consider the variation:
\begin{equation}\label{eq:WeakForm}
0=\left.\frac{d}{dt}\left[ \mathcal{V}[u+t\eta]\right]\right|_{t=0} 
\quad \Longrightarrow \quad
\int_\Omega 
\left[L_{\overline{g}}^* \eta \right]^{ij}
\left[L_{\overline{g}}^* u \right]_{ij}\wei\,
\d{\mu_{\overline{g}}}
= \int_\Omega \eta \tilde{\deform}\wei\,\d{\mu_{\overline{g}}},
\end{equation}
where the compact support of $\eta$ (or alternatively a presumed decay rate for
$\omega$ towards $\partial\Omega$) enables us to drop all boundary
terms. Equation (\ref{eq:WeakForm}) is the so-called
weak-formulation~\cite{Partial1998Evans,:ref:multipleintegrals2009morrey,%
  :ref:elliptic2015gilbarg}
of the following strong-form problem~\cite{scalarcurvature2000corvino}):
\begin{equation}\label{eq:linSF}
L_{\overline{g}}\left[\wei L_{\overline{g}}^*\left[u\right] \right] 
= \tilde{\deform}\wei 
= \deform, \quad \left(\delta \in C^{k,\,\alpha}_{\omega^{-1}}(\Omega)\right),
\end{equation}
where $C^{k,\,\alpha}_{\omega^{-1}}(\Omega)$ is a weighted H\"older
space~\cite{scalarcurvature2000corvino}.  In light of this equivalence and later
use of Eq.\eqref{eq:WeakForm} to iteratively construct $h$ we shall refer to $u$
as a ``potential function''.  In order to ensure future enforcement of $u\in
H{}^2_{\omega}(\Omega)$ in Eq.\eqref{eq:WeakForm} at the numerical level ---
indeed allowing for controlled growth of $u$ towards $\partial\Omega$ --- an
explicit rewriting exploiting the decay properties of the weight term $\omega$
can be made and viewed as a solution ansatz. 

Suppose $x$ is a boundary defining function in a neighbourhood of
$\partial\Omega$, i.e., $x\ge0$ with $x=0$ and $\mathrm{d}x \ne0$  on
$\partial\Omega$. Suppose $\omega\sim x^{2N}$ for $N$ sufficiently large. The condition
$\partial{}^k_x [u]\in L^2_\omega(\Omega)$ where $k\in\{0,\,1,\,2\}$ 
leads to:
\begin{equation}\label{eq:potAnsatzu}
  \begin{aligned}
    u &= \mathcal{N}\tilde{u} \omega^\beta,&
    \beta &= - \frac{1}{2};
  \end{aligned}
\end{equation}
where $\mathcal{N}$ is a function with quadratic decay in $x$ towards
$\partial\Omega$ and $\tilde{u}$ shall be assumed to be bounded and smooth. We
shall defer explicit specification of $\omega$ to \S\ref{sec:ProPro}.

\subsection{Nonlinear local \texorpdfstring{$\mathcal{R}$}{R} deformation and gluing}
\label{sec:nonlinloc}
The problem we would now like to solve is: On $\Sigma$ fix a choice of 
$\overline{g}$ and hence 
$\overline{\mathcal{R}}$ and $L{}_{\overline{g}}$ together with $\delta$ of 
compact support on $\Omega$.
Assume that $\overline{\mathcal{R}}$ is non-constant.
Find a symmetric $2$-tensor $h\in\mathcal{S}_2(\Sigma)$ 
with compact support on $\Omega$
such that $\mathcal{R}[\overline{g}+h] = \mathcal{R}[\overline{g}] + \delta$.

Theorem $1$ of~\cite{scalarcurvature2000corvino} allows us to proceed as
follows: Set 
${}^{(0)}\deform:=\mathcal{R}[g] - \mathcal{R}[\overline{g}]$, solve
Eq.\eqref{eq:WeakForm} for ${}^{(0)}u$, and hence construct 
${}^{(0)}h_{ij}=\omega L{}_{\overline{g}}^*[{}^{(0)}u]_{ij}$. This yields
${}^{(1)}g_{ij} = \overline{g}_{ij} + {}^{(0)}h_{ij}$ with
${}^{(0)}h_{ij}$ small in an appropriate H\"{o}lder space. Now, it would be
natural to apply Newton's method and linearise about the new metric
${}^{(1)}g_{ij}$. However, it turns out~\cite{scalarcurvature2000corvino} that
this would apparently result in a loss of differentiability.
Instead\footnote{It would be interesting to see whether this analytical problem
  manifests itself also on the numerical level. However, we have not pursued
  this any further, yet.} the proof technique leverages Picard iteration with
the linearisation fixed at the background $\overline{g}_{ij}$ and the
approximate solution is iteratively improved via:
\begin{equation}\label{eq:nonlinearrefinement}
  {}^{(k)}\deform = \mathcal{R}[g_{ij}] - \mathcal{R}\left[\overline{g}_{ij}
    +\omega\sum^{k-1}_{m=0}{}^{(m)}\tilde{h}_{ij} \right],
\end{equation}
where ${}^{(m)}\tilde{h}_{ij}:=L_{\overline{g}}^*\left[{}^{(m)}u\right]$ and
the solution (metric) is given by $g_{ij}=\overline{g}_{ij} +
\lim_{N\rightarrow\infty}\omega \sum_{m=0}^N {}^{(m)}\tilde{h}_{ij}$.

An issue remains when $\overline{\mathcal{R}}$ is constant and correspondingly a
non-trivial kernel $\mathscr{K}[\overline{g}]:=\ker(L_{\overline{g}}^*)$ exists.
This must be addressed if exterior asymptotic gluing (EAG) to a time-symmetric
slice of Schwarzschild $g_S$ is to be achieved and a solution to the constraints
found.  Consider $\overline{g}_{\Omega_\rho}$ of Eq.\eqref{eq:cutnpaste}.  The
dimension of $\mathscr{K}[\overline{g}_{\Omega_\rho}]$ is related to the
underlying (approximate) symmetries of $\overline{g}_{\Omega_\rho}$ and due to
the assumption of $\overline{g}_{\Omega_\rho}$ being asymptotically
Euclidean must approach that of $\mathscr{K}[\delta_\mathrm{Euc}]$ in the
asymptotic regime (described below)
\cite{:ref:mathematicalsampler2010chrusciel,:ref:scalar2011corvino}.  It is also
known that in linearisation of the full vacuum constraints (no longer at an MIT)
a further contribution to the kernel of the corresponding formal linear adjoint
arises, which is comprised of the generators of translation and rotation of
$\mathbb{R}^3$
\cite{:ref:mathematicalsampler2010chrusciel,:ref:scalar2011corvino}.
Collectively, elements of the non-trivial kernel in this latter case are called
Killing initial data (KID) due to the one-to-one correspondence with Killing
vectors in the vacuum space-time obtained by evolving the initial data set
\cite{:ref:spacetime1975moncrief}.

Thus for EAG on Schwarzschild we clearly encounter an obstruction as $\rho$ is
increased due to the fall-off properties of $\overline{g}_{\Omega_\rho}$ and
approach to the non-trivial $\mathscr{K}[g_S]$.  As a preliminary, notice 
that we
can identify
$\mathscr{K}_0:=\mathscr{K}[\delta_{\mathrm{Euc}}]=\mathrm{span}(1,\,x^1,\,x^2,\,x^3)$.
This can be seen by observing Eq.\eqref{eq:LstformalAdjoint} implies:
\begin{equation}
    L_{\delta_{\mathrm{Euc}}}^*[f]{}^i{}_j = -\delta{}^i{}_j
    \partial^k[\partial_k[f]] + \partial{}^{i} \partial{}_{j}[f],
\end{equation}
and hence $L_{\delta_{\mathrm{Euc}}}^*[\cdot]_{ij}$ annihilates affine 
functions of the form $f=a + b_k x^k$ where $a\in\mathbb{R}$ and 
$b_k\in\mathbb{R}^3$.
That these are the only possible functions then follows from considering the
finite-dimensional space of initial data for Eq.\eqref{eq:lstSurfODE}.

Returning to EAG on Schwarzschild, note that we are only approximately 
approaching a non-trivial kernel.
To account for this~\cite{scalarcurvature2000corvino} proceeds by introducing
an approximating kernel $\mathscr{K}_*:= \zeta \mathscr{K}_0$ where $\zeta$
is a smooth, spherically symmetric bump function of compact support on 
$\Omega_\rho$.
The idea is then to solve a projected nonlinear local deformation problem
based on $\overline{g}_{\Omega_\rho}$ as above but
working with functions in the $\mathrm{L}^2(\Omega_\rho)$ orthogonal complement
of $\mathscr{K}_*$. When carried out at sufficiently large $\rho$ this yields a 
glued solution $g$ with $\mathcal{R}[g]\in \mathscr{K}_*$
and a choice of $(M_{\mathrm{ADM}},\,C^i)$ is shown to exist (though it is not 
demonstrated how to select these parameters \emph{a priori}) such that a $g$ 
with $\mathcal{R}[g]=0$ may be found.

For our numerical approach a slightly different strategy shall be pursued. 
An alternative way to construct an appropriately projected problem that 
treats non-trivial $\mathscr{K}_0$ indirectly is provided by linear 
algebraic techniques. 
The idea here is to consider evaluation of the weak-formulation statement
of Eq.\eqref{eq:WeakForm} with a suitably chosen dense, approximating
collection of test space and solution (trial) space functions. 
A singular value decomposition (SVD) of the ensuing linear system may then be
inspected and any (approximate) kernel directly removed 
\cite{:ref:appliedlinearalgebra2007shores}. 

During numerical calculations involving EAG on Schwarzschild (to be performed 
in 
\S\ref{sec:numgluebbh})
symmetry conditions shall be imposed. A precise identification of 
the dimension of the non-trivial kernel in this context may be motivated
as follows: Consider the affine functions $f$ as annihilated by
$L{}^*_{\delta_\mathrm{Euc}}[\cdot]_{ij}$. According to 
\cite{generalrelativity2015Ashtekar}
the parameters $a$ and $b_k$ entering $f$ as above primarily affect how 
$M_{\mathrm{ADM}}$ and centre of mass $C^i$ should be chosen in the composite 
(numerical) solution. Thus in a context with a high degree of symmetry the 
effective dimension of the kernel may be reduced.

\section{Frame-formalism treatment of \texorpdfstring{$\Sigma$}{Sigma}-geometry}
\label{sec:GeomPre}
With the physical problem and geometric preliminaries outlined in
\S\ref{sec:CorvMeth} we now turn our attention to concretising the formulation
for numerical work.  Given a $\Sigma$ with underlying symmetries a chart
selection exploiting this property allows for a description of geometric
quantities that can lead to more efficient numerical schemes (see
\S\ref{sec:NumPre}). In order to accomplish this in a robust fashion, such that
issues of regularity do not arise from the choice of coordinatisation we adopt a
frame based approach that leverages the so-called $\eth$-formalism.  It shall be
assumed that $\Sigma$ is endowed with metric $g_{ij}$ and that
$(\Sigma,\,g_{ij})$ may be smoothly foliated by a one-parameter family of
non-intersecting topological $2$-spheres $\mathbb{S}{}^2_{\rho}$ which are to be
viewed as the level surfaces of a smooth function
$\rho:\Sigma\rightarrow\mathbb{R}$.  Denote the Levi-Civita connection
associated with $g_{ij}$ by ${\nabla}$.

Following the standard prescription of ADM decomposition adapted to a spatial
manifold, the normalised 1-form $n_i=N{\nabla}_i[\rho]$ provides a normal to
$\mathbb{S}{}^2_{\rho}$. Recall that the ambient metric induces the metric
$\gamma_{ij}$ on the submanifolds $\mathbb{S}{}^2_{\rho}$ via
$g_{ij}=\gamma_{ij}+n_i n_j$ and gives rise to the projector
$\mathcal{P}{}^i{}_{j}:= \gamma{}^i{}_j-n^i n_j$.  Supplementation with
$\mathcal{N}{}^i{}_j:=n^i n_j$ allows for decomposition of type $(q,\,r)$ tensor
fields, collectively denoted $\mathcal{T}{}^q_r(\Sigma)$, into intrinsic and
normal parts. Introduce a smooth vector field $\rho^i\in\mathfrak{X}(\Sigma)$
satisfying $\rho^i {\nabla}{}_i[\rho]=1$. Then $\rho^i=N n^i+N^i$ where
$N^i=\mathcal{P}{}^i{}_j \rho^j$ and consequently the ambient metric may be
decomposed via:
\begin{equation}\label{eq:ambientMetricDec}
  g_{ij} = (1-N^{-2}N_K N^K)n_i n_j + 2 N^{-1} n{}_{(i} N_{j)}
  +\delta{}^I_i \delta{}^J_j \gamma_{IJ},
\end{equation}
where capital Latin indices take values in $\{2,\,3\}$ and here $\delta$ is the
Kronecker delta. 
\subsection{Intrinsic \texorpdfstring{$2$}{2}-geometry and spin-weight}
\label{sec:intrinsicsw}
In order to further adapt the intrinsic $\mathbb{S}{}^2_{\rho}$ part of 
the geometry we take the view 
of~\cite{beyer::2014::numerical,spectral2015Beyer,:ref:unifiedwigner1986dray}. 
Without going into too much detail we mention that the $\eth$-formalism is based
on the fundamental relationships between the unit 2-sphere $\mathbb{S}^2$, its 
frame bundle and the group $\mathrm{SO}(3)$\footnote{Strictly speaking its 
simply connected cover $SU(2)$ is more fundamental because it allows us to also
  describe spinorial quantities but for the present purpose it is enough to
  consider the vectorial aspects related to the rotation group
  $\mathrm{SO}(3)$.}. In what follows we regard the 2-sphere as the unit-sphere
equipped with the usual Euclidean metric. The bundle of frames over
$\mathbb{S}^2$ is diffeomorphic to the rotation group since every rotation
matrix consists of three orthonormal vectors which form an oriented basis of
$\mathbb{R}^3$. Interpreting the first vector as a point $\mathbf{e}$ on
$\mathbb{S}^2$, the other two vectors yield an orthonormal basis in the tangent
space $T_{\mathbf{e}}\mathbb{S}^2$. Keeping $\mathbf{e}$ fixed we see that all
frames at $\mathbf{e}$ are related by a 2-dimensional rotation, i.e., an element
of $\mathrm{SO}(2)$. It is easily seen that this correspondence between frames
on the 2-sphere and a rotation matrix is bijective and that it allows us to
regard the $2$-sphere as the factor space
$\mathrm{SO}(3) / {\mathrm{SO}(2)}$. The projection map
$\pi : \mathrm{SO}(3)\rightarrow \mathbb{S}^2$ is called the Hopf map.

Every tensor field defined at a point $\mathbf{e} \in \mathbb{S}^2$ can be
decomposed into components with respect to a basis in
$T_{\mathbf{e}}\mathbb{S}^2$ and we may regard these components as functions
defined at a particular point on $\mathrm{SO}(3)$. Since they are components of
a tensor field they change in a very characteristic way when we change the basis
in $T_{\mathbf{e}}\mathbb{S}^2$. In this way we can describe every tensor field
on the sphere by a set of functions with special behaviour under change of
basis. By regarding this set as a whole we have eliminated the need for
referring to a particular choice of basis on the 2-sphere. This is the main
advantage in this formalism since it is well known that there are no globally
well defined frames on $\mathbb{S}^2$ --- a fact, which creates many problems 
for numerical simulations involving the 2-sphere.

Next, we introduce appropriate Euler angles for rotations and polar coordinates
on the 2-sphere so that we can express these well defined global relationships
in local coordinates. The Hopf map can be expressed in these coordinates as
$\pi: (\theta,\,\psi,\,\phi)\mapsto (\vartheta,\,\varphi)=(\theta,\,\phi)$.

Consider the open subset $U\subset\mathbb{S}^2$ away from the poles
($\vartheta=0,\,\pi$) such that the Hopf map with respect to the given
coordinates is well-defined. A smooth (real) orthonormal frame
$\hat{\mathbf{e}}_{(I)}$ on $U$ may be introduced where the parentheses indicate
distinct frame fields. Define the complex field
$\mathbf{m}:=\left(\hat{\mathbf{e}}_{(2)} + i \hat{\mathbf{e}}_{(3)}\right)/\sqrt{2}$.
In terms of this complex linear combination we can express the action of
$\mathrm{SO}(2)$ as multiplication with a phase
$\mathbf{m}\mapsto\mathbf{m}' = e^{i\psi} \mathbf{m}$ inducing rotation of the
complex frame $(\mathbf{m},\,\overline{\mathbf{m}})$ together with its
dual coframe ($\boldsymbol{\omega},\,\overline{\boldsymbol{\omega}})$ which
leads to the notion of spin-weight~\cite{beyer::2014::numerical}. Given a
smooth tensor field $T\in\mathcal{T}{}^q_r(U)$ its (equivalent)
spin-weighted representation is provided by:
\begin{equation}\label{eq:swrepr}
  \sw{s}{T} := T(
  \underbrace{\boldsymbol{\omega},\,\cdots \boldsymbol{\omega}}_{
    \mbox{\small$q_1$ times}};\,
  \underbrace{\overline{\boldsymbol{\omega}},\,\cdots,\,
              \overline{\boldsymbol{\omega}}}_{
    \mbox{\small$q_2$ times}};\,
  \underbrace{\mathbf{m},\,\cdots \mathbf{m}}_{
    \mbox{\small$r_1$ times}};\,
  \underbrace{\overline{\mathbf{m}},\,\cdots,\,
              \overline{\mathbf{m}}}_{
    \mbox{\small$r_2$ times}}
  ),
\end{equation}
where $s=r_1-r_2-q_1+q_2$ is the spin-weight. 

The unit-sphere metric $\overset{\circ}{\gamma}$ on $\mathbb{S}^2$ when
expressed in these coordinates acquires the form:
\begin{equation}\label{eq:2sphmetric}
    \begin{aligned}
    \overset{\circ}{\gamma} &= \overset{\circ}{\gamma}_{IJ}
    \d{x}^I\otimes \d{x}^J =
    \d{\vartheta}\otimes \d{\vartheta} +
    \sin^2\vartheta \,\d{\varphi}\otimes \d{\varphi},
    \end{aligned}
\end{equation}
where the choice of coordinatisation\footnote{This selection is made here 
to align with the numerical scheme
we utilise in \S\ref{sec:NumPre}. 
An equivalent construction may be performed in (for example) complex 
stereographic coordinates 
\cite{:ref:ethformalism1997gomez,:ref:solvingconstraints2016racz,%
  :ref:spin1967goldberg,:ref:spinorsandspacetimevoli1987penrose}.} entails that
orthonormal frame vectors may be selected as
$\hat{\mathbf{e}}_{(2)}=\partial_\vartheta$
and $\hat{\mathbf{e}}_{(3)}=\csc\vartheta\,\partial_\varphi$. The 
associated complex reference (co)frame becomes:
\begin{equation}\label{eq:stdCplFrame}
    m^I=\frac{1}{\sqrt{2}}\left(
        \partial_\vartheta^I - i\csc\vartheta\,\partial_\varphi^I
    \right)
    \Longrightarrow
    \omega_I=
        \frac{1}{\sqrt{2}}\left(
        \d{\vartheta}_I + i \sin\vartheta\,\d{\varphi}_I
    \right),
\end{equation}
subject to the complex orthonormality conditions\footnote{To avoid later
confusion we shall keep $\mathbf{m}$ and $\boldsymbol{\omega}$ distinct
as $\overset{\circ}{\gamma}$ shall be demoted from the status of a metric.}:
\begin{equation}\label{eq:orthoCond}
    \begin{aligned}
        m^I\overline{\omega}_I =\,& 0, &
        m^I \omega_I =& 1; &
        m^I m^J \overset{\circ}{\gamma}{}_{IJ} =&0, &
        \omega_I \omega_J \overset{\circ}{\gamma}{}^{IJ} =\,&0; &
        m^I \overline{m}^J \overset{\circ}{\gamma}{}_{IJ} =\,&1, &
        \omega_I \overline{\omega}_J \overset{\circ}{\gamma}{}^{IJ} =\,&1.
    \end{aligned}
\end{equation}
On account of \cref{eq:2sphmetric,eq:stdCplFrame,eq:orthoCond} we thus have:
\begin{equation}
  \begin{aligned}
    \overset{\circ}{\gamma}{}_{IJ} =&
    2\omega_{(I}\overline{\omega}_{J)},&
    \overset{\circ}{\gamma}{}^{IJ} =&
    2 m{}^{(I} \overline{m}{}^{J)}, &
    \overset{\circ}{\gamma}{}^I{}_J =&
    \overline{m}{}^I \overline{\omega}{}_J
    + \omega{}_J m{}^I = \delta{}^I_J.
  \end{aligned}
\end{equation}
Denote the Levi-Civita connection associated with $\overset{\circ}{\gamma}$ by
$\nabo$. We will now use this to define derivative operators which are adapted
to the notion of spin-weight, mapping spin-weighted quantities to spin-weighted
quantities. Suppose $\sw{s}{f}$ is a spin-weighted quantity in the sense of
Eq.\eqref{eq:swrepr}. We define the $\eth$ operators as components of the
corresponding tensor field in the direction of $\mathbf{m}$ and 
$\overline{\mathbf{m}}$ as follows:
\begin{equation}\label{eq:dirFraDer}
    \begin{aligned}
        \mathbf{m}\left[\sw{s}{f}\right] =\,&
        m^I \nabo_I\left[\sw{s}{f}\right] =
        \frac{1}{\sqrt{2}}\edth{\sw{s}{f}} +
        \sw{s}{f}s\overset{\circ}{\Gamma},&
        \overline{\mathbf{m}}\left[\sw{s}{f}\right] =\,&
        \overline{m}^I \nabo_I\left[\sw{s}{f}\right] =
        \frac{1}{\sqrt{2}}\edtb{\sw{s}{f}} -
        \sw{s}{f} s \overset{\circ}{\Gamma};
    \end{aligned}
\end{equation}
where $\overset{\circ}{\Gamma}=m^I\overline{m}^J\nabo_J\left[\omega_I\right]$
and for the choice of \cref{eq:2sphmetric,eq:stdCplFrame} we have 
$\overset{\circ}{\Gamma} = \cot\vartheta / \sqrt{2}$ together with:
\begin{equation}\label{eq:coordinatEthSW}
    \begin{aligned}
        \edth{\sw{s}{f(\vartheta,\varphi)}} 
        =\,&
        (\sin\vartheta)^s
        \left(\partial_\vartheta-i\csc\vartheta\,\partial_\varphi \right)
        \left[(\sin\vartheta)^{-s}\sw{s}{f(\vartheta,\varphi)}\right],\\
        \edtb{\sw{s}{f(\vartheta,\varphi)}} 
        =\,&
        (\sin\vartheta)^{-s}
        \left(\partial_\vartheta + i\csc\vartheta\,\partial_\varphi \right)
        \left[(\sin\vartheta)^s \sw{s}{f(\vartheta,\varphi)}\right].
    \end{aligned}
\end{equation}
Explicit translation formulae for covariant derivatives may be arrived at
by virtue of Eq.\eqref{eq:dirFraDer}
(see appendix of~\cite{:ref:solvingconstraints2016racz},
but note conventions differ):
\begin{equation}\label{eq:symEdthProj}
    \begin{aligned}
        \sqrt{2}m^J m^{I_1}\cdots m^{I_n} \nabo_J\left[
            W_{(I_1\cdots I_n)}
        \right] =\,&
        \edth{\sw{n}{W}},&
        \sqrt{2}\overline{m}^J m^{I_1}\cdots m^{I_n} \nabo_J\left[
            W_{(I_1\cdots I_n)}
        \right] =
        \edtb{\sw{n}{W}}.
    \end{aligned}
\end{equation}
Finally, we note that if the tensor field $W$ is real then under complex 
conjugation $\sw{+s}{W^*}=\sw{-s}{W}$ and furthermore the operator actions
$\eth\leftrightarrow \overline{\eth}$ are interchanged.

\subsection{Topological \texorpdfstring{$2$}{2}-spheres}\label{sec:topotwosph}
In order to relax our treatment to more general geometries the
assumption of \S\ref{sec:intrinsicsw} shall be modified and instead we shall 
consider working with a manifold $\mathbb{S}^2_\rho$ which is diffeomorphic to
$\mathbb{S}^2$ but is equipped with a different metric. The approach we follow 
is based on~\cite{:ref:ethformalism1997gomez,:ref:solvingconstraints2016racz} 
and hence we shall only briefly summarise the idea here.

Consider the manifold $(\mathbb{S}{}^2_{\rho},\,\gamma{}_{IJ})$ endowed with
metric:
\begin{equation}\label{eq:topoS2metric}
  \gamma{}_{IJ} =
  \sw{-2}{\gamma} \overline{\omega}_I \overline{\omega}_J
  + 2\,\sw{0}{\gamma} \omega_{(I}\overline{\omega}_{J)}
  + \sw{+2}{\gamma} \omega_I \omega_J,
\end{equation}
where the coframe is that of Eq.\eqref{eq:stdCplFrame} 
and the expression follows
from consideration of the irreducible decomposition of a type $(0,\,2)$
tensor field
\cite{:ref:ethformalism1997gomez,:ref:spinorsandspacetimevoli1987penrose}.
The $\overset{\circ}{\gamma}$ of Eq.\eqref{eq:2sphmetric} is now demoted to
the status of an auxiliary field. Thus, to be explicit, while the conditions of 
Eq.\eqref{eq:orthoCond} continue to hold, indicial manipulations of tensorial
quantities are now to be 
performed with $\gamma{}_{IJ}$. The inverse
metric is given by:
\begin{equation}\label{eq:topoS2metricInv}
    \gamma^{IJ}=\sw{0}{\tilde{\gamma}}
    \left(
    -\sw{-2}{\gamma}m^I m^J +
    2\sw{0}{\gamma}m^{(I}\overline{m}^{J)}
    -\sw{+2}{\gamma}\overline{m}^I \overline{m}^J
    \right),
\end{equation}
where~\cite{:ref:ethformalism1997gomez}:
\begin{equation}\label{eq:topoSWdetTerm}
    \sw{0}{\tilde{\gamma}} = 
    \left(\sw{0}{\gamma}^2 - \sw{-2}\gamma\sw{+2}\gamma\right)^{-1}
    = \det(\overset{\circ}{\gamma}_{IJ}) /
    \det\left(\gamma_{IJ}\right).
\end{equation}
Note that:
\begin{equation}\label{eq:covFrTopo}
    \begin{aligned}
        \overline{\omega}^I=\gamma^{IJ}\omega_J =\,& \sw{0}{\tilde{\gamma}}
        \left( -\sw{-2}{\gamma}m^I  + \sw{0}{\gamma} \overline{m}^I\right), &
        \omega^I=
        \gamma^{IJ}\overline{\omega}_J =\,& \sw{0}{\tilde{\gamma}}
        \left(\sw{0}{\gamma} m^I - \sw{+2}{\gamma}\overline{m}^I \right).
    \end{aligned}
\end{equation}
If coordinate components $(\vartheta,\,\varphi)$ are specified as:
\begin{equation}\label{eq:protoTopoMetr}
    \gamma_{IJ} =
    \begin{bmatrix}
        \Gamma_1 & \Gamma_2\\
        \Gamma_2 & \Gamma_3 \sin^2\vartheta
    \end{bmatrix},
\end{equation}
then comparison with Eq.\eqref{eq:stdCplFrame} and Eq.\eqref{eq:topoS2metric}
yields:
\begin{equation}\label{eq:swTopoMetrCoeff}
    \begin{aligned}
        \sw{-2}{\gamma} =\,&
        \frac{1}{2}(\Gamma_1-\Gamma_3+2i\Gamma_2\csc\vartheta),&
        \sw{0}{\gamma} =\,&
        \frac{1}{2}(\Gamma_1 + \Gamma_3),&
        \sw{+2}{\gamma} =\,&
        \frac{1}{2}(\Gamma_1-\Gamma_3-2i\Gamma_2\csc\vartheta);
    \end{aligned}
\end{equation}
and $\Gamma_i\in\mathbb{R}$ implies
$\sw{-2}{\gamma} = \sw{+2}{\gamma^*}$. In order to complete the
ingredients for a manifestly
regular treatment of quantities in $\mathcal{T}{}^q_r(\mathbb{S}^2_\rho)$
we require description of the covariant derivative operator in this new setting.
Denote the Levi-Civita connection associated with
$(\mathbb{S}{}^2_{\rho},\,\gamma_{IJ})$ by $\nab$. Then
$\nab$ and $\nabo$ may be
uniquely related 
by introduction of a $(1,\,2)$ tensor field
$C{}^K{}_{IJ}$~\cite{:ref:generalrelativity1984wald}:
\begin{equation}\label{eq:gamDiff}
    C^K{}_{IJ} = C^K{}_{(IJ)} = \frac{1}{2}\gamma^{KL}
    \left(
        \nabo_I[\gamma_{LJ}] + \nabo_J[\gamma_{IL}] - \nabo_L[\gamma_{IJ}]
    \right).
\end{equation}
The tensor field $C^K{}_{IJ}$ arises as the difference between Christoffel 
symbols associated with each connection and consequently in evaluating the
action of $\nab$ on a given field the pattern of additional terms matches
that of the usually required $\Gamma{}^K{}_{IJ}$. For example,
let 
$V\in\mathcal{T}{}^0_1(\mathbb{S}^2_{\rho})$
and
$T\in\mathcal{T}{}^0_2(\mathbb{S}^2_{\rho})$
then:
\begin{equation}\label{eq:covCor}
    \begin{aligned}
        \nab_I[V_J] =\,& \nabo_I[V_J] - C^K{}_{IJ}V_K,\\
        \nab_I[T_{JK}] =\,& \nabo_I[T_{JK}] - C^L{}_{IJ}T_{LK} - 
        C^L{}_{IK}T_{JL}.
    \end{aligned}
\end{equation}
Furthermore,
$C{}^K{}_{IJ}=\gamma{}^{KL}C{}_{LIJ}$
may itself be described in terms of spin-weighted
components of the metric $\gamma$ and $\eth$ derivatives thereof
via projection exploiting Eq.\eqref{eq:swrepr} and Eq.\eqref{eq:symEdthProj}:
\begin{equation}\label{eq:ethCorrCoeff}
    \begin{aligned}
        2\sqrt{2}C_{KIJ} =\,&
        \left(
            \edth{\sw{+2}{\gamma}}\omega_K +
            \left(2\edth{\sw{0}{\gamma}} -
            \edtb{\sw{+2}{\gamma}}\right)\overline{\omega}_K
        \right)\omega_I \omega_J
        +2\left(
            \edtb{\sw{+2}{\gamma}}\omega_K +
            \edth{\sw{-2}{\gamma}}\overline{\omega}_K
        \right)\omega_{(I}\overline{\omega}_{J)}\\
        \,&
        \left(
            \left(2\edtb{\sw{0}{\gamma}} - \edth{\sw{-2}\gamma}\right)\omega_K+
            \edtb{\sw{-2}{\gamma}}\overline{\omega}_K
        \right)\overline{\omega}_I\overline{\omega}_J.
    \end{aligned}
\end{equation}
Thus translation formulae for construction of manifestly
regular expressions (in the sense of coordinates) may also be derived in the
present context for
description of Eq.\eqref{eq:covCor} and more general
geometric quantities. In particular, see~\cite{:ref:ethformalism1997gomez}
for explicit calculations and expressions involving the scalar curvature 
$\mathcal{R}[\gamma]$.

\subsection{\texorpdfstring{$\Sigma$}{Sigma} decomposition}\label{sec:sigdecomp}
A frame formalism based description of geometric quantities
exploiting the preferred selections made in
\S\ref{sec:intrinsicsw} and \S\ref{sec:topotwosph}
may now be constructed as follows. An element of the foliation of $\Sigma$ by
$\mathbb{S}{}^2_{\rho}$ is fixed by selecting some $\rho_0$ wherein local
coordinates $x^I=(\vartheta,\,\varphi)$ may be chosen. These coordinates
may then be Lie dragged along the integral curves of $\rho^i$ to other leaves
of the foliation 
\cite{:ref:solvingconstraints2016racz,:ref:numericalrelativity2010baumgarte}
resulting in $x^i=(\rho,\,\vartheta,\,\varphi)$.
The preferred orthonormal complex (co)frame introduced in
Eq.\eqref{eq:stdCplFrame} is extended as
$\omega_i:=\delta^I_i \omega_I$ and $m^i:=\delta^i_I m^I$
and further supplemented with
$\hat{\mathbf{e}}_{(1)}=\partial_\rho$ which allows for spin-weighted 
decomposition of fields in $\mathcal{T}{}^q_r(\Sigma)$.

We briefly demonstrate how this pieces together in decomposition of the
ambient metric of $(\Sigma,\,g_{ij})$.
The normalisation condition on $n_i$ together with the fact 
that $N^i$ is an intrinsic vector leads to:
\begin{equation}
    \begin{aligned}\label{eq:normShiftAdapted}
        n^i =\, & \left(N^{-1},\, -N^{-1} N^I \right), &
        n_i =\, & \left(N,\,\mathbf{0} \right), &
        N^i =\, & \left(0,\,N^I\right), &
        N_i =\,&
        \left(\gamma_{IJ} N^I N^J,\, \gamma_{IJ} N^J \right),
    \end{aligned}
\end{equation}
which may be written by virtue of \cref{eq:swrepr,eq:stdCplFrame,eq:orthoCond} 
as:
\begin{equation}
    \begin{aligned}
        n^i =&
        -\frac{1}{\sw{0}{N}}\left(
            -1,\,
        \sw{+1}{\tilde{N}} \overline{m}^I
        + \sw{-1}{\tilde{N}} m^I
        \right),&
        n_i =&
        \left(\sw{0}{N},\,\mathbf{0}\right);
    \end{aligned}
\end{equation}
where we have set $\sw{0}{N}:=N$.
Similarly,
\begin{equation}\label{eq:shiftDec1}
    \begin{aligned}
        N_I =\,& \sw{-1}{N} \overline{\omega}_I + \sw{+1}{N}\omega_I,&
        N^I =\,& \sw{+1}{\tilde{N}} \overline{m}^I 
        + \sw{-1}{\tilde{N}} m^I;
    \end{aligned}
\end{equation}
where:
\begin{equation}\label{eq:swshiftadapted}
    \begin{aligned}
        \sw{-1}{N} =\,& N_I \overline{m}^I, &
        \sw{+1}{N} =\,& N_I m^I;\\
        \sw{-1}{\tilde{N}} :=\,&
        \sw{0}{\tilde{\gamma}}\left(
            \sw{-1}{N}\sw{0}{\gamma} - \sw{+1}{N}\sw{-2}{\gamma}
        \right),&
        \sw{+1}{\tilde{N}} :=\,&
        \sw{0}{\tilde{\gamma}}\left(
            -\sw{-1}{N}\sw{+2}{\gamma} + \sw{+1}{N}\sw{0}{\gamma}
        \right).
    \end{aligned}
\end{equation}
We expand $g_{ij}$ (or indeed any covariant symmetric tensor field) with 
respect to the 
coframe as:
\begin{equation}\label{eq:symtwotenproj}
  g_{ij} = \mathfrak{g}_{\rho\rho}\,
  \d{\rho}_i\d{\rho}_j
  +2\sw{-1}{\mathfrak{g}}\,\d{\rho}_{(i} \overline{\omega}_{j)}
  +2\sw{+1}{\mathfrak{g}}\,\d{\rho}_{(i} \omega_{j)}
  +\sw{-2}{\mathfrak{g}}\,\overline{\omega}_i \overline{\omega}_j
  +2 \sw{0}{\mathfrak{g}}\,\omega_{(i} \overline{\omega}_{j)}
  +\sw{+2}{\mathfrak{g}}\,\omega_{i}\omega_{j},
\end{equation}
and with Eq.\eqref{eq:ambientMetricDec} it is found that:
\begin{equation}\label{eq:metrspF}
  \begin{aligned}
  \mathfrak{g}_{\rho\rho} =&
    \sw{0}{N}^2 + \sw{-1}{\tilde{N}}\sw{+1}{N}
    + \sw{+1}{\tilde{N}}\sw{-1}{N}, &
    \sw{\pm2}{\mathfrak{g}} &= \sw{\pm2}{\gamma}, &
    \sw{\pm1}{\mathfrak{g}} &= \sw{\pm1}{N}, &
    \sw{0}{\mathfrak{g}} &= \sw{0}{\gamma}.
  \end{aligned}
\end{equation}
A similar, though more laborious approach of projection may be used to find
explicit decompositions for more general elements of 
$\mathcal{T}{}^q_r(\Sigma)$. In particular, 
Eq.\eqref{eq:covCor} and Eq.\eqref{eq:ethCorrCoeff} lead to a representation
of the action of the ambient Levi-Civita connection 
$\nabla:\mathcal{T}{}^q_r(\Sigma)\rightarrow \mathcal{T}{}^q_{r+1}(\Sigma)$ in 
terms of the (complex) frame and thus spin-weighted components together with
terms involving the extrinsic curvature 
$K_{IJ}:=\frac{1}{2}\pounds_{\mathbf{n}}[\gamma_{IJ}]$. 
Consequently manifestly regular, frame representations of the formal, linear 
adjoint 
$L_{\overline{g}}^*[\cdot]_{ij}$ 
appearing in Eq.\eqref{eq:LstformalAdjoint} and indeed all related, required 
quantities for the scalar curvature deformation problem may be constructed
\cite{:ref:numericalscalar2018daszuta}.

\section{Numerical method}\label{sec:NumPre}
In considering the numerical solution of the deformation problem described in
\S\ref{sec:CorvMeth} as adapted to the frame-formalism of \S\ref{sec:GeomPre} we
exploit (pseudo)-spectral (PS) methods
\cite{:ref:spectralmatlab2000trefethen,:ref:spectralmethods2007canuto,%
  :ref:spectralmethods2007hesthaven,:ref:chebyshev2013boyd}
as they give rise to highly efficient techniques for solution approximation as
the differentiability class of functions increases. In brief, the idea is that
given a square-integrable function $f\in L^2(\Omega)$, global approximation of
$f$ over $\Omega$ is made by truncating a representation of $f$ in terms of a
suitably chosen complete orthonormal basis $(\Phi_n){}_{n=0}^\infty$ of
$L^2(\Omega)$ as $\tilde{f}:= \sum_{n=0}^N f_n \Phi_n$. Numerical derivatives
may thus be evaluated directly through their action on basis functions or
embedded via recursion relations involving the expansion coefficients
$(f_n){}_{n=0}^N$. The details of how the approximation is enforced and hence
how the aforementioned coefficients are to be selected is controlled through a
choice of test functions $(\Psi_n){}_{n=0}^N$ and the inner product associated
with the natural function space for $f$.

\subsection{Function approximation on \texorpdfstring{$\mathbb{S}{}^2_\rho$}{S2rho}}\label{sec:fcnapprsig}
In order to numerically treat functions over $\Sigma$ we begin by considering
(as in \S\ref{sec:topotwosph}) the submanifold with metric
$(\mathbb{S}^2_{\rho^*},\,\gamma_{IJ})$ of $\Sigma$ where $\rho^*$ has been
fixed. We shall assume square integrability with respect to the measure induced
by $\overset{\circ}{\gamma}$ (Eq.\eqref{eq:2sphmetric}) for sufficiently regular
scalar fields $\sw{0}{f}:=f$ or more generally, upon projection via
Eq.\eqref{eq:swrepr} and Eq.\eqref{eq:stdCplFrame}, spin-weighted components
$\sw{s}{f}$ of tensor fields $f\in\mathcal{T}{}^q_r(\mathbb{S}^2_{\rho^*})$.
Leveraging the well-known spin-weighted spherical harmonics (SWSH)
$(\sw{s}{Y_{lm}})_{l,m}$ allows us to write
\cite{:ref:spin1967goldberg,:ref:spinorsandspacetimevoli1987penrose}:
\begin{equation}\label{eq:sfappr}
  \sw{s}{f(\vartheta,\,\varphi)}
  = \lim_{L\rightarrow \infty}\sum_{l=|s|}^L \sum_{m=-l}^{l}
  \sw{s}{f_{lm}}\,\sw{s}{Y_{lm}(\vartheta,\,\varphi)},
\end{equation}
which converges in the $L^2$ sense described in~\cite{beyer::2014::numerical}. 
The band-limit $L$ appearing in Eq.\eqref{eq:sfappr} is fixed at some finite
value to provide a truncated approximation $\sw{s}{\tilde{f}}$. On account of
the SWSH orthonormality relation (note commensurate $s$)
\cite{:ref:spin1967goldberg,:ref:spinorsandspacetimevoli1987penrose}:
\begin{equation}\label{eq:sYlmOrth}
    \left\langle {{}_sY_{l_1m_1}},\,{{}_sY_{l_2m_2}}\right\rangle
    =
    \int_0^{2\pi}\int_0^\pi
    \sw{s}{Y_{l_1m_1}(\vartheta,\varphi)} 
    \overline{\sw{s}{Y_{l_2m_2}(\vartheta,\varphi)}} 
    \sin\vartheta\,\d{\vartheta}\,\d{\varphi}
    =
    \delta_{l_1l_2}\delta_{m_1m_2}.
\end{equation}
an invertible map $\mathcal{F}: \sw{s}{f} \mapsto \sw{s}{f_{lm}}$ may be
constructed allowing one to transform between nodal (sampled function) and modal
(coefficient) descriptions of an approximated function. Due to
$\sw{0}{Y}_{00}(\vartheta,\,\varphi)=1/\sqrt{2}$ \cite{beyer::2014::numerical},
the relation of Eq.\eqref{eq:sYlmOrth} together with $\mathcal{F}$ also may be
viewed as supplying a general quadrature rule for functions of spin-weight $0$.

During the course of our numerical work, the fast Fourier transformation (FFT)
based algorithm of~\cite{beyer::2014::numerical} is used to compute
$\mathcal{F}$ and its inverse with an overall algorithmic complexity of
$\mathcal{O}(L^3)$. In this approach the function
$\sw{s}{f}(\vartheta,\,\varphi)$ is sampled on a finite, product grid in
$\vartheta$ and $\varphi$ of uniform spacing, and subsequently, data is
periodically extended -- the details of which are controlled by the value of $s$
and the choice of $L$. With a view towards later imposition of axisymmetry in
\S\ref{sec:axisymdeform} note that if the $\varphi$ dependence appearing in
$\sw{s}{f(\vartheta,\,\varphi)}$ is trivial\footnote{Or equivalently only the
  $m=0$ mode need be considered in Eq.\eqref{eq:sfappr} and the angular
  dependence of the SWSH becomes trivial~\cite{:ref:spin1967goldberg} motivating
  the definition
  $\sw{s}{Y}_{l}(\vartheta):=\sw{s}{Y_{l0}(\vartheta,\,\varphi)}$.} then
algorithmic complexity may be further improved in accordance with the usual
$\mathcal{O}(L\log L)$ scaling associated with a one-dimensional FFT
\cite{:ref:fourier1965cooley}. While it is straightforward to modify the SWSH
transformation algorithm of~\cite{beyer::2014::numerical} such that the sampling
in the $\varphi$ direction is reduced or varied adaptively, a few
points\footnote{With the $\varphi$ sampling of~\cite{beyer::2014::numerical} the
  sum over $m$ in Eq.\eqref{eq:sfappr} may be reduced to
  $|m|\leq \min(L_\varphi,\,l)$ where $L_\varphi:=|s|$.} must be sampled on
account of various auxiliary quantities appearing in the implementation.

The transformation gives us the freedom to describe functions in two ways. This
freedom is crucial for our scheme insofar as the $\eth$ and $\overline{\eth}$
operators introduced in \S\ref{sec:intrinsicsw} when evaluated on SWSH reduce to
an algebraic action~\cite{beyer::2014::numerical} (see also Eq.(4.15.122)
of~\cite{:ref:spinorsandspacetimevoli1987penrose}):
\begin{equation}\label{eq:ethswshreduction} \begin{aligned}
    \eth\left[\sw{s}{Y_{lm}(\vartheta,\varphi)}\right]
    &=-\sqrt{(l-s)(l+s+1)} \sw{s+1}{Y_{lm}(\vartheta,\varphi)},\\
    \overline{\eth}\left[\sw{s}{Y_{lm}(\vartheta,\varphi)}\right] 
    &=
    \sqrt{(l+s)(l-s+1)} \sw{s-1}{Y_{lm}(\vartheta,\varphi)};
\end{aligned}
\end{equation}
which in turn allows for numerical derivative calculation to be embedded in the
modal representation of a numerically sampled spin-weighted function.
Consequently, by making use of the SWSH in a truncated approximation
$\sw{s}{\tilde{f}}(\vartheta,\,\varphi)$ based on Eq.\eqref{eq:sfappr} the
adapted action of Eq.\eqref{eq:ethswshreduction} shunts away any issues that may
have arisen due to apparent singularities introduced by our choice of
coordinatisation. Given two spin-weighted functions $\sw{s_1}{f}$ and
$\sw{s_2}{g}$ where $s_1$ and $s_2$ may be distinct the SWSH transformation also
allows for decomposition of the nodal, point-wise product
$\sw{s_1}{f}\sw{s_2}{g}$ in terms of a linear combination of SWSH with
spin-weight $s=s_1+s_2$~\cite{beyer::2014::numerical}.

A final remark with respect to efficiency mentioned in the section introduction
when working with spin-weighted functions is in order. Define the averaged
coefficient $F_l:=\langle \sw{s}{f_{lm}} \rangle_m$ where $\sw{s}{f}_{lm}$ is as
in Eq.\eqref{eq:sfappr}. Suppose $\sw{s}{f(\vartheta,\,\varphi)}$ is smooth,
then there exist $\mathcal{A},\,\mathcal{B}>0$ such that for sufficiently large
$l$ we have $|F_l| \sim \mathcal{A}\exp(-\mathcal{B} l)$
\cite{:ref:harmonicanalysis2004katznelson,:ref:chebyshev2013boyd}.

\subsection{Function approximation on \texorpdfstring{$\Sigma\simeq \mathbb{R}\times \mathbb{S}{}^2_\rho$}{Sigma}}
\label{sec:funcSigAppr}
Since our goal is the numerical solution of the deformation problem outlined in
\S\ref{sec:CorvMeth} and formulated with respect to $\Omega\Subset\Sigma$ we
shall now focus on the domain $\Omega:=[\rmin,\,\rmax]\times \mathbb{S}^2_\rho$.
In order to approximate fields in $\mathcal{T}{}^q_r(\Omega)$ the expansion of
Eq.\eqref{eq:sfappr} is generalised by allowing $\rho^*$ to vary over the closed
interval $[\rmin,\,\rmax]$ such that the expansion coefficients acquire an
additional univariate $\rho$ dependence for each fixed $l$ (and $m$ if
axisymmetry is not assumed). As a preliminary we map the interval
$[\rmin,\,\rmax]$ to the standard interval $[-1,1]$ with coordinate $\nu$ via:
\begin{equation}\label{eq:fundDomMap} \rho(\nu)=
  \frac{1}{2}\left[ \left(\rmax-\rmin\right)\nu + \left(\rmin+\rmax \right)
  \right] \Longleftrightarrow \nu(\rho) = \frac{2\rho - \left(\rmin+\rmax
    \right)}{\left(\rmax-\rmin\right)}.
\end{equation}
If a function $f$ is continuous and either of bounded-variation or satisfies a
Dini-Lipschitz condition on $[-1,\,1]$ then the Chebyshev series converges
uniformly~\cite{:ref:chebyshev2002mason,:ref:chebyshev2013boyd}:
\begin{equation}\label{eq:unifChebSum}
    \begin{aligned}
    f_N(\nu):=\,&\sum_{n=0}^{N-1} f_n T_n(\nu)
    ,&
    \lim_{N\rightarrow \infty}\left\Vert f(\nu) - f_N(\nu)
    \right\Vert_2=\,&0.
    \end{aligned}
\end{equation}
Note that in Eq.\eqref{eq:unifChebSum}
a factor of $1/2$ has been absorbed into $f_0$:
\begin{equation}\label{eq:chebCoeffConv}
    \begin{aligned}
    \tilde{f}_n =& \frac{2}{\pi}
    \int_{-1}^1 \frac{f(\nu) T_n(\nu)}{\sqrt{1-\nu^2}}\,\d{\nu},
    &
    f_n:=& \frac{1}{1+\delta_{0n}} \tilde{f}_n,
    \end{aligned}
\end{equation}
where $\delta_{0n}=1$ if $n=0$ and is $0$ otherwise.
The rate of convergence of the truncated approximant $f_N$ is controlled by 
function differentiability. For $f\in C^{m+1}([-1,\,1])$ the bound 
 $\left|f(\nu) - f_N(\nu) \right| = \mathcal{O}(N^{-m})$
for all $\nu \in [-1,\,1]$ holds
\cite{:ref:chebyshev2002mason}. Combining 
Eq.\eqref{eq:sfappr} and Eq.\eqref{eq:unifChebSum}
allows for the spin-weighted representation of 
$X\in\mathcal{T}{}^q_r(\Omega)$ to be approximated as:
\begin{equation}\label{eq:swXexpaRad}
    \sw{s}{X}(\rho,\,\vartheta,\,\varphi) =
    \sum_{l=|s|}^{L_{\vartheta}}
    \sum_{m=-\mathrm{min}(l,\,L_\varphi)}^{\mathrm{min}(l,\,L_\varphi)}
    \sum_{n=0}^{N_\rho-1}
    \sw{s}{X}_{lmn} T_n(\nu(\rho)) 
    \,\sw{s}{Y}_{lm}(\vartheta,\,\varphi),
\end{equation}
where we have allowed for the possibility of adaptivity in $\varphi$ (indeed 
axisymmetry) when using~\cite{beyer::2014::numerical}. 
As we shall actually restrict to axisymmetry in \S\ref{sec:axisymdeform} 
it is convenient to further rewrite Eq.\eqref{eq:swXexpaRad} as:
\begin{equation}\label{eq:swXexpaAxi}
    \sw{s}{X}(\rho,\,\vartheta) =
    \sum_{l=|s|}^{L_{\vartheta}}
    \sum_{n=0}^{N_\rho-1}
    \sw{s}{X}_{ln} T_n(\nu(\rho)) 
    \,\sw{s}{Y}_{l}(\vartheta),
\end{equation}
where $\sw{s}{Y_l(\vartheta)}$ are real functions~\cite{beyer::2014::numerical}
and this expansion is to be understood as implicitly evaluated via 
Eq.\eqref{eq:swXexpaRad}.

\subsection{Complex analytic tools}\label{sec:cplxAnSec}

In principle we now have the ingredients required to turn Eq.\eqref{eq:WeakForm}
subject to the ansatz on $u$ offered by Eq.\eqref{eq:potAnsatzu} into a
numerical, linear-algebraic problem. However, imposing
$u=\mathcal{N}\tilde{u}\omega^\beta$ requires various ratios of weight function
terms (and their derivatives) to be computed. Additionally, a method is required
for accurate determination of $\mathcal{N}\delta / \sqrt{\omega}$ when only the
numerical result of the product $\delta=f\omega$ is known.  Recall that
$\omega \rightarrow 0$ as $\partial\Omega$ is approached (see
\S\ref{sec:linweak}). Consequently division of two quantities that vanish
towards $\partial \Omega$ in a manner that is known from analytical results to
yield a quotient of well-defined (finite) value must be computed using only
numerical data.  A further issue occurs in that high-order derivatives (up to
fourth order in Eq.\eqref{eq:linSF} for example) are required which is known to
be ill-posed\footnote{Ill-posed in the sense that small perturbations in the
  function to be differentiated may lead to large errors in the differentiated
  result~\cite{:ref:variational1995knowles,:ref:condition1985miel,%
    :ref:introductionapproximation2003rivlin}.}
when formulae are restricted to finite precision calculations with real
arithmetic.

The concern of derivative accuracy for analytic functions may be mitigated by
transformation to integrals in the complex plane through the use of the Cauchy
representation formula (CRF)
\cite{:ref:numerical1967lyness,:ref:numerical1981fornberg,%
  :ref:accuracy2011bornemann,:ref:numerical2011gautschi}.
This approach also potentially provides a solution to the division problem.  To
concretise the idea suppose that the real function $f$ possesses a complex 
analytic extension such that $f:U\rightarrow \mathbb{C}$ is holomorphic on an 
open set $U\subset \mathbb{C}$. Suppose the closed disc of radius $R$ satisfies
$\overline{D}_{R}\subset U$. Recall the CRF allows for the value of $f$ (and
complex derivatives thereof) to be calculated at a base point $z_0\in D_R$
through integration over a piecewise $C^1$ closed curve $\Gamma$ equipped with
counter-clockwise orientation in $U\setminus\{z_0\}$ that can be continuously
deformed in $U\setminus\{z_0\}$ to $\partial D_R$
\cite{:ref:functiontheory2006greene}.  In particular, if $\Gamma$ circumscribes
a base point $z_0$ on the real line then $f$ may also be described implicitly at
$z_0$ without recourse to direct sampling at the point. Immediately this
provides a potential mechanism to avoid the numerically unstable division
process.

Pursuing the problem of derivative conditioning further,
Bornemann~\cite{:ref:accuracy2011bornemann} investigates stability properties of
computing the Taylor series coefficients $(\tilde{f}_n){}_{n=0}^\infty$ of $f$,
which when evaluated by the CRF on an origin centered, circular contour
$\Gamma=\Gamma_C$ of radius $r>0$ take the form:
\begin{equation}\label{eq:TayCplxCoeff}
    \tilde{f}_n=\frac{1}{2\pi r^n} \int_0^{2\pi}
    f(r e^{i\theta})e^{-in\theta}\,\d{\theta}.
\end{equation}
Since the integrand in Eq.\eqref{eq:TayCplxCoeff} is periodic and analytic
its approximation via the $m$-point trapezoidal rule:
\begin{equation}\label{eq:trapFFTfn}
\tilde{f}_n(m,\,r) = \frac{1}{mr^n}\sum_{j=0}^{m-1}
\exp\left(-2\pi i \frac{jn}{m}\right)
f\left(r  \exp\left(2\pi i \frac{j}{m}\right)\right),
\end{equation}
converges exponentially~\cite{:ref:exponentially2014trefethen}
(cf. the final remark of \S\ref{sec:fcnapprsig}).  This approach for calculating
$\tilde{f}_n$ was advocated for by~\cite{:ref:numerical1967lyness} together with
the identification of $r^n \tilde{f}_n$ as got from Eq.\eqref{eq:trapFFTfn}
being readily evaluated via the FFT~\cite{:ref:algorithm1971lyness}.  A delicate
question of how to select\footnote{We shall assume that for fixed $n$ the value
  of $m$ has been selected to satisfy the Nyquist condition so as to avoid
  spurious aliasing.}  the order dependent $r(n)\in (0,\,R)$ now arises. From
the perspective of the CRF any choice is valid however numerical stability
degrades in the limits $r\rightarrow 0$ and $r\rightarrow R$
\cite{:ref:accuracy2011bornemann}.
While an early algorithm exists for determination of $r$ by a search procedure 
\cite{:ref:numerical1981fornberg,:ref:algorithm1981fornberg} it has 
disadvantages due to
the assumption that $(r^n \tilde{f}_n)_{n=0}^{m-1}$ be approximately 
proportional to a geometric sequence (which may not be the case generally) and a
requirement for judicious selection of starting value in the search
\cite{:ref:algorithm1981fornberg}. 

The stability issues of evaluating Eq.\eqref{eq:trapFFTfn} when working at
finite precision arise from small, finite error in evaluation of $f$ amplifying
to large error in evaluation of the sum. To analyse this
\cite{:ref:accuracy2011bornemann} examines both the absolute and relative error
associated with calculating the coefficients $\tilde{f}_n$. By considering a
perturbation $\hat{f}$ of $f$ within a bound of the absolute error $\epsilon$
with respect to the $L^\infty$ norm over the contour $\Gamma_C$ it is found that
the normalised coefficients $r^n \tilde{f}_n$ of Eq.\eqref{eq:TayCplxCoeff} and
their approximations $r^n \tilde{f}_n(m,r)$ by Eq.\eqref{eq:trapFFTfn} remain
within the same $\epsilon$ bound which follows from noting that the integral and
sum are both rescaled mean values of $f$.  Thus normalised Taylor coefficients
are well conditioned with respect to absolute error.
It is the relative error of coefficients that is shown to be crucial
\cite{:ref:accuracy2011bornemann}; indeed the relative condition number $\kappa$
of the CRF\footnote{For our purposes this may be computed via
Eq.\eqref{eq:TayCplxCoeff} with the definition 
of~\cite{:ref:numerical2012hohmann}.}  evaluated over $\Gamma_C$ for each
coefficient is considered and through minimisation of $\kappa$ the existence of
and a method for identification of an optimal $r_*(n)\in(0,\,R)$ is
provided. Optimal here entails selection of $r_*(n)$ such that round-off error
is minimised during numerical work.

The recent work of~\cite{:ref:fastaccuratechebyshev2016wang} extends this
analysis to the case of Chebyshev expansion coefficients which may be considered
to be embedded as the Taylor coefficients of a particular integral
transformation there described. As a preliminary, define the Bernstein ellipse
$\Gamma_E$ with foci at $\pm1$ and major and minor semi-axis lengths summing to
the ``radius'' parameter $r_B$:
\begin{equation}\label{eq:GammaBernstein}
  \begin{aligned}
  \Gamma_E(r_B):=&\left\{
  z\in\mathbb{C}\,\vphantom{\frac{1}{2}}\right| \left.
  \, z=\frac{1}{2}\left(r_B e^{i\theta} +
  r_B^{-1} e^{-i\theta} \right),\,
  (0\leq \theta \leq 2\pi)
  \right\},
  \end{aligned}
\end{equation}
where it is assumed that $r_B \geq 1$. Set $u=r_B e^{i\theta}$ then for
$z\in \Gamma_E(r_B)$ and $|u|\geq 1$
the relation
$u(z) = z + \sqrt{z^2-1}$ holds~\cite{:ref:chebyshev2002mason}.
The Chebyshev polynomials of the first kind
of degree $n$ may be defined 
by~\cite{:ref:chebyshev2002mason}:
\begin{equation}\label{eq:ChebStdDefn}
  \begin{aligned}
    T_n(\cos(\theta))&:=\cos(n\theta), &
    &(n\geq0),
  \end{aligned}
\end{equation}
Introducing $w:=e^{i\theta}$ and making use of Eq.\eqref{eq:ChebStdDefn}
yields:
\begin{equation}\label{eq:ChebCplxDefn}
  T_n(w) =\frac{1}{2}\left(w^n + w^{-n}\right).
\end{equation}
Which motivates extension of the domain of definition for $T_n$ 
to $\Gamma_E(r_B)$ as provided by
\cite{:ref:chebyshev2002mason,:ref:fastaccuratechebyshev2016wang}:
\begin{equation}\label{eq:cplxChebPoly}
  \begin{aligned}
    T_n(z(\theta;\,r_B)) &= \frac{1}{2}\left(r_B^n e^{i n \theta} + 
    r_B^{-n} e^{-in\theta} \right),&
    z&\in \Gamma_E(r_B).
  \end{aligned}
\end{equation}
Suppose that $f$ is analytic on the domain interior to $\Gamma_E(r_B)$, i.e.,
$\mathrm{int}(\Gamma_E(r_B))$. Then the Chebyshev series
(Eq.\eqref{eq:unifChebSum}) is convergent on $\mathrm{int}(\Gamma_E(r_B))$
\cite{:ref:approximation2013trefethen} and Chebyshev coefficients $f_n$ can be
given a complex analytic representation
\cite{:ref:fastaccuratechebyshev2016wang}:
\begin{equation}\label{eq:ChebCoeffBern}
    f_n = \frac{1}{\pi r_B^n}\int_0^{2\pi}
    f\left(
    \frac{1}{2}\left(r_B e^{i\theta} + r_B^{-1} e^{-i\theta}\right)
    \right)e^{-in\theta}\,\d{\theta}.
\end{equation}
Periodicity and analyticity of the integrand 
again allow for efficient approximation of $f_n$ through the $m$-point 
trapezoidal rule (cf. Eq.\eqref{eq:trapFFTfn}):
\begin{equation}\label{eq:trapChebfn}
    f_n(m,\,r_B) = \frac{2}{m r_B^n}
    \sum_{j=0}^{m-1}
    \exp\left(
    -2\pi i \frac{j n}{m}
    \right)
    f\left(\frac{1}{2}\left(
    r_B \exp\left(2\pi i \frac{j}{m}\right)
    + r_B^{-1} \exp\left(-2\pi i \frac{j}{m}\right)
    \right) \right).
\end{equation}
In complete analogy to~\cite{:ref:accuracy2011bornemann} the absolute and 
relative stability of the above approximation are then considered by
\cite{:ref:fastaccuratechebyshev2016wang}. It is shown that evaluation of
the Chebyshev coefficients is absolutely stable. However, the relative error
depends on the $r_B$ selected and is controlled by the
relative condition number $\kappa(\Gamma_E(r_B),\,n)$.

The determination of $r_*(n)$ for general $f$ which optimises the relative 
stability is crucial for the computation of approximations to derivatives of 
(the truncated approximation of) $f$ in terms of the coefficients directly which
involves the evaluation of the recursion relation~\cite{:ref:chebyshev2013boyd}:
\begin{equation}\label{eq:chebCoeffRec}
    \begin{aligned}
    f_{n-1}^{(k)} =\,& f_{n+1}^{(k)} + 2 n f_{n}^{(k-1)}, &
    n \in& \{N-k + 1,\,\dots,\, 1\};
    \end{aligned}
\end{equation}
initialised as $f_n^{(0)}:= f_n$ ($0\leq n \leq N$) and subject to the condition
$f_{N-k+2}^{(k)} = f_{N-k+1}^{(k)} = 0$. Unfortunately, the search for $r_*(n)$
requires extensive use of asymptotic approximations to infer the condition
number $\kappa$ directly
\cite{:ref:accuracy2011bornemann,:ref:fastaccuratechebyshev2016wang}.

In order to provide a practical, numerical method for the approximate
determination of $\kappa$ we propose to instead approximate $\kappa$ on some
$\Gamma_E(r_B)$ via the $m$-point trapezoidal rule of Eq.\eqref{eq:trapChebfn}.
The convexity of $\log(\kappa)$ together with monotonicity of
$r_*(n)$~\cite{:ref:accuracy2011bornemann,:ref:fastaccuratechebyshev2016wang}
allows for us to employ the downhill simplex minimisation
algorithm~\cite{:ref:numericalrecipes2007press}.  We initialise the search at
$n=0$ with $r_B=1$ constructing $r_*(0)$.  For $n\geq1$ the search is
initialised with $r_*(n-1)$, which once complete yields $r_*(n)$. Thus an
order-dependent sequence $(r_*(n))_{n=0}$ is iteratively generated.

If the replacement $r_B\rightarrow r_*(n)$ is made in the expression for ${f}_n$
of Eq.\eqref{eq:trapChebfn} then 
usage of the FFT to simultaneously compute the result for all orders $n$
is precluded
-- thus we propose a compromise: on account of the rapid
convergence of Chebyshev series for smooth functions we consider instead the
average $\langle r_*(n)\rangle_{n\leq n_\sigma}$ where:
\begin{equation}\label{eq:nsigmaDeterm}
    n_\sigma:= \argmax_{n\leq N} \left(\frac{{f}_n}{|\max_n\tilde{f}_n|}
    > \sigma\right),
\end{equation}
and $\sigma$ is a tolerance corresponding to a normalised coefficient
magnitude. As only the scaled, absolute value of ${f}_n$ is
required in Eq.\eqref{eq:nsigmaDeterm} we may determine
$n_\sigma$ by making use of Eq.\eqref{eq:trapChebfn} with $r_B=1$ and
subsequently recalculate for an improved (relative) accuracy. 

\section{Prototype problems and EAG for interior BBH data}\label{sec:ProPro}
We are now in a position to numerically carry out scalar curvature
deformation and provide composite, scalar-flat, initial data.

Explicit expressions for the cut-off functions $\chi$ appearing in
\S\ref{sec:CorvMeth}, together with prototype weight functions $\omega$ are
required. Based on the discussion in~\cite{:ref:smoothmanifolds2013nestruev},
define $\chi:[0,\,1]\rightarrow [0,\,1]$ by:
\begin{equation}
  \chi(x) = \frac{f(x)}{f(x)+f(1-x)},
\end{equation}
where in order to avoid steep numerical gradients $f(x):=x^N$ $(N>0)$ shall be
selected here\footnote{It is also possible to select (for example)
  $f(x):=\exp(-1/x)$ however this may potentially degrade numerical properties
  of the scheme.}. This serves (approximately) the role of a cut-off function.
Let $\rho\in\Omega_\rho:=[\rho_{\mathrm{min}},\,\rho_{\mathrm{max}}]$ and define
$\Delta\rho:=\rho_{\mathrm{max}}-\rho_{\mathrm{min}}$. For later convenience, we
immediately (linearly) map so as to introduce $\chi_L:\Omega_\rho\to\mathbb{R}$
growing from $0$ to $1$ over
$[\rho_{\mathrm{min}},\,\rho_{\mathrm{min}} + \mathfrak{f}\Delta\rho]$ and,
similarly, $\chi_R$ decaying from $1$ to $0$ over
$[\rho_{\mathrm{max}} - \mathfrak{f}\Delta\rho,\,\rho_{\mathrm{max}}]$ where
$\mathfrak{f}>0$. Consequently, we may model a univariate, normalised, weight
function $\hat{\omega}_C$ through: \begin{equation}\label{eq:defomC}
  \begin{aligned}
    \omega_C(\rho)&:=\chi_L(\rho)\chi_R(\rho), &
    \hat{\omega}_C(\rho) :=& \omega_C(\rho) / \max_{\rho} \omega_C(\rho).
  \end{aligned}
\end{equation}
Thus, explicit selection of $N$, together with $\mathfrak{f}$ allows for
implicit control on the behaviour of the potential $u$ in the vicinity of
$\partial\Omega_\rho$ when Eq.\eqref{eq:WeakForm} is solved numerically.

In order to close the details required to specify the ansatz of 
Eq.\eqref{eq:potAnsatzu} we also introduce:
\begin{equation}\label{eq:quadrPoly}
  \mathcal{N}(\rho;\,\alpha):=
  \frac{(2\alpha+1)!}{(\rho_{\mathrm{max}}
  -\rho_{\mathrm{min}})^{2\alpha+1}(\alpha!)^2}
  (\rho-\rho_{\mathrm{min}})^\alpha (\rho_\mathrm{max}-\rho)^\alpha,
\end{equation}
where the prefactor choice is motivated through integration
of the polynomial terms over 
$\rho\in\Omega_\rho$ so as to mitigate the
dependence of the overall magnitude of $\mathcal{N}$ on the extent of
$\Omega_\rho$. Unless otherwise stated, $\alpha=2$ will be selected
in Eq.\eqref{eq:quadrPoly} throughout.

To demonstrate the numerical properties of our scheme we begin by considering 
the simpler setting of spherical symmetry in \S\ref{sec:sphsymred}, which 
allows for self-consistent, convergence tests during numerical 
construction of the potential in \S\ref{sec:sphsymSCCT} to be performed. 
The more physically interesting case of axisymmetry is
described in \S\ref{sec:axisymdeform} and a test problem investigated in
\S\ref{sec:axiToyProblem}. In \S\ref{sec:numgluebbh} 
we demonstrate the gluing construction numerically in axisymmetry.

\subsection{Spherical symmetry reduction}\label{sec:sphsymred}
We now fix the region over which the deformation takes place as 
$\Omega:=[\rho_{\mathrm{min}},\,\rho_{\mathrm{max}}]
\times\mathbb{S}^2$. 
Spherical symmetry is imposed via the metric ansatz:
\begin{equation}\label{eq:sphmetr}
  \overline{g}_{ij} =
  \mathrm{diag}(\overline{F}(\rho),\, \overline{G}(\rho),\,
  \overline{G}(\rho) \sin^2(\vartheta)).
\end{equation}
One finds that upon inserting this $\overline{g}_{ij}$ into
Eq.\eqref{eq:WeakForm} (i.e., the weak-formulation)
together with the assumption that $u$ and $\eta$ have a univariate
dependence on $\rho$ an effective, one-dimensional
problem results due to angular dependence integrating out.
This observation motivates formal expansion of test and trial space 
functions respectively through:
\begin{equation}\label{eq:testTrisph}
    \begin{aligned}
        \sw{0}{\eta}(\rho) &= \sum_{n=0}^{L_\rho-1}\sw{0}{\eta}_{n}
        \Psi_{n}(\rho), &
        \sw{0}{u}(\rho) &= \sum_{n=0}^{L_\rho-1} \sw{0}{u}_n \Phi_n(\rho). &
    \end{aligned}
\end{equation}
In order to incorporate the solution ansatz of Eq.\eqref{eq:potAnsatzu} 
the function families are taken to be:
\begin{equation}\label{eq:testtrialsmash}
  \Psi_n(\rho)=\Phi_n(\rho)=\mathcal{N}(\rho;\,\alpha)\omega(\rho)^\beta
  T_n(\nu(\rho)),
\end{equation}
where $\mathcal{N}$ is defined in Eq.\eqref{eq:quadrPoly},
$T_n$ is a Chebyshev polynomial and $\nu(\rho)$ is the grid mapping of 
Eq.\eqref{eq:fundDomMap}.

Description of $L^*_{\overline{g}}[\cdot]_{ij}$ in the frame formalism 
(conventions of Eq.\eqref{eq:symtwotenproj}) 
gives rise to coframe coefficient terms $\mathfrak{l}_{\rho\rho}$ and
$\sw{s}{\mathfrak{l}}$ with integer $s$ satisfying $|s|\leq2$. With the 
$\overline{g}_{ij}$ of Eq.\eqref{eq:sphmetr} fixed during evaluation of 
$L^*_{\overline{g}}[\cdot]_{ij}$
only the coefficients\footnote{Explicit expressions for which are
provided in~\cite{:ref:numericalscalar2018daszuta}.}
$\mathfrak{l}_{\rho\rho}$ and $\sw{0}{\mathfrak{l}}$
are non-zero.
Schematically the weak formulation subject to Eq.\eqref{eq:testTrisph} becomes:
\begin{equation}\label{eq:weakFormNumerSphA}
  \begin{aligned}
  &\begin{aligned}
    \sum_{j=0}^{L_\rho-1} A_{ij}\,\sw{0}{u}_j =\,& \tilde{\deform}_i;
  \end{aligned}\\
  &\begin{aligned}
  A_{ij} &:= \int_{\rho_{\mathrm{min}}}^{\rho_{\mathrm{max}}}
    \left(
    \mathfrak{m}[\Phi_i] \omega^{|\beta|}
    \mathfrak{m}[\Phi_j] \omega^{|\beta|}
    +
    \mathfrak{n}[\Phi_i] \omega^{|\beta|}
    \mathfrak{n}[\Phi_j] \omega^{|\beta|}
    \right)\,\d{\rho},&
  \tilde{\delta}_i &:=
  \int_{\rho_{\mathrm{min}}}^{\rho_{\mathrm{max}}}
  \Phi_i \tilde{\delta}\omega^{2|\beta|}\,\d{\rho};
  \end{aligned}
  \end{aligned}
\end{equation}
where $\mathfrak{m}[\cdot]$ and $\mathfrak{n}[\cdot]$ are linear functionals
depending on $\overline{g}_{ij}$ and contain up to second order derivative
operators in $\rho$.  If the deformation is constructed based on a potential via
$\delta[u]=\tilde{\delta}[u]\omega=L_{\overline{g}}\left[\omega
  L{}_{\overline{g}}^*[u]\right]$ then up to fourth order derivatives in $\rho$
are also required.

Once $A_{ij}$ and $\tilde{\deform}_i$
are assembled the solution coefficients $\sw{0}{u}_j$ may be extracted via
standard, numerical, linear-algebraic techniques. Unfortunately the function
family $(\Phi_n)_{n=0}$ involves $\omega(\rho)^\beta$ with $\beta<0$ and hence
some care is required in the assembly process itself so as to preserve numerical
stability during the course of evaluation.  Define the weighted operator:
\begin{equation}\label{eq:sphWeightOp}
  D{}^n_\rho[\zeta,\,\eta,\,\theta][\cdot] :=
  \omega(\rho)^\zeta
  (\omega'(\rho))^\eta
  (\omega''(\rho))^\theta
  \partial{}^n_\rho[\cdot].
\end{equation}
Substitution of Eq.\eqref{eq:testtrialsmash} into
$\mathfrak{m}[\Phi_i]\omega^{|\beta|}$ 
(or $\mathfrak{n}[\Phi_i]\omega^{|\beta|}$)
appearing in $A_{ij}$ of Eq.\eqref{eq:weakFormNumerSphA} and expansion 
allows for a refactoring of expressions into
products of $D{}^n_\rho[\zeta,\,\eta,\,\theta][\cdot]$ with manifestly
regular functions involving the background metric coefficient terms 
and polynomials but excluding $\omega^{(n)}(\rho)$. Though involved, the
manipulations are straightforward and provide for a mechanism to individually 
regularise terms containing the weight function.

During solution of the (local) nonlinear deformation problem as described in
\S\ref{sec:nonlinloc} the background metric of Eq.\eqref{eq:sphmetr} 
is fixed and $g$ satisfying 
${\mathcal{R}}[g] - {\mathcal{R}}[\overline{g}]=\deform$
for a given choice of $\deform$ is sought. The iterative scheme
of \S\ref{sec:nonlinloc} is implemented through construction of a sequence of 
solutions ${}^{(k)}_{\hphantom{(}0}u_i$ to Eq.\eqref{eq:weakFormNumerSphA} with
$(\overline{F}(\rho),\,\overline{G}(\rho))$ fixed throughout. At each iterate 
the replacement $\tilde{\deform}\omega=\delta \rightarrow {}^{(k)}\deform$ is
made, where ${}^{(k)}\deform$ is defined in accordance with
Eq.\eqref{eq:nonlinearrefinement}.
Corrections to the potential allow for updated metric functions to be
formed through:
\begin{equation}\label{eq:sphUpdaCoeff}
\begin{aligned}
    {}^{(k+1)}F(\rho) =\,& \overline{F}(\rho) + \omega(\rho)
    \mathfrak{l}_{\rho\rho}\left[
    \sum_{l=0}^{k}{}^{(l)}u(\rho);\,\overline{F},\,\overline{G}
    \right],\\
    {}^{(k+1)}G(\rho) =\,& \overline{G}(\rho) + \omega(\rho)
    \sw{0}{\mathfrak{l}}\left[
    \sum_{l=0}^{k}{}^{(l)}u(\rho);\,\overline{F},\,\overline{G}
    \right];
\end{aligned}
\end{equation}
where we have emphasised the background dependence of 
$\mathfrak{l}_{\rho\rho}$ and $\sw{0}{\mathfrak{l}}$.

During numerical construction of an update it is
${}^{(m)}\deform(\rho)={}^{(m)}\tilde{\deform}(\rho) \omega(\rho)$ that is known
and hence a term of the form ${}^{(m)}\deform(\rho)\omega(\rho)^\beta$ with
$\beta<0$ must be explicitly evaluated. This may potentially lead to numerical
instability as $\rho\rightarrow \partial\Omega$ on account of the behaviour of
$\omega$ in this limit. One method to alleviate this is provided in the tools of
\S\ref{sec:cplxAnSec}.  Numerical calculation of the truncated family
$\left(\partial_\rho^m\left[T(\nu(\rho)) \right] \right)_{n=0}^{L_\rho -1}$ we
continue to perform with real arithmetic based on recursion.  The background
metric coefficients $(\overline{F},\,\overline{G})$ however will be represented
by sampling on a mapped Bernstein ellipse $\Gamma_E$ (see
Eq.\eqref{eq:GammaBernstein}) so as to provide a spectral representation (as in
Eq.\eqref{eq:trapChebfn}) with radius parameter $r_B$ selected for each function
according to the averaged, optimal radius
$\langle r_*(n)\rangle_{n\leq n_\sigma}$.
Derivatives of the background metric coefficients are to be prepared via the
recursion relation of Eq.\eqref{eq:chebCoeffRec}.  Products of weight function
terms appearing in $D{}^n_\rho[\zeta,\,\eta,\,\theta]$ together with polynomials
shall be evaluated on $\Gamma_E$ with a radius parameter $r_\circ$ selected 
(uniformly for all basis function orders). 
During construction of terms such as the corrected
metric coefficients $\left({}^{(i+1)}F(\rho(z)),\,{}^{(i+1)}G(\rho(z)) \right)$
appearing in Eq.\eqref{eq:sphUpdaCoeff} or updated scalar curvature
$\dimd{\mathcal{R}}\left[{}^{(i+1)}g\right]$ individual terms may initially be
sampled on contours with distinct radii. In order to combine such terms an
initial transformation to their respective modal representations is made, which
allows for a subsequent, complex, nodal representation on a single, contour of
commensurate radius (i.e., $r_\circ$) to be computed.  When required, numerical
quadrature is computed based on the real nodal representation of functions via
the Clenshaw-Curtis rule~\cite{:ref:spectralmatlab2000trefethen} with the number
of samples selected as $2L_\rho+2$.
\subsection{Spherical symmetry: SCCT and local nonlinear deformation}
\label{sec:sphsymSCCT}
We now perform self-consistent convergence tests (SCCT) on prototype
problems. At the linear level, this entails selection
of a background metric $\overline{g}_{ij}$, weight function $\omega$, and
a ``seed'' potential function $u$ which allows for generation of a deformation
$\delta$ analytically via Eq.\eqref{eq:linSF}. We now demonstrate that our
numerical scheme is robust by showing that solution of the weak formulation
yields $\tilde{u}$ which converges to $u$ as resolution is increased.

Introduce the background metric functions:
\begin{align}\label{eq:protoAbgFcns}
\overline{F}_A(\rho;\,M,\,P) =\, & 1+M\sin^2(P\pi \rho), &
\overline{G}_A(\rho) =\, & \rho^2;
\end{align}
the selection of which is motivated by both simplicity and construction of
a prototype problem with non-constant background scalar curvature such that
for $M\neq0$ and $P\neq0$ non-triviality of the kernel
of $L_{\overline{g}}^*$ is avoided.

Define the seed potentials:
\begin{equation}\label{eq:potI}
\begin{aligned}
u_1(\rho)
=&
\frac{1}{2\times 10^5} \left(\cos(4\pi\rho)\rho^2 - \sin(6\pi\rho)\rho \right)
,&
u_2(\rho)
=&
15 \cos^4\left(\frac{\pi}{16}\left(2\rho - 6\right)\right)
\exp\left(\sin\left(\frac{\pi\rho}{8}\right) \right).
\end{aligned}
\end{equation}

\begin{figure}[H]
\begin{subfigure}
  \centering
  \includegraphics[width=.49\linewidth]{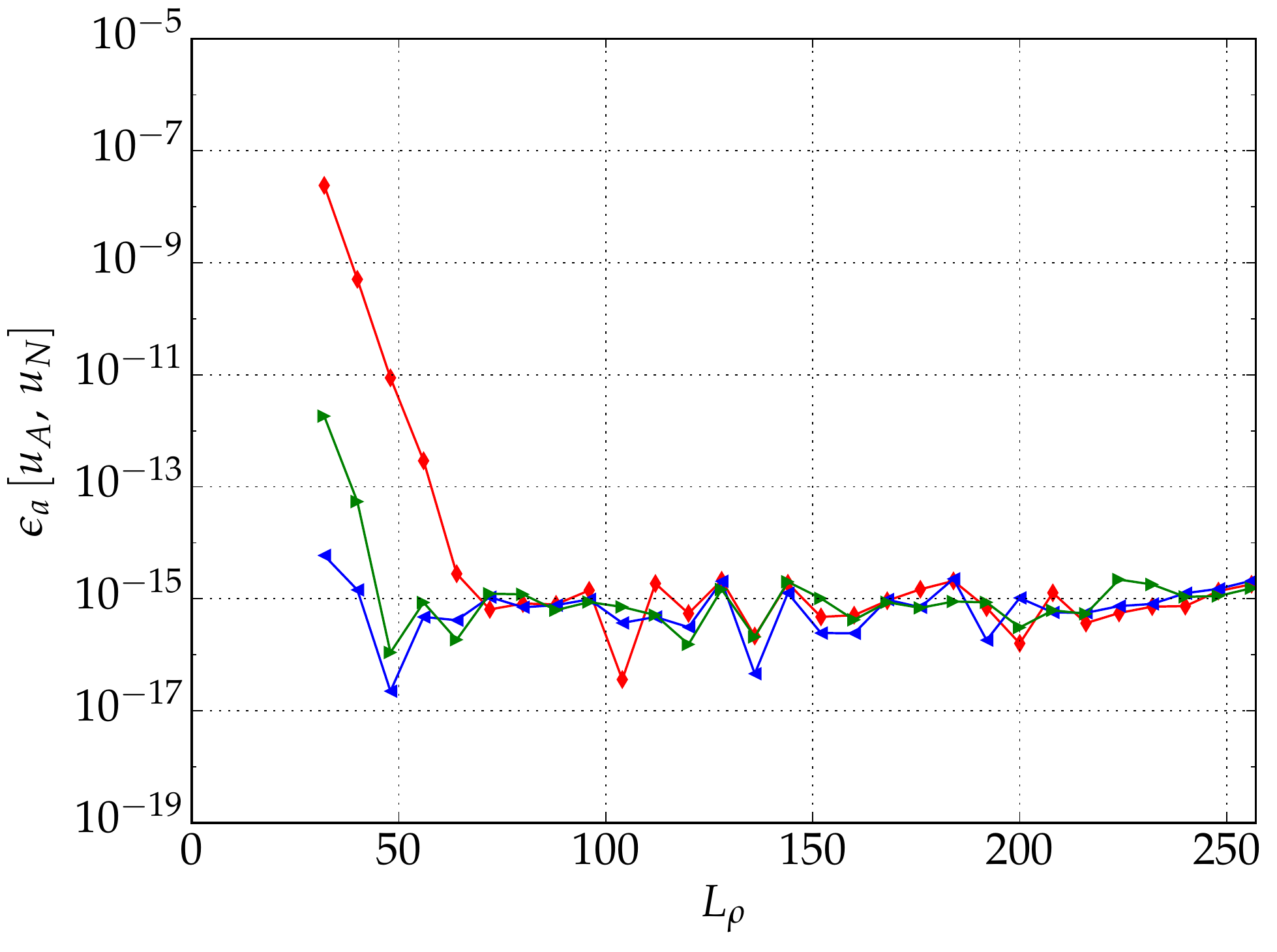}
\end{subfigure}%
\begin{subfigure}
  \centering
  \includegraphics[width=.49\linewidth]{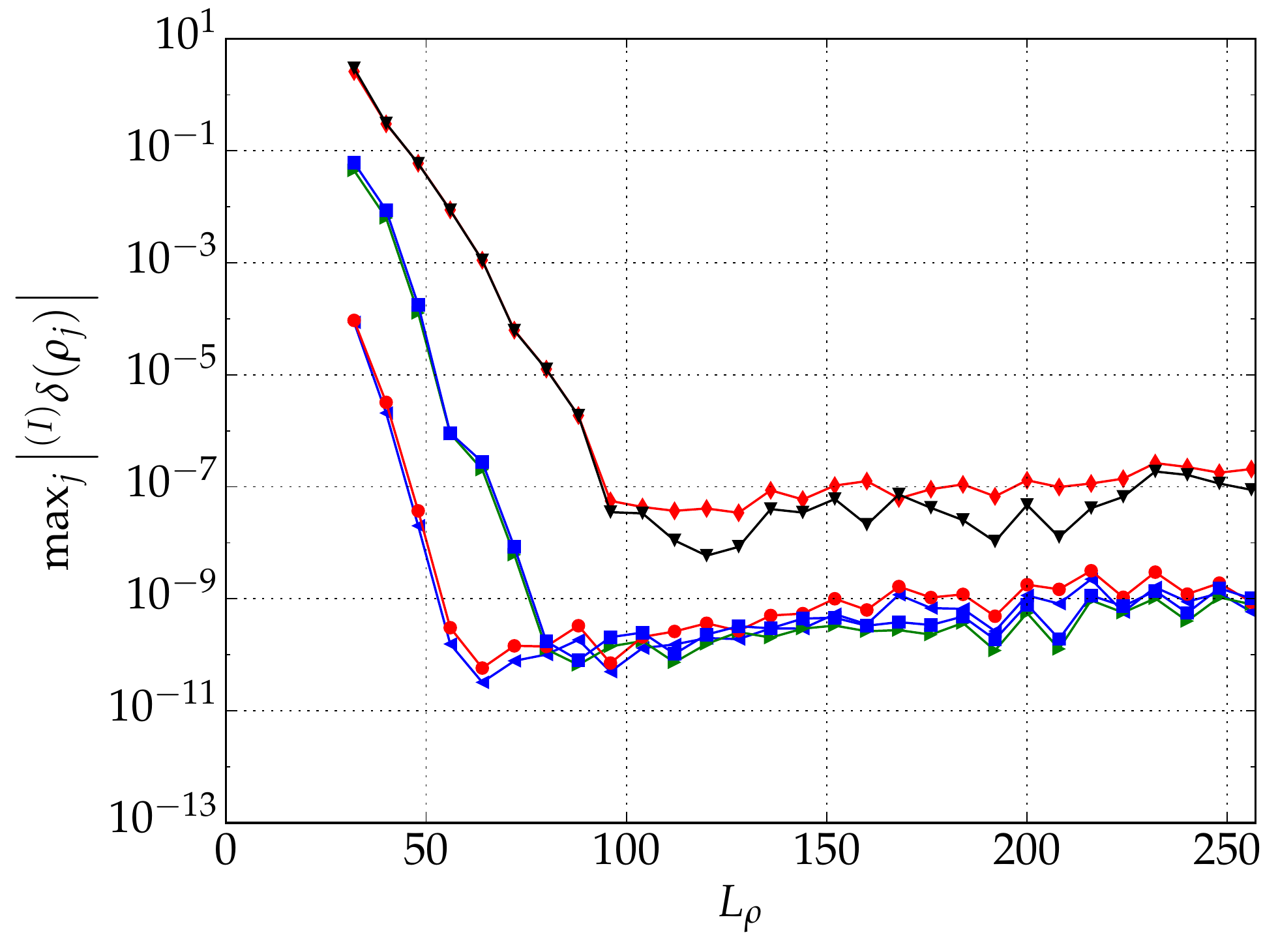}
\end{subfigure}
\caption{
Numerical complex-analytic approach applied to calculating scalar curvature 
deformation. Parameters selected according to $\mathcal{P}_C$ of 
Eq.\eqref{eq:parammapsOneDimCplx} and $r_\circ=1.05$.
The physical domain is selected with
$\rho\in[1,\,2]$. Background metric coefficients
$\left(\overline{F}(\rho(z)),\,\overline{G}(\rho(z)\right)$
are as in Eq.\eqref{eq:protoAbgFcns}
with $M=1=P$ wherein we determine optimal radii parameters as
$\langle r_*(F;\,n)\rangle_{n\leq n_\sigma} = 3.3$ and 
$\langle r_*(G;\,n)\rangle_{n\leq n_\sigma} = 1.2$ with
$\sigma=10^{-10}$ (see~Eq.\eqref{eq:nsigmaDeterm}).
(\textbf{Left})
Linear SCCT based on the seed potential $u_1$ defined in Eq.\eqref{eq:potI}
where evaluation of Eq.\eqref{eq:weakFormNumerSphA} is based on refactoring
involving the operators $D^m_\rho[\zeta,\,\eta,\,\theta][\cdot]$ and individual
weight function terms which are calculated on an elliptic contour with $r_\circ$
as described in \S\ref{sec:sphsymred}. Optimal radius of the potential function
is given by $\langle r_*(u_1;\,n)\rangle_{n\leq n_\sigma} = 1.9$. Remaining
parameter choices are denoted by: (red ``{\footnotesize$\blacklozenge$}''):
$\mathcal{P}_C(0)$; (blue ``{\footnotesize$\blacktriangleleft$}''):
$\mathcal{P}_C(1)$; (green ``{\footnotesize$\blacktriangleright$}''):
$\mathcal{P}_C(2)$. Even at moderate band-limit ($L_\rho\simeq 64$) we find that
convergence to numerical round-off is attained.
(\textbf{Right})
Absolute maximum of deformation over the real grid when the target 
$\dimd{\mathcal{R}}[g]$ is provided by Eq.\eqref{eq:directTarDefTerms} and 
Eq.\eqref{eq:directTarDefn} with number of iterations taken as $I=25$ 
(saturation in convergence verified by doubling).  Deformation function 
$R_i(\rho(z))$ (Eq.\eqref{eq:directTarDefTerms}) optimal radii determined as 
$\langle r_*(R_1;\,n)\rangle_{n\leq n_\sigma} = 1.9$ and 
$\langle r_*(R_2;\,n)\rangle_{n\leq n_\sigma} = 6.4$. 
Denoted by: (red ``{\footnotesize$\blacklozenge$}''): $R_1[\rho(z)]$ and 
$\mathcal{P}_C(0)$; (blue ``{\footnotesize$\blacktriangleleft$}''): 
$R_1[\rho(z)]$ and $\mathcal{P}_C(1)$; 
(green ``{\footnotesize$\blacktriangleright$}''): 
$R_1[\rho(z)]$ and $\mathcal{P}_C(2)$; 
(black ``{\footnotesize$\blacktriangledown$}''): $R_2[\rho(z)]$ and 
$\mathcal{P}_C(0)$; (red ``{$\bullet$}''): $R_2[\rho(z)]$ and 
$\mathcal{P}_C(1)$; (blue ``{\tiny$\blacksquare$}''): $R_2[\rho(z)]$ and 
$\mathcal{P}_C(2)$. In order that the deformation sequence maintains stability 
we apply an Orszag-style low-pass filter via ${}^{(i)}u_n=0$ 
($n>\frac{2}{3}L_\rho$)~\cite{:ref:spectralmethods2007hesthaven,%
  :ref:chebyshev2013boyd}. 
Note: in both subfigures prior to saturation linear tails clearly indicate the 
property of exponential convergence.}
\label{fig:p1alterCplx}
\end{figure}

Furthermore, we supplement the usual linear SCCT with direct specification 
of a target scalar curvature so as to provide prototypical scalar curvature 
deformation problems by introducing:
\begin{equation}\label{eq:directTarDefTerms}
\begin{aligned}
R_1(\rho) =& \frac{1}{10}\sin(4\rho),&
R_2(\rho) =& \frac{1}{10}\left(
\frac{1}{20}\rho - \frac{1}{8}\cos(2\rho)
\right);
\end{aligned}
\end{equation}
where with Eq.\eqref{eq:directTarDefTerms} the target scalar curvature 
becomes:
\begin{equation}\label{eq:directTarDefn}
\dimd{\mathcal{R}}_k[g] :=
\dimd{\mathcal{R}}[\overline{g}] + \underbrace{R_k(\rho) \omega(\rho)}_{
=:\delta\left[\overline{g},\,R_k,\,\omega \right]}.
\end{equation}
For convenience, remaining parameters are collected into the map:
\begin{equation}\label{eq:parammapsOneDimCplx}
    \mathcal{P}_C: j \mapsto
    \begin{cases}
    \left(\mathfrak{f},\,N,\,\beta\right)
    \mapsto \left(1.2,\, 4,\, -1/2\right),
    & j = 0;\\
    \left(\mathfrak{f},\,N,\,\beta\right)
    \mapsto \left(1.2,\, 2,\, -1\right),
    & j = 1;\\
    \left(\mathfrak{f},\,N,\,\beta\right)
    \mapsto \left(0.8,\, 2,\, -1\right),
    & j = 2;
    \end{cases}
\end{equation}
Results of numerical calculations involving a variety of numerical parameters
with the complex analytic approach are provided in Fig.\ref{fig:p1alterCplx}.
We find that while linear SCCT may be carried out with excellent accuracy the
sequence of linear solutions entering the deformation problem is far more
susceptible to instability. We ascribe this latter to the
numerical division process involved in the calculation of
${}^{(m)}\tilde{\delta} /\omega^{|\beta|}$ where (small) local error in the
vicinity of $\partial\Omega$ accumulates and is represented by spuriously
populating high-order modes which in turn grow in scale with each iterate and
gradually pollute low-order modes.  Suppression of this is provided by
filtering. While it is the case that either choice of $\beta=1$ or $\beta=-1/2$
in the ansatz on the potential $u$ appears to lead to convergence,
unfortunately, as can be seen in Fig.\ref{fig:p1alterCplx} (right) saturation in
convergence still presents before numerical round-off.
As an alternative we pursue a hybrid scheme where terms involving
$\partial_{\rho}^n[\omega]$ and $\partial_\rho^n[\mathcal{N}]$ that enter the
factorisation of the integrand describing $A_{ij}$ in
Eq.\eqref{eq:weakFormNumerSphA} are computed using arbitrary precision. All
other quantities are calculated using standard, complex, floating-point
arithmetic with the techniques of \S\ref{sec:NumPre}.  An upshot of this
approach is that for more generic weight functions such as:
\begin{equation}
  \label{eq:weibump}
  C^\infty_c([-1,\,1])\ni\omega_B(\nu) =
  \begin{cases}
    \exp\left(1 - [1-\nu^2]^{-1}\right), & \nu\in(-1,\,1);\\
    0, & \nu \notin (-1,\,1);
  \end{cases}
\end{equation}
entering the deformation term $\delta$ no inconveniences due to complex 
analytic extensions or essential singularities arise.

We introduce further metric coefficient 
functions (cf. Eq.\eqref{eq:protoAbgFcns}):
\begin{equation}\label{eq:new1dmetrcoeff}
\begin{aligned}
&\begin{aligned}
(\overline{F}_A(\rho),\,\overline{G}_A(\rho))
:=&
\left(
1+\sin^2(\pi\rho),\,\rho^2
\right),&
(\overline{F}_B(\rho),\,\overline{G}_B(\rho))
:=&
\left(1 + \sin^2\left(\frac{\pi\rho}{3} \right),\,
\rho^2
\right);
\end{aligned}\\
&\begin{aligned}
(\overline{F}_C(\rho),\,\overline{G}_C(\rho))
:=&
\left(
2 + \rho + 2 \cos^2(4\rho),\,
1 + \rho^4 \exp(-\rho)
\right);
\end{aligned}
\end{aligned}
\end{equation}
to which the metrics $\overline{g}_A$, $\overline{g}_B$ and
$\overline{g}_C$ are associated.
For convenience, set:
\begin{equation}\label{eq:parammapsOneDim}
    \mathcal{P}_H: j \mapsto
    \begin{cases}
    \left(\mathfrak{f},\,N,\,\beta\right)
    \mapsto \left(1.2,\, 4,\, -1/2\right),
    & j = 0;\\
    \left(\mathfrak{f},\,N,\,\beta\right)
    \mapsto \left(1.2,\, 2,\, -1\right),
    & j = 1;\\
    \left(\mathfrak{f},\,N,\,\beta\right)
    \mapsto \left(0.8,\, 4,\, -1/2\right),
    & j = 2;\\
    \left(\mathfrak{f},\,N,\,\beta\right)
    \mapsto \left(0.4,\, 4,\, -1/2\right),
    & j = 3.
    \end{cases}
\end{equation}

\begin{figure}[H]
  \centering
  \includegraphics[width=.49\linewidth]{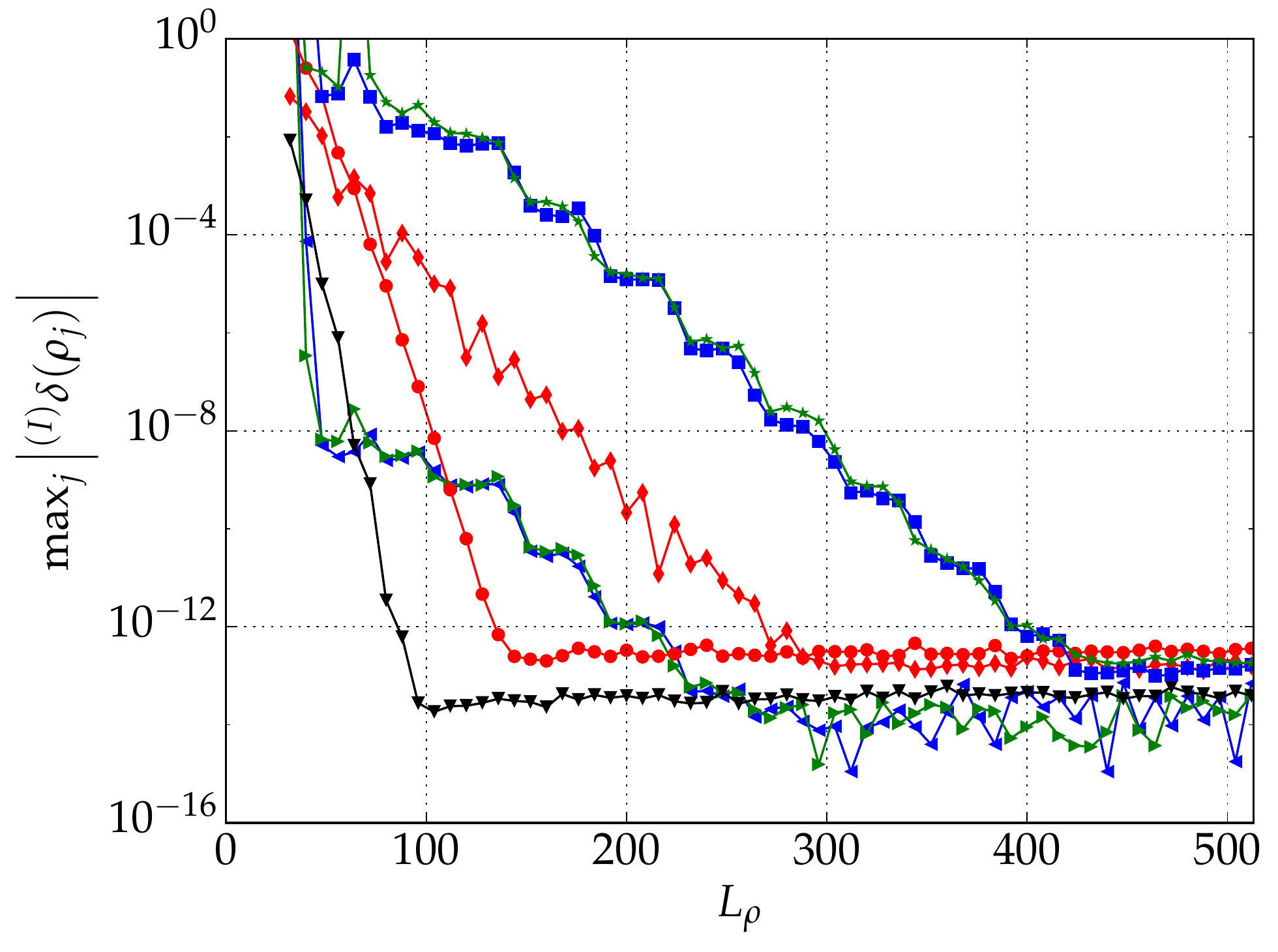}
\caption{
Numerical hybrid approach applied to calculating scalar curvature deformation.
The maximum of the absolute value of the deformation on the real grid
after $I=50$ iterations have been performed (saturation verified by doubling).
In
(red ``{\footnotesize$\blacklozenge$}''):
Target deformation $\delta$ constructed based on the seed potential
$u_1$ of Eq.\eqref{eq:potI} and $\omega_B$ of Eq.\eqref{eq:weibump}
with background metric coefficients those of Eq.\eqref{eq:protoAbgFcns}
where $(M,\,P)=(1,\,1)$ and $\rmin=1$ and $\rmax=2$.
Parameters of $\hat{\omega}_C$ are selected
via $\mathcal{P}_H$ of Eq.\eqref{eq:parammapsOneDim} where for this case
$\mathcal{P}_H(0)$ is chosen.
Now consider
$\delta[{u}_2]$ constructed based on Eq.\eqref{eq:potI}
with background metric $\overline{g}_C$ (see Eq.\eqref{eq:new1dmetrcoeff}).
The physical domain is selected with $\rho\in[10,\,20]$.
Remaining parameters chosen as:
(blue ``{\footnotesize$\blacktriangleleft$}''):
$\mathcal{P}_H(0)$;
(green ``{\footnotesize$\blacktriangleright$}''):
$\mathcal{P}_H(1)$.
Now consider $\delta[\overline{g}_B,\,R_1,\,\hat{\omega}_C]$
constructed based on Eq.\eqref{eq:directTarDefTerms} and 
Eq.\eqref{eq:directTarDefn}.
Taking $\rho\in[5,\,10]$ where in:
(black ``{\footnotesize$\blacktriangledown$}''):
$\mathcal{P}_H(0)$;
(red ``{$\bullet$}''):
$\mathcal{P}_H(2)$.
Finally consider $\delta[\overline{g}_C,\,R_2,\,\hat{\omega}_C]$
with $\rho\in[10,\,20]$ then in:
(blue ``{\tiny$\blacksquare$}''):
$\mathcal{P}_H(1)$;
(green ``{\footnotesize$\bigstar$}''):
$\mathcal{P}_H(3)$.
Note: In all cases tested approximate exponential convergence
is a clear feature.
}
\label{fig:p1alterHybrid}
\end{figure}

Results of numerical calculations making use of the hybrid scheme
are shown in Fig.\ref{fig:p1alterHybrid} for various test deformation
problems.
As in Fig.\ref{fig:p1alterCplx} we find that the results of calculations based
on the hybrid approach presented in Fig.\ref{fig:p1alterHybrid} all lead to
exponential convergence with a saturation in
$\max_j\left|{}^{(I)}\delta(\rho_j) \right|$ that is near numerical
round-off. In contrast we find that no filtering is required and there does not
appear to be much sensitivity with respect to how parameters of $\hat{\omega}_C$
are selected. Indeed, even with $\delta$ prepared such that $\omega_B$ is
utilised we find convergence in the hybrid approach that does not degrade with
increasing $L_\rho$. Due to these properties we henceforth
shall only make use of this hybrid scheme and fix $\beta=-1/2$.
We emphasise however that in the case of assembling $A_{ij}$
(see Eq.\eqref{eq:weakFormNumerSphA}) explicit refactoring of the integrand as
described previously is required in order for numerical solutions to be found
(linear or otherwise) based on both of the approaches investigated.

\subsection{Axisymmetric deformation}\label{sec:axisymdeform}
Having investigated our numerical approach under the imposition of spherical
symmetry in \S\ref{sec:sphsymred} we now turn our attention to scalar curvature
deformation when the underlying metric is axisymmetric.  It shall be assumed
that this metric is of the form of Eq.\eqref{eq:symtwotenproj} and
that the coefficients appearing in Eq.\eqref{eq:metrspF} carry no $\varphi$
dependence. On account of the success of the mixed complex-analytic
floating-point and arbitrary precision arithmetic hybrid approach a similar
strategy shall be pursued here.  As metric coefficients now carry a $\vartheta$
dependence that does not integrate out decompositions of fields shall be made by
leveraging the SWSH functions and transformation algorithm described in
\S\ref{sec:fcnapprsig}.

Set $\Omega_\rho:=[\rmin,\,\rmax]$, $\Omega_\vartheta:=[0,\,\pi]$ and
$\Omega_\varphi:=\mathbb{S}^1$ then the full domain of interest is
$\Omega_\rho\times \Omega_\vartheta \times \Omega_\varphi$; however, the
$\varphi$ dependence is trivial and shall henceforth be suppressed.
Taking the view that $\omega$ serves to impose boundary conditions by
inducing decay towards $\partial \Omega_\rho$ on salient fields we shall 
continue to assume the dependence $\omega=\omega(\rho)$. 

Thus, immediately we expand test and trial space functions respectively:
\begin{equation}\label{eq:testtrialExpa}
    \begin{aligned}
    \sw{0}{\eta}(\rho,\,\vartheta) =&
    \sum_{m=0}^{L_\rho-1}\sum_{k=0}^{L_\vartheta}
    \sw{0}{\eta}_{mk}
    \sw{0}{\Psi_{mk}}(\rho,\,\vartheta),&
    \sw{0}{u}(\rho,\,\vartheta)=&
    \sum_{n=0}^{L_\rho-1}\sum_{l=0}^{L_\vartheta}
    \sw{0}{u}_{nl}\sw{0}{\Phi_{nl}}(\rho,\,\vartheta),
\end{aligned}
\end{equation}
with expansion functions of both spaces treated symmetrically:
\begin{equation}\label{eq:axiTestTrial}
    \sw{0}{\Psi_{nl}}(\rho,\,\vartheta)=
    \sw{0}{\Phi_{nl}}(\rho,\,\vartheta)=
    \Phi_n(\rho)
    \sw{0}{Y_l(\vartheta)},
\end{equation}
where $\Phi_n(\rho)$ is defined in Eq.\eqref{eq:testtrialsmash}
and $\sw{0}{Y}_l:=\sw{0}{Y}_{l0}$ is an axisymmetric SWSH function
as in \S\ref{sec:fcnapprsig}. 
In the present context, the weak formulation of Eq.\eqref{eq:WeakForm} becomes:
\begin{equation}\label{eq:weakLinFormAxiProb}
    \sum_{n=0}^{L_\rho-1}\sum_{l=0}^{L_\vartheta} A_{mknl}\,\sw{0}{u}_{nl}
    =
    \tilde{\delta}_{mk},
\end{equation}
where we have defined:
\begin{equation}\label{eq:weakAxiExpa}
    \begin{aligned}
    A_{mknl}&:=\int_\Omega
    \left(L^*_{\overline{g}}\left[
        \sw{0}{\Psi_{mk}}(\rho,\,\vartheta)
    \right]
    \right)^{ij}
    \left(L^*_{\overline{g}}\left[
        \sw{0}{\Phi_{nl}}(\rho,\,\vartheta)
    \right]
    \right)_{ij}
    \omega(\rho)
    \sqrt{\overline{g}(\rho,\,\vartheta)}\,\d{\rho}\d{\vartheta},\\
    \tilde{\delta}_{nl}
    &:=
    \int_\Omega
    \sw{0}{\Psi_{nl}}(\rho,\,\vartheta)
    \tilde{\delta}(\rho,\,\vartheta)
    \omega(\rho)
    \sqrt{\overline{g}(\rho,\,\vartheta)}\,\d{\rho}\d{\vartheta};
  \end{aligned}
\end{equation}
and $\overline{g}(\rho,\,\vartheta)$ is the determinant of the background
metric. To evaluate the internal contraction between $L_{\overline{g}}^*$
operators and implement a regularisation scheme analogous to that of
\S\ref{sec:sphsymred} define the vector operator:
\begin{equation}\label{eq:veccontrAxi}
  \begin{aligned}
    \boldsymbol{\mathcal{L}}[\sw{0}{u}]:=\left(
    \sw{0}{u},\,
    \partial_\rho\left[\sw{0}{u}\right],\,
    \partial_\rho^2\left[\sw{0}{u}\right],\,
    \edtb{\edtb{\sw{0}{u}}},\,
    \edtb{\sw{0}{u}},\,
    \partial_\rho\left[\edtb{\sw{0}{u}}\right],\,
    \edth{\edtb{\sw{0}{u}}}
    \right);
\end{aligned}
\end{equation}
and introduce:
\begin{equation}\label{eq:contraExpa}
    \mathcal{C}[\sw{0}{\eta},\,\sw{0}{u}] = 
    \sum_{q=1}^7\sum_{r=1}^7
    \boldsymbol{\mathcal{L}}[\sw{0}{\eta}]_q\,
    \sw{s}{\mathcal{C}}_{qr}
    \boldsymbol{\mathcal{L}}[\sw{0}{u}]_r;
\end{equation}
where the ($\omega$ independent) functions $\sw{s}{\mathcal{C}}_{qr}$ now
complete specification of the contraction: they are the coefficients in front of
all possible products of the derivatives of $\sw{0}{u}$. Each
$\sw{s}{\mathcal{C}}_{qr}$ carries a spin-weight $s$ such that when combined
with both $\boldsymbol{\mathcal{L}}[\cdot]$ the product has resultant
spin-weight $0$.

The aforementioned factoring serves an additional purpose beyond numerical
regularisation in the assembly of $A_{mknl}$.
As the expansions of Eq.\eqref{eq:testtrialExpa} are truncated such that for
$m$ and $n$ together there is a storage requirement of $L_{\rho}^2$ elements, 
each of which in turn requires $k$ and $l$ to be specified, 
the number of elements appearing in $A_{mknl}$ (ignoring symmetry) scales 
as $\mathcal{O}(L_\rho^2 (L_\vartheta + 1)^2)$. 
Thus, if all elements are immediately constructed and sampled then naive 
intermediate calculations involving $A_{mknl}$ result in a storage requirement 
scaling as
$\mathcal{O}\left(2L_\rho^2(2L_\rho+2)(L_\vartheta+1)^2)(L_\vartheta+2)\right)\sim\mathcal{O}(L_\rho^3 L_\vartheta^3)$.

Embedding a quadrature evaluation at the intermediate stage is more efficient.
To accomplish this we perform a further regrouping of individual terms in 
the integrand of $A_{mknl}$. On account of the tensor product basis utilised
the action of the weighted operator 
$\tilde{\boldsymbol{\mathcal{L}}}[\cdot]:=\omega(\rho)^{1/2}\boldsymbol{\mathcal{L}}[\cdot]$ 
may be decoupled to $\rho$ and $\vartheta$ specific subspaces where with 
Eq.\eqref{eq:testtrialsmash} and Eq.\eqref{eq:axiTestTrial}:
\begin{equation}
\tilde{\boldsymbol{\mathcal{L}}}\left[\sw{0}{\Phi_{nl}}\right]
=
\sum_{q=1}^7
\boldsymbol{\mathcal{L}}_\rho[\Phi_n]_q
\boldsymbol{\mathcal{L}}_\vartheta[\sw{0}{Y_l}]_q;
\end{equation}
and the individual components of 
$\mathcal{\boldsymbol{\mathcal{L}}}_\rho[\Phi_n]_q$ are of the
form of $D^m_\rho\left[\frac{1}{2},\,0,\,0\right][\Phi_n]$
(see Eq.\eqref{eq:sphWeightOp}).
Set:
\begin{equation}
\sw{s}{\tilde{\mathcal{C}}_{qr}}(\rho,\,\vartheta):=
\sqrt{\overline{g}(\rho,\,\vartheta)}
\sw{s}{\mathcal{C}_{qr}}(\rho,\,\vartheta)\csc{\vartheta}=
\sw{0}{\overline{N}}(\rho,\,\vartheta)
\sqrt{\sw{0}{\overline{\gamma}}(\rho,\,\vartheta)^2 - 
\sw{-2}{\overline{\gamma}}(\rho,\,\vartheta)
\sw{+2}{\overline{\gamma}}(\rho,\,\vartheta)}
\sw{s}{\mathcal{C}_{qr}}(\rho,\,\vartheta),
\end{equation}
then $A_{mknl}$ of Eq.\eqref{eq:weakAxiExpa} becomes:
\begin{equation}
A_{mknl}=\sum_{q=1}^7\sum_{r=1}^7\int_{0}^\pi
\underbrace{\left[\int_{\rmin}^{\rmax}
\boldsymbol{\mathcal{L}}_\rho[{\Psi_{m}}]_q\,
\sw{s}{\mathcal{\tilde{C}}}_{qr}
\boldsymbol{\mathcal{L}}_\rho[{\Phi_{n}}]_r\,\d{\rho}\right]}_{
=:\sw{s}{A}_{mn}(\vartheta)}
\boldsymbol{\mathcal{L}}_\vartheta[\sw{0}{Y_{k}}]_q 
\boldsymbol{\mathcal{L}}_\vartheta[\sw{0}{Y_{l}}]_r
\sin\vartheta \,\d{\vartheta}.
\end{equation}
The inner quadrature $\sw{s}{A_{mn}(\vartheta)}$ is numerically 
evaluated with a Clenshaw-Curtis rule~\cite{:ref:spectralmatlab2000trefethen} 
whereupon expansion
with the family $(\sw{s}{Y_j}(\vartheta))_{j=0}^{L_\vartheta}$
allows for evaluation of the outer quadrature.

Linear SCCT requires evaluation of
$\deform(\rho,\,\vartheta)=\tilde{\deform}(\rho,\,\vartheta)\omega(\rho)$
appearing in the integrand of $\tilde{\delta}_{nl}$ of Eq.\eqref{eq:weakAxiExpa}
for a given choice of seed potential 
$u(\rho,\,\vartheta)=\mathcal{N}(\rho)\tilde{u}(\rho,\,\vartheta)\omega(\rho)^{-1/2}$.
This we accomplish by expressing
$\partial_\rho^m\left[\omega(\rho)L_{\overline{g}}^*[u(\rho,\,\vartheta)]\right]$
for $m=0,\,1,\,2$
via the (non-zero) spin-weighted terms
$\partial_\rho^m\left[\omega(\rho)\mathfrak{l}_{\rho\rho}[u(\rho,\,\vartheta)]\right]$
and
$\partial_\rho^m\left[\omega(\rho)\sw{s}{\mathfrak{l}}[u(\rho,\,\vartheta)]\right]$  
based on the decomposition technique described in \S\ref{sec:sigdecomp}.
In accordance with Eq.\eqref{eq:linSF}, $\deform(\rho,\,\vartheta)$ 
is formed by application of the frame representation of 
$L_{\overline{g}}[\cdot]$.
Finally, resultant terms are expanded and regrouped
such that all $\omega$ containing terms are collected and represented
solely via the weighted operators $D^n_\rho$ 
introduced in Eq.\eqref{eq:sphWeightOp}. This is possible due to the assumption 
of the univariate $\rho$ dependence of $\omega$.

To close this section, we provide an update rule for metric coefficient
functions when represented in terms of the spin-weighted components
$\left(\sw{0}{\overline{N}},
  \,\sw{\pm1}{\overline{N}},\,\sw{0}{\overline{\gamma}},
  \,\sw{\pm2}{\overline{\gamma}}
\right)$. 
On account of the underlying axisymmetry we may drop the distinction
between $\pm|s|$. This is a consequence of the particular properties of the
coordinate representation of the SWSH and the $\eth$ operators: in axisymmetry
the representations of ${}_{\pm s}Y_{l0}$ are real and agree even though,
abstractly, these quantities have different spin-weight and therefore lie in
different spaces.

Given a sequence of potential function solutions ${}^{(k)}u_{nl}$ define the
update functional: 
\begin{equation}
  \mathcal{U}[\overline{f},\,\mathfrak{l};\,i](\rho,\,\vartheta):=
  \overline{f}(\rho,\,\vartheta) + \omega(\rho)\mathfrak{l}
  \left[\sum_{j=0}^i{}^{(j)}u(\rho,\,\vartheta);\, \sw{0}{\overline{N}},
    \,\sw{-1}{\overline{N}},\,\sw{0}{\overline{\gamma}},
    \,\sw{-2}{\overline{\gamma}}
  \right],
\end{equation}
where $\mathfrak{l}$ is a general component of
$L^*_{\overline{g}}[\cdot]_{ij}$ in the frame formalism (conventions of
Eq.\eqref{eq:symtwotenproj}). We may now write:
\begin{equation}
  \begin{aligned}
    \psbrc{(i+1)}{s}{\gamma}(\rho,\,\vartheta) =\,&
    \mathcal{U}[\sw{s}{\overline{\gamma}},\,
    \sw{s}{\mathfrak{l}};\,i](\rho,\,\vartheta),
    & \psbrc{(i+1)}{-1}{N}(\rho,\,\vartheta) =\,&
    \mathcal{U}[\sw{-1}{\overline{N}},
    \,\sw{-1}{\mathfrak{l}};\,i](\rho,\,\vartheta).
  \end{aligned}
\end{equation}
In order to update $\sw{0}{\overline{N}}$ first compute:
\begin{equation}
  {}^{(i+1)}\mathfrak{g}_{\rho\rho}(\rho,\,\vartheta) =
  \mathcal{U}[\overline{\mathfrak{g}}_{\rho\rho},
  \,\mathfrak{l}_{\rho\rho};\,i](\rho,\,\vartheta),
\end{equation}
together with:
\begin{equation}
  \psbrc{(i+1)}{0}{\tilde{\gamma}}(\rho,\,\vartheta)= \left(
    \psbrc{(i+1)}{0}{\gamma}(\rho,\,\vartheta)^2-
    \bigl|\psbrc{(i+1)}{-2}{\gamma}(\rho,\,\vartheta)\bigr|^2 \right)^{-1}.
\end{equation}
Finally, based on Eq.\eqref{eq:swshiftadapted} 
and Eq.\eqref{eq:metrspF} set:
\begin{equation}\label{eq:axishiftUpd}
\begin{aligned}
  \psbrc{(i+1)}{0}{N}(\rho,\,\vartheta)^2 =\,&
    {}^{(i+1)}\mathfrak{g}_{\rho\rho}(\rho,\,\vartheta)
    +2 \psbrc{(i+1)}{0}{\tilde{\gamma}}(\rho,\,\vartheta)\\
    &\times
    \left(
      \psbrc{(i+1)}{-1}{N}(\rho,\,\vartheta)^2
      \psbrc{(i+1)}{+2}{\gamma}(\rho,\,\vartheta)
      -\left|\psbrc{(i+1)}{-1}{N}(\rho,\,\vartheta)\right|^2
      \psbrc{(i+1)}{0}{\gamma}(\rho,\,\vartheta)
    \right),
\end{aligned}
\end{equation}
whereupon the positive root is taken.

\subsection{Axisymmetric deformation: Test problem}\label{sec:axiToyProblem}

On account of the restriction $\omega:=\omega(\rho)$ convergence properties in
the axisymmetric case are largely controlled by the resolution selected in
$\rho$. Essentially, for a sufficiently large, fixed $L_\vartheta$ behaviour as
in \S\ref{sec:sphsymSCCT} was observed. Hence, for the sake of expediency we
will only provide an illustrative test problem here.

Introduce the background metric:
\begin{equation}
  \overline{g}=
  \frac{\rho}{25} F(\vartheta) 
  \d{\rho}\otimes\d{\rho}
  + \frac{\rho^2}{100}\left[
  F(\vartheta) (\omega\otimes\overline{\omega} + \overline{\omega}\otimes\omega)
  + G(\vartheta) \omega\otimes\omega +
  \overline{G}(\vartheta) \overline{\omega}\otimes\overline{\omega}
  \right],
\end{equation}
where:%
\begin{equation}
F(\vartheta) := 1 + \sw{0}{Y_1}(\vartheta), \qquad
G(\vartheta):= \sw{2}{Y_2}(\vartheta).
\end{equation}
We now represent $\overline{g}$ in terms of the spin-weighted components
$\left(\sw{0}{\overline{N}},
\,\sw{-1}{\overline{N}},\,\sw{0}{\overline{\gamma}},\,\sw{-2}{\overline{\gamma}}
\right)$. 
According to the decomposition of Eq.\eqref{eq:symtwotenproj}
and Eq.\eqref{eq:metrspF} we may immediately take $\sw{-1}{\overline{N}}=0$,
and it follows that 
$\sw{0}{\overline{N}}(\rho,\,\vartheta) = \rho\sqrt{F(\vartheta)}/5$.
The intrinsic metric expression provided by Eq.\eqref{eq:protoTopoMetr}
together with the maps of Eq.\eqref{eq:swTopoMetrCoeff} yields:
\begin{equation}\label{eq:protogamsw}
\begin{aligned}
\sw{-2}{\overline{\gamma}}(\rho,\,\vartheta)=\,&
\frac{\rho^2}{50}\sw{-2}{Y_2}(\vartheta), &
\sw{0}{\overline{\gamma}}(\rho,\,\vartheta)=\,&
\frac{\rho^2}{100}\left(1 + \sw{0}{Y_1}(\vartheta) \right).
\end{aligned}
\end{equation}
The target scalar curvature shall be defined by:
\begin{equation}\label{eq:scRtarAxi}
  {\mathcal{R}}[g]:=
  {\mathcal{R}}[\overline{g}] + \deform(\rho,\,\vartheta;\,\omega),
\end{equation}
where in this section we take $\delta$ as:
\begin{equation}\label{eq:axiDeformProto}
  \begin{aligned}
    \deform(\rho,\,\vartheta;\,\omega) =& 
    \frac{401}{47} \sin(4\rho) \sw{0}{Y}_4(\vartheta) \omega(\rho).
  \end{aligned}
\end{equation}
Spin-weighted metric coefficients shall be sampled in $\Omega_\rho$ with mapped
$\Gamma_E$ at fixed $r_B=1.6$ in order to numerically determine partial
derivatives in $\rho$ based on the techniques discussed in \S\ref{sec:cplxAnSec}
which are then sampled back to the real, mapped Chebyshev-Gauss-Lobatto 
grid (see~\cite{:ref:spectralmatlab2000trefethen} for a definition).
Approximation in $\Omega_\vartheta$ is based on the axisymmetric SWSH algorithm
discussed in \S\ref{sec:funcSigAppr}.

\begin{figure}[H]
\begin{subfigure}
  \centering
  \includegraphics[width=.32\linewidth]{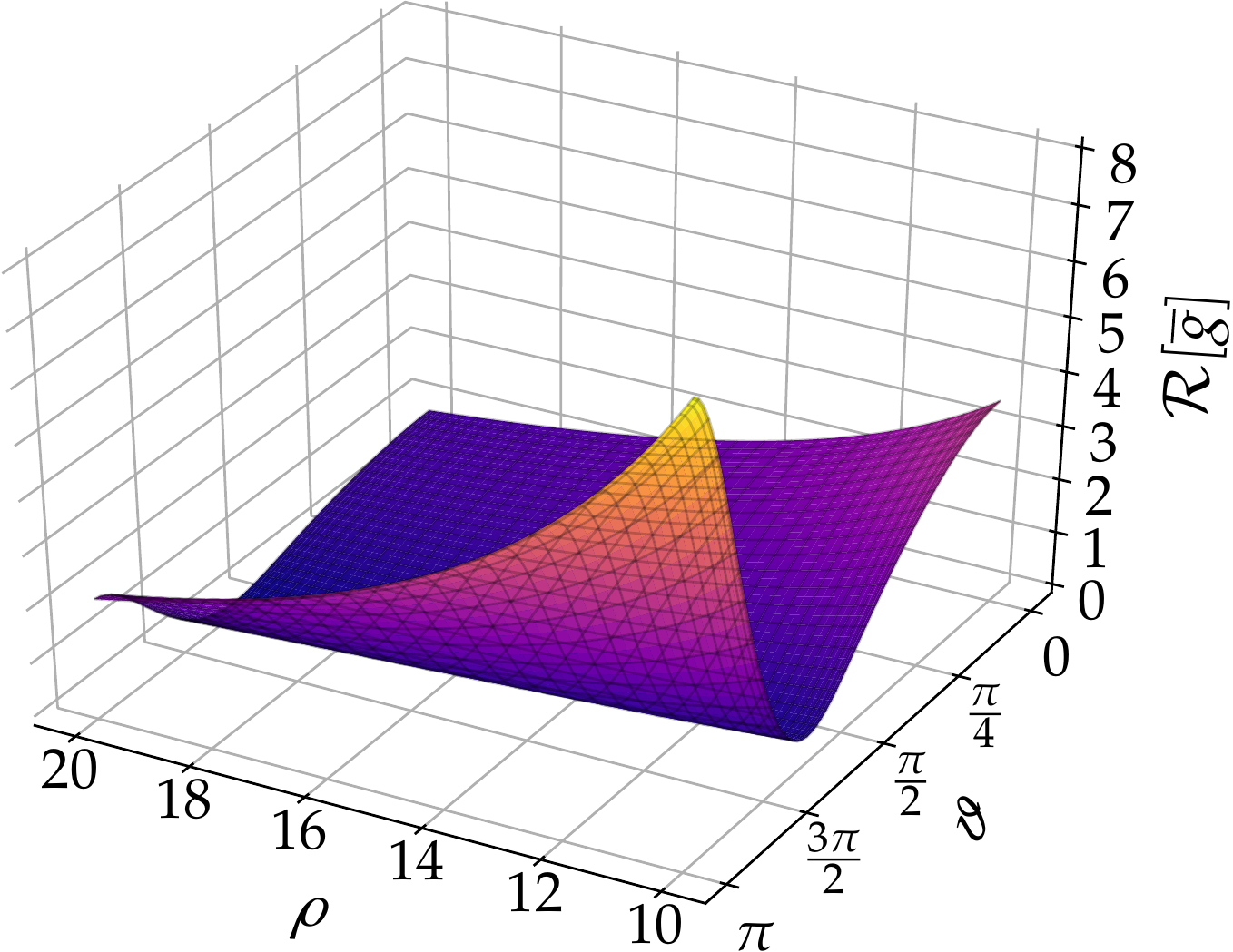}
\end{subfigure}%
\begin{subfigure}
  \centering
  \includegraphics[width=.32\linewidth]{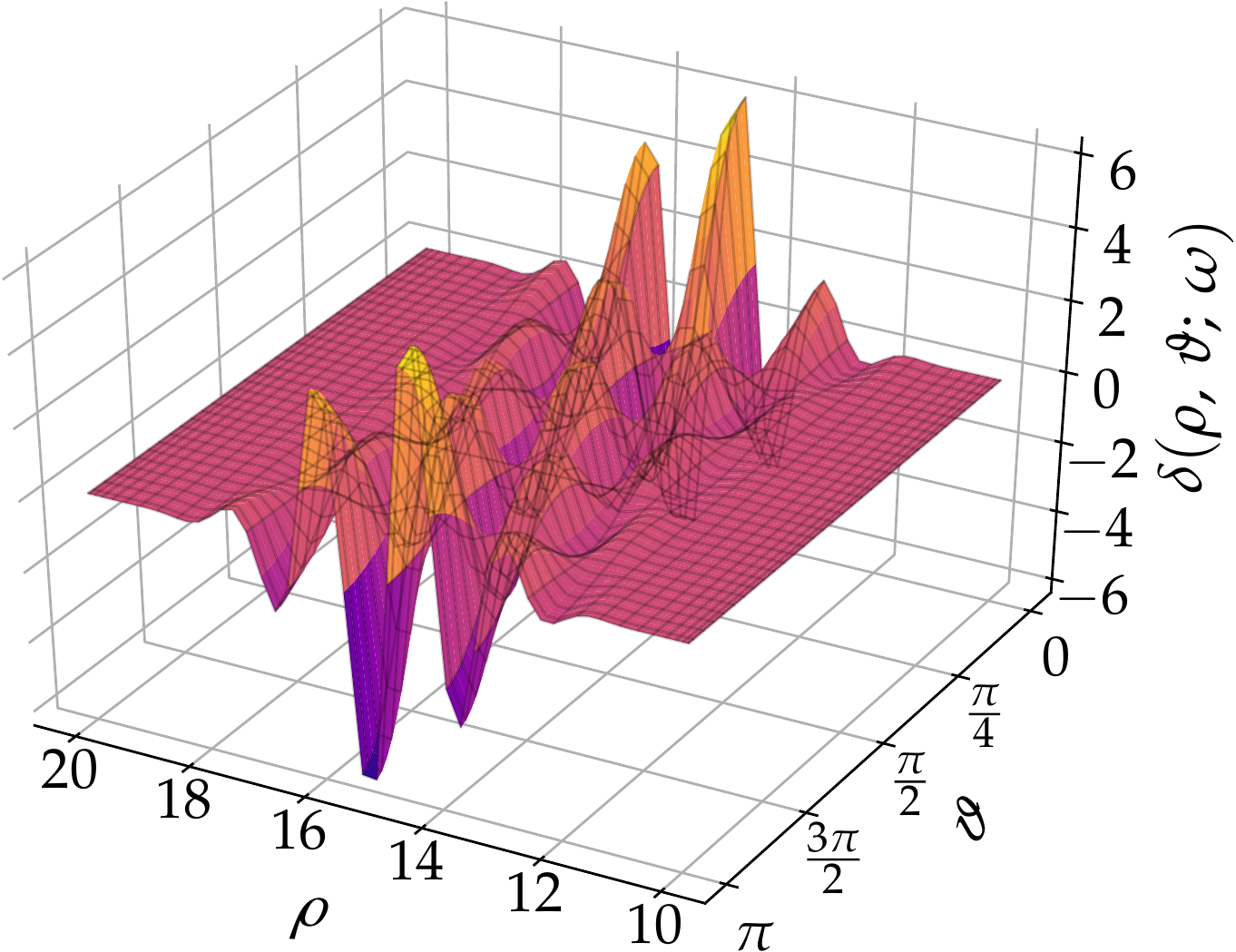}
\end{subfigure}
\begin{subfigure}
  \centering
  \includegraphics[width=.32\linewidth]{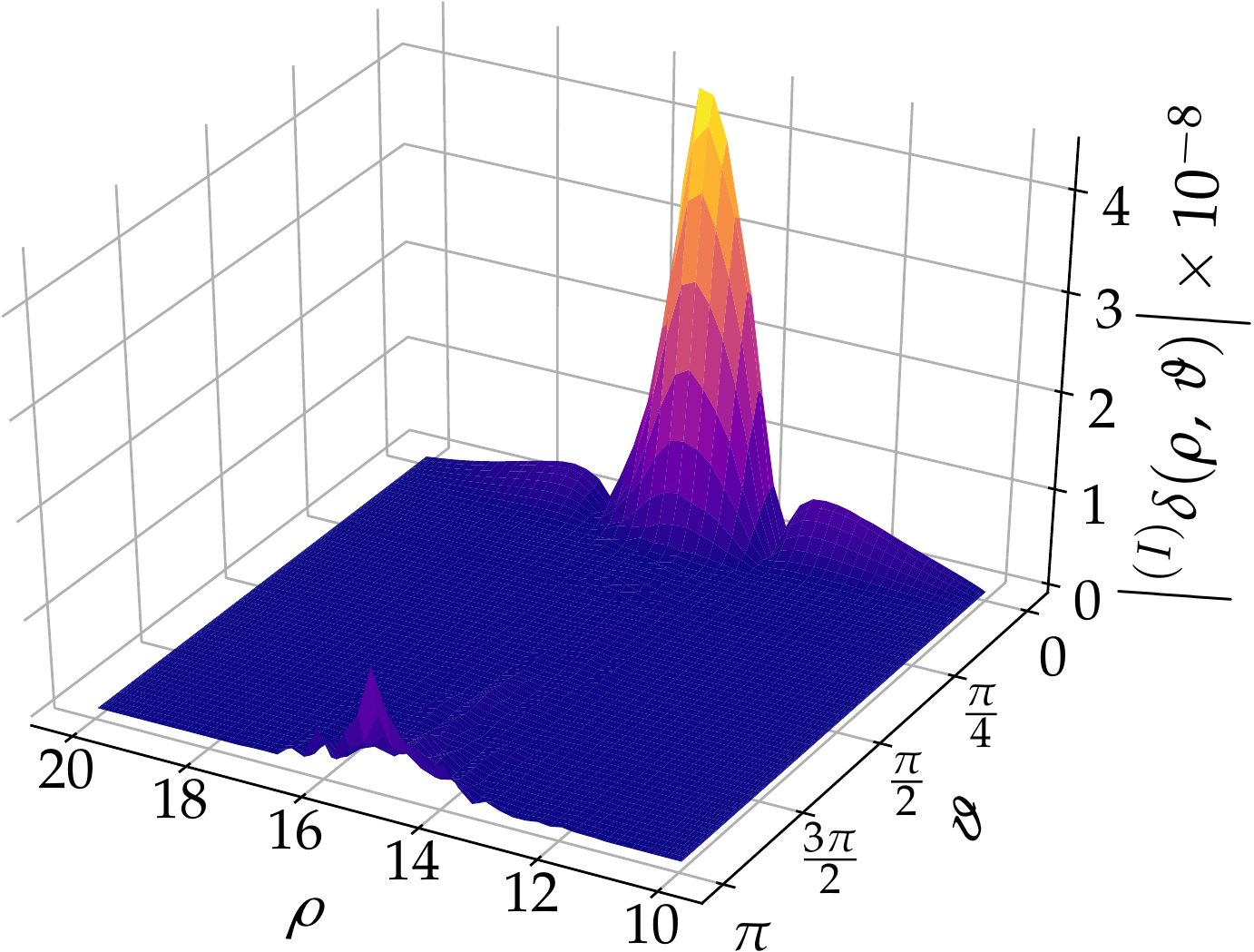}
\end{subfigure}
\caption{
Background scalar curvature ${\mathcal{R}}[\overline{g}]$ associated with
spin-weighted metric coefficients of $\overline{g}$ as in the text.
The
deformation $\deform(\rho,\,\vartheta)$ of Eq.\eqref{eq:axiDeformProto}
used to generate the target 
${\mathcal{R}}[g]$ defined by Eq.\eqref{eq:scRtarAxi}
and magnitude of updated deformation
$\left|{}^{(I)}\delta(\rho,\,\vartheta)\right|$
where $I=80$ iterations have been taken (saturation verified by doubling) is
also depicted.
The weight function
is $\hat{\omega}_C(\rho)$ of Eq.\eqref{eq:defomC} with parameters provided via 
$\mathcal{P}_H(0)$ of Eq.\eqref{eq:parammapsOneDim}; this enforces 
$\delta\rightarrow 0$ as $\rho\rightarrow \partial\Omega_\rho$.
Band-limits are selected as $L_\rho=128$ and $L_\vartheta=64$. Note the 
comparable magnitudes of ${\mathcal{R}}[\overline{g}]$ and $\delta$. Colouring 
of all subfigures corresponds to function values.
}
\label{fig:axiInsp}
\end{figure}

\begin{figure}[htb]
\begin{subfigure}
  \centering
  \includegraphics[width=.32\linewidth]{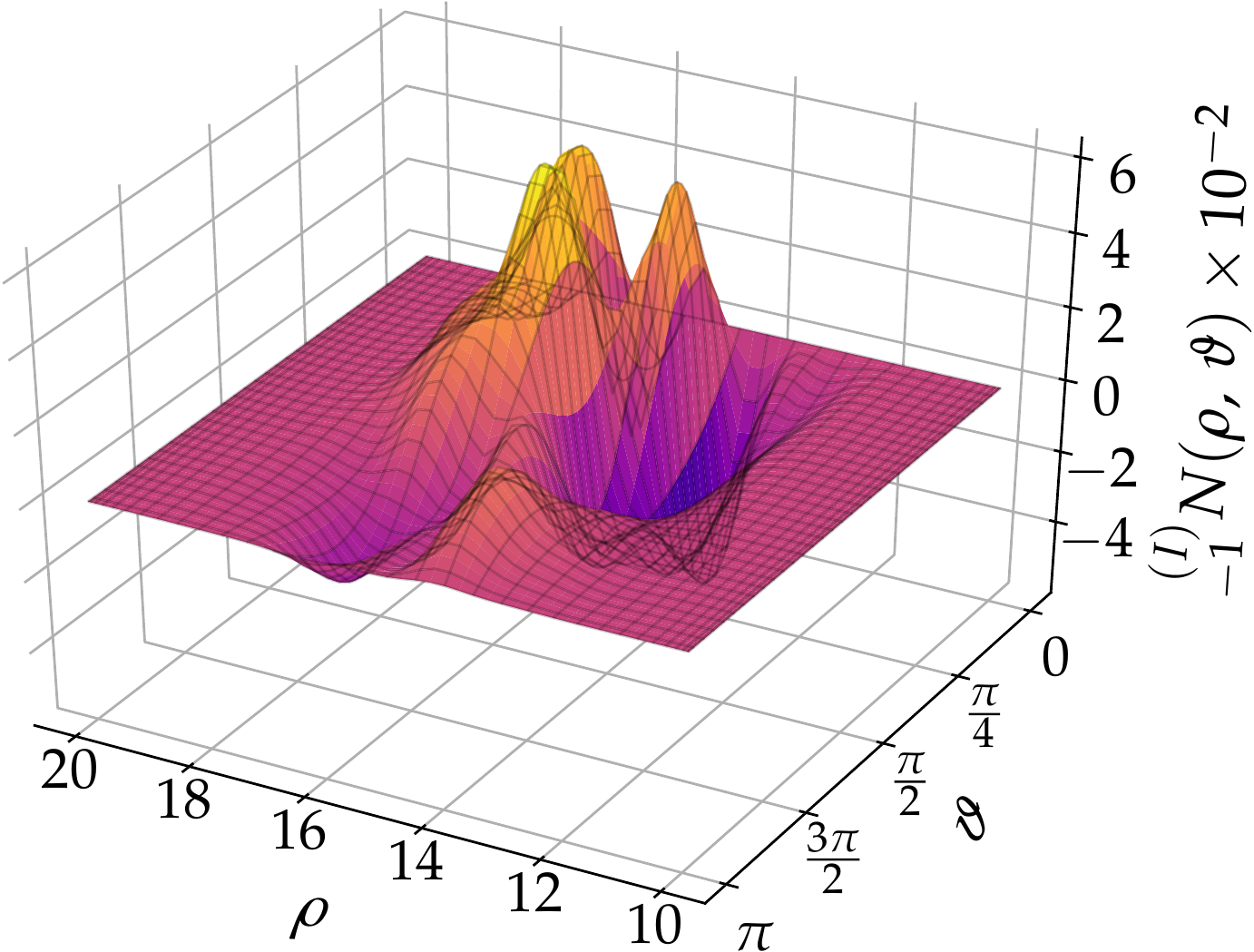}
\end{subfigure}%
\begin{subfigure}
  \centering
  \includegraphics[width=.32\linewidth]{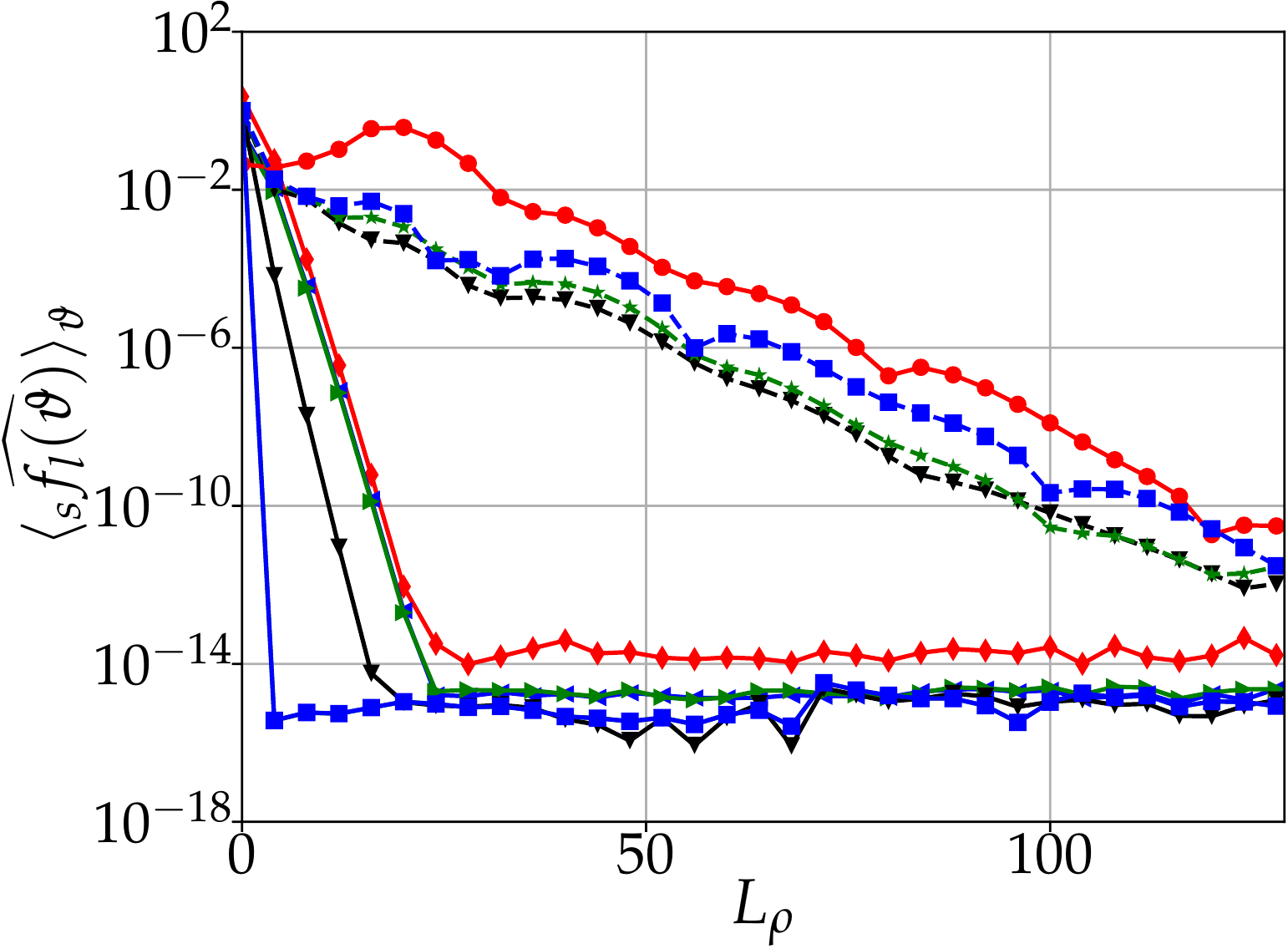}
\end{subfigure}
\begin{subfigure}
  \centering
  \includegraphics[width=.32\linewidth]{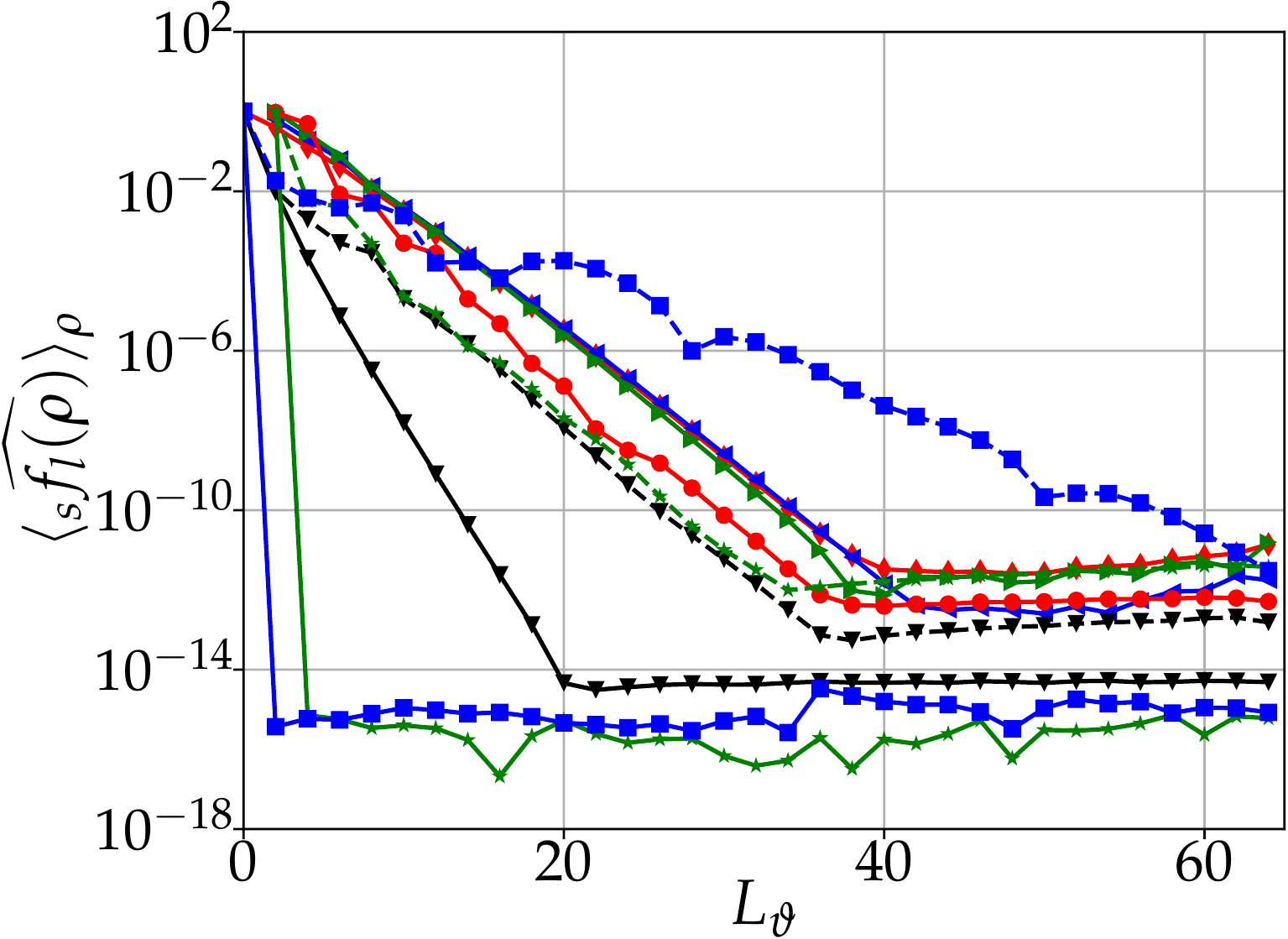}
\end{subfigure}
\caption{
(\textbf{Left}) Resulting ${}^{(I)}_{-1}N(\rho,\,\vartheta)$ where $I=80$
for scalar curvature deformation with $\overline{g}$ and target $\delta$ 
of the text. Color selected according to ${}^{(I)}_{-1}N(\rho,\,\vartheta)$
value.
(\textbf{Middle}) Normalised (by maximum absolute value) modal representations
of functions with RMS over nodal $\vartheta$ samples where the
function $\sw{s}{f}$ in
(green ``{\footnotesize$\blacktriangleright$}'',
blue ``{\footnotesize$\blacktriangleleft$}'',
red ``{\footnotesize$\blacklozenge$}''): 
$\sw{-2}{\mathcal{C}}_{mn}$, $\sw{-1}{\mathcal{C}}_{mn}$
and $\sw{0}{\mathcal{C}}_{mn}$ are selected respectively;
(black solid, dashed ``{\footnotesize$\blacktriangledown$}''):
$\sw{0}{\overline{N}}$
and
${}^{(I)}_{\;\,0}{\overline{N}}$ respectively;
(red ``{$\bullet$}''):
${}^{\;(I)}_{-1}{N}$;
(green solid, dashed ``{\footnotesize$\bigstar$}''):
$\sw{-2}{\overline{\gamma}}$
and
${}^{\;(I)}_{-2}{\gamma}$ respectively;
(blue solid, dashed ``{\tiny$\blacksquare$}''):
$\sw{0}{\overline{\gamma}}$
and
${}^{(I)}_{\;\,0}{\gamma}$ respectively.
(\textbf{Right}) Legend as before; normalised, 
RMS is now performed over nodal $\rho$ samples.
Note: While construction of the Picard iteration damps the original exponential
decay we find that coefficient magnitudes are sufficiently small at larger
band-limits to accurately represent functions.
}
\label{fig:axiInsp2}
\end{figure}

A representative calculation for local scalar-curvature deformation 
is inspected in Fig.\ref{fig:axiInsp} where
geometric quantities are updated as described at the end of 
\S\ref{sec:axisymdeform}.
A further remark is in order: while we have selected $\overline{g}$ such that
$\sw{-1}{\overline{N}}=0$ there is no reason \emph{a priori} to inhibit
non-zero updates to this quantity during the iterative construction of
${\mathcal{R}[g]}$. Indeed we find this is the case for the present example (see
Fig.\ref{fig:axiInsp2} (left)). Furthermore, on account of
$\left|{}^{(i)}\deform(\rho,\,\vartheta) \right|$ accumulating towards
$\partial\Omega_\vartheta$ when $\rho\simeq \frac{1}{2}(\rmax-\rmin)$ we inspect
how modal representations in $\rho$ decay when averaging is performed over
$\vartheta$ and vice versa for background and updated metric coefficients
together with the spin-weighted contraction coefficients
$\sw{s}{\mathcal{C}_{mn}}$ of Eq.\eqref{eq:contraExpa}. In order to compactly
represent $\sw{s}{\mathcal{C}_{mn}}$ an additional average over $m$ and $n$ is
taken over all coefficients of fixed $s$. Coefficient decay is displayed in
Fig.\ref{fig:axiInsp2} (middle, right). It is clear that on average coefficients
do not display any spurious growth -- this was also verified by inspecting
individual $\sw{s}{\mathcal{C}}_{mn}$.

\subsection{Gluing: Internal binary black holes and external Schwarzschild}
\label{sec:numgluebbh}

We finally turn our attention to a problem of physical interest, namely the
gluing of binary black hole
(Brill-Lindquist~\cite{:ref:interactionenergy1963brill} and
Misner~\cite{:ref:methodimages1963misner}) data to an exterior asymptotic
Schwarzschild end. As in~\cite{:ref:numericalconstruction2016doulis} our
approach shall be construction of initial data on a spherical shell
$\Omega=\Omega_\rho\times\Omega_\vartheta\times\Omega_\varphi$. On the interior
ball bounded by $\Omega$ for which $\rho<\rmin$ a vacuum constraint (at a MIT
symmetry) satisfying, asymptotically Euclidean metric $g_E$ is prescribed
whereas to the exterior of $\Omega$ where $\rho>\rmax$ Schwarzschild initial
data are chosen. For $g_\Omega$ where $\rho\in\Omega_\rho$ we select a suitably
truncated combination of these choices (see Eq.\eqref{eq:sshellComb}). In
contrast to~\cite{:ref:numericalconstruction2016doulis} our numerical scheme
does not follow the proposal of~\cite{corvino2005Giulini}. We rather attempt to
follow the construction of Corvino's proof~\cite{scalarcurvature2000corvino} as
closely as possible.

For convenience, recall the Euclidean metric with $\dim(\Sigma)=3$
in spherical coordinates:
\begin{equation}
    \delta_{\mathrm{Euc}} = \d{\rho}\otimes \d{\rho} +
    \rho^2\left(\d{\vartheta}\otimes\d{\vartheta} +
      \sin^2\vartheta\,\d{\varphi}\otimes \d{\varphi} \right).
\end{equation}
We can use conformal transformations so as to provide an interesting $g_E$ by
rescaling $\delta_{\mathrm{Euc}}$ via the factor (function) $\psi$
as~\cite{:ref:numericalrelativity2010baumgarte,%
  :ref:introduction3p1numerical2012alcubierre}:
\begin{equation}\label{eq:internalgI}
  g_E = \psi^4 \delta_{\mathrm{Euc}}.
\end{equation}
Selection of initial data that can be interpreted as corresponding to a
quantity of $\Xi$ black holes is provided by the Brill-Lindquist (BL)
choice~\cite{:ref:interactionenergy1963brill,%
  :ref:numericalrelativity2010baumgarte}:
\begin{equation}\label{eq:internalpsiBL}
  \begin{aligned}
    \psi &= 1 + \sum_{\xi=1}^{\Xi} \frac{m_\xi}{2 r_\xi}, &
    r_\xi &= |x^i - C^i_\xi|;
\end{aligned}
\end{equation}
where $r_\xi$ is the (coordinate) separation from the centre $C^i_\xi$ of the
$\xi^{\mathrm{th}}$ black hole. In order to compare
with~\cite{:ref:numericalconstruction2016doulis} we work within the context of
axisymmetry where symmetrically spaced, on-axis, equal mass, binary black hole
data, i.e., $\Xi=2$, $m=m_1=m_2$ and (in Cartesian coordinates)
$C_1^i= (0,\,0,\,d/2) = -C_2^i$ is chosen. 

Free parameters appearing in $g_E$ are fixed as $m=2$ and $d=10$ 
so as to facilitate comparison with~\cite{:ref:numericalconstruction2016doulis}.
A further reason for this selection is to have a scenario where
the two interior black hole horizons do not intersect and to avoid
the formation of a tertiary outer horizon which is the case if
the inequality $m/d \lesssim 0.64$ is satisfied 
(see also~\cite{:ref:interactionenergy1963brill}).

External to $\Omega$ we follow~\cite{scalarcurvature2000corvino} and select
Schwarzschild initial data in isotropic
form~\cite{:ref:numericalrelativity2010baumgarte}:
\begin{equation}
    g_S = \left(1 + \frac{M_{\mathrm{ADM}}}{2r} \right)^4\delta_{\mathrm{Euc}},
\end{equation}
where $M_{\mathrm{ADM}}$ is the mass of the full $g$ on $\Sigma$ which is to
satisfy ${\mathcal{R}}[g]=0$. The underlying axisymmetry together with
invariance under $z\rightarrow-z$ for $g_E$ entails that $g_S$ need not be
shifted from the origin and the only physical parameter to be adjusted for the
gluing construction is $M_{\mathrm{ADM}}$.

On $\Omega$ put:
\begin{equation}\label{eq:sshellComb}
    \overline{g}_\Omega = \chi_R g_E + (1 - \chi_R) g_S,
\end{equation}
where $\chi_R$ is the mapped cut-off function described at the start of
\S\ref{sec:ProPro}.  The $\omega$ entering the metric coefficient update
formulae together with the weak formulation of Eq.\eqref{eq:weakAxiExpa} is
selected as $\hat{\omega}_C$ of Eq.\eqref{eq:defomC} with $\mathfrak{f}=6/5$ and
in both $\chi_R$ and $\hat{\omega}_C$ polynomial decay with $N=4$ is chosen.

In the present context a further complication resulting from constant (i.e.
zero) ${\mathcal{R}}$ exists. Specifically, if
$g\rightarrow \delta_{\mathrm{Euc}}$ then as explained in \S\ref{sec:nonlinloc}
the kernel of the formal adjoint $L_{g}^*[\cdot]_{ij}$ becomes non-trivial and a
modified, projected problem must instead be treated. In order to avoid extensive
changes to our numerical scheme we propose to solve
Eq.\eqref{eq:weakLinFormAxiProb} via truncated singular-value decomposition
(TSVD)~\cite{:ref:appliedlinearalgebra2007shores}. For a more general choice of
axisymmetric $g_E$ (no longer invariant under $z\rightarrow -z$) the centre of
$g_S$ may also need adjustment during the gluing process thus we have two
degrees of freedom if we view $g_S$ as a parameterised family of candidate
solutions -- accordingly all but the two smallest singular values $\sigma_i$
shall be retained.

\begin{figure}[h]
\begin{subfigure}
  \centering
  \includegraphics[width=.48\linewidth]{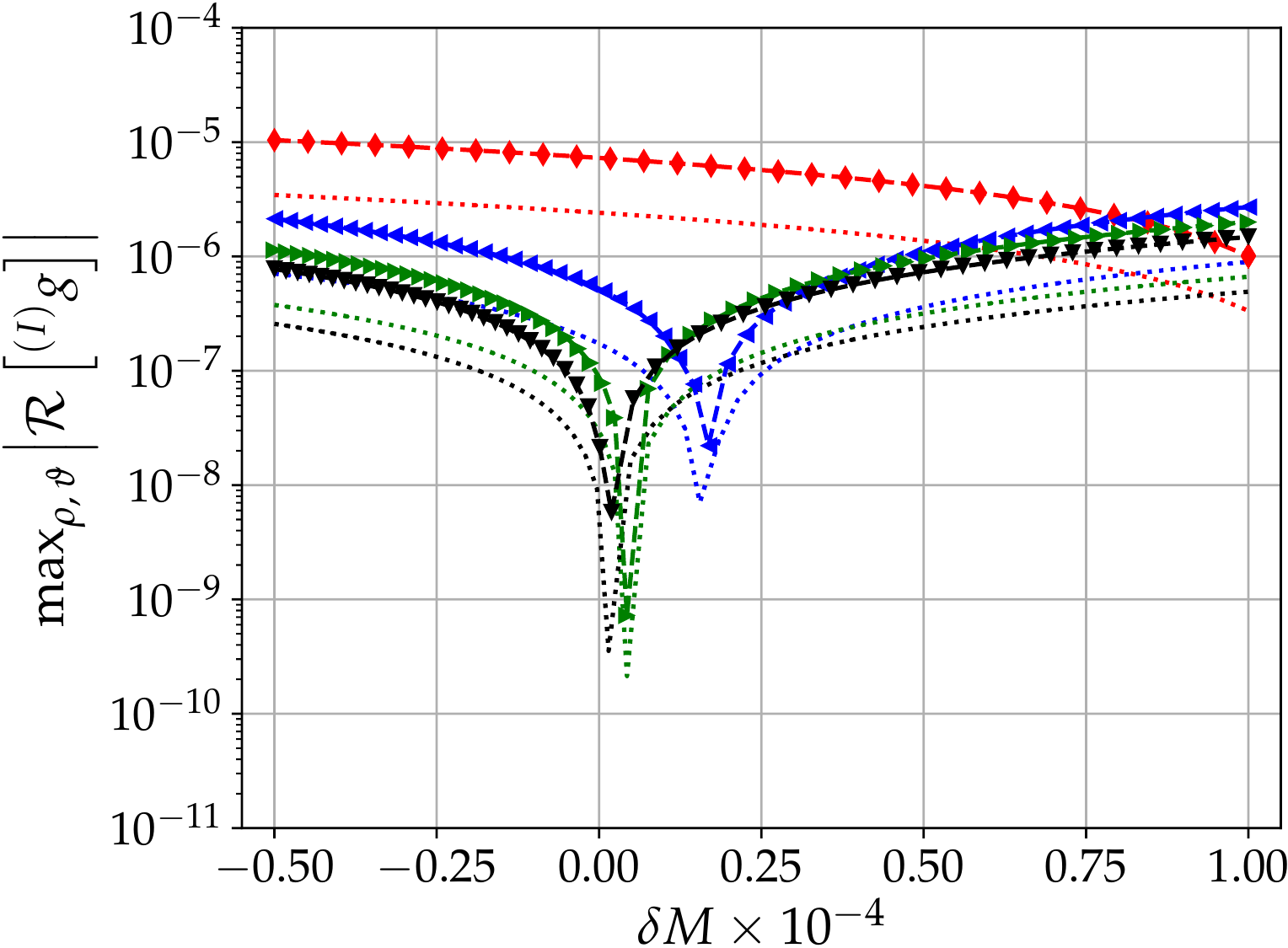}
\end{subfigure}%
\begin{subfigure}
  \centering
  \includegraphics[width=.48\linewidth]{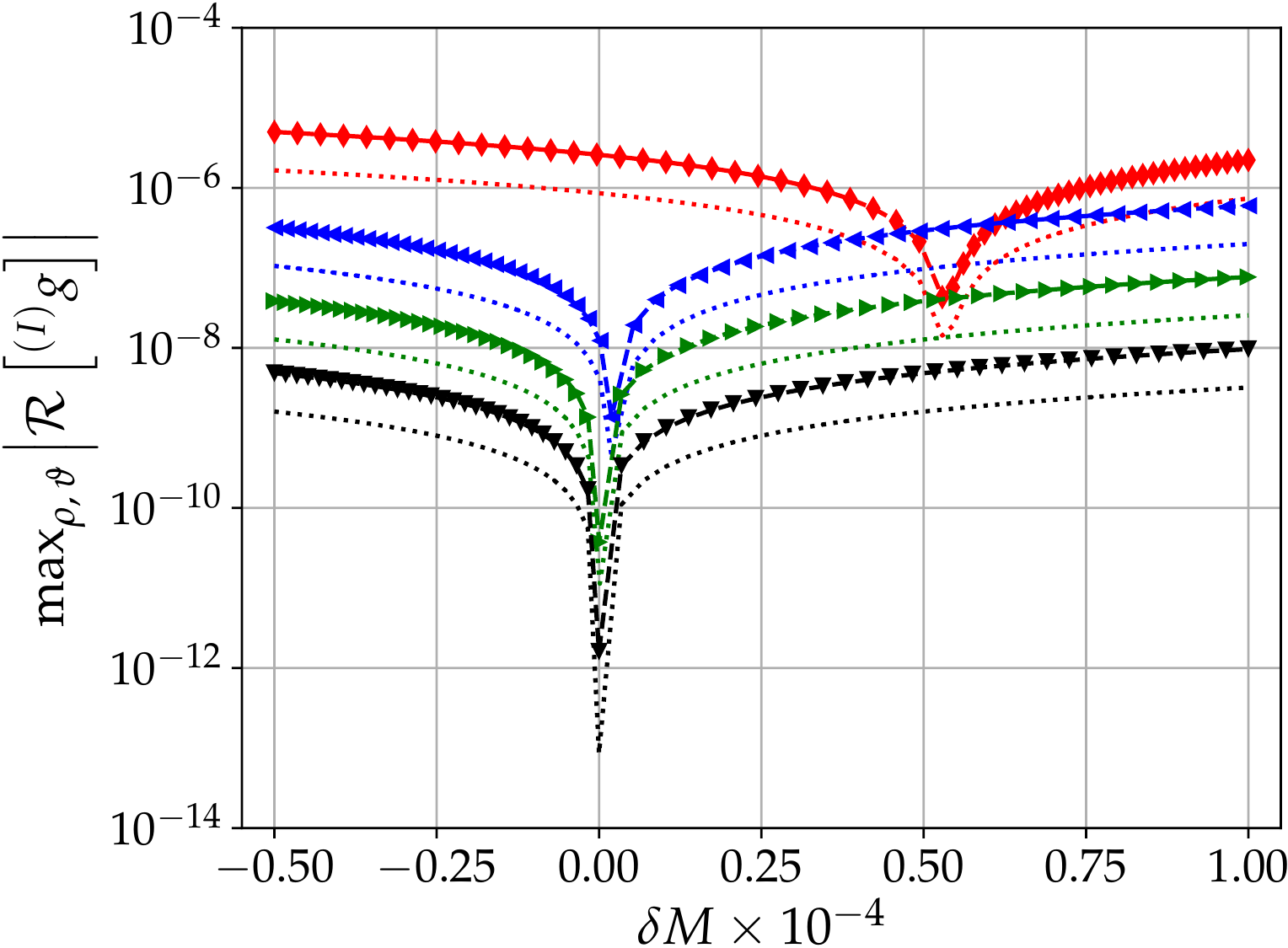}
\end{subfigure}
\caption{
Maximum absolute value of scalar curvature after deformation
for gluing of BL data. The domain $\Omega_\rho$ and
$M_{\mathrm{ADM}}$ of $g_S$ is varied
$M_{\mathrm{ADM}}\rightarrow 2m + \delta M$ where the background $\overline{g}_\Omega$
is that of Eq.\eqref{eq:sshellComb}.
In both subfigures the number of iterations taken is $I=20$ 
(saturation verified by doubling) where solid lines
correspond to $(L_\rho,\,L_\vartheta)=(128,\,32)$ and dotted lines
to $(L_\rho,\,L_\vartheta)=(64,\,16)$.
For $(L_\rho,\,L_\vartheta)=(128,\,32)$ with $\delta M$ selected such that 
the resulting scalar curvature is minimised put 
$\overline{r}:=\max_{\rho,\,\vartheta}\left|{\mathcal{R}}[\overline{g}_\Omega]\right|$.
(\textbf{Left}) Internal value $\rmin$ fixed at $25$. Set
$\Omega_\rho=\Omega_\rho(\mu):=[25,\,30+15\mu]$. Denoted in
(red ``{\footnotesize$\blacklozenge$}''): 
$\Omega_\rho(1)$, $\overline{r}=7.6\times10^{-5}$;
(blue ``{\footnotesize$\blacktriangleleft$}''): 
$\Omega_\rho(2)$, $\overline{r}=1.5\times10^{-5}$;
(green ``{\footnotesize$\blacktriangleright$}''): 
$\Omega_\rho(3)$, $\overline{r}=4.9\times10^{-6}$;
(black ``{\footnotesize$\blacktriangledown$}''): 
$\Omega_\rho(4)$, $\overline{r}=2.0\times10^{-6}$.
(\textbf{Right}) Scaling applied to both end-points in the radial extent
of $\Omega$. Set $\Omega_\rho=\Omega_\rho(\mu):=2^{\mu-1}[25,\,50]$.
Denoted in
(red ``{\footnotesize$\blacklozenge$}''): 
$\Omega_\rho(0)$, $\overline{r}=4.1\times10^{-5}$;
(blue ``{\footnotesize$\blacktriangleleft$}''): 
$\Omega_\rho(1)$, $\overline{r}=1.5\times10^{-6}$;
(green ``{\footnotesize$\blacktriangleright$}''): 
$\Omega_\rho(2)$, $\overline{r}=4.8\times10^{-8}$;
(black ``{\footnotesize$\blacktriangledown$}''): 
$\Omega_\rho(3)$, $\overline{r}=1.6\times10^{-9}$.
In both subfigures it is clear that $2m$ is approached from
above with $M_{\mathrm{ADM}} \rightarrow 2m$ as $\Omega_\rho$ is enlarged.
Furthermore, increasing band-limit does not significantly alter the value
of $\delta M$ that minimises the resultant scalar curvature.
}
\label{fig:axiGlue}
\end{figure}

In Fig.\ref{fig:axiGlue} we display the results of a numerical calculation where
the gluing constructed is implemented as previously described. In addition
to determination of all updated geometric quantities we must further fix
the ``optimal'' $M_{\mathrm{ADM}}$ in the sense that the resulting
scalar curvature is minimised (and ideally $0$). This is accomplished by varying 
about $2m$.
In agreement with~\cite{:ref:numericalconstruction2016doulis} we find that
the mass parameter entering $g_S$ must satisfy $M_{\mathrm{ADM}}\geq2m$ for
the gluing construction to proceed. Qualitatively, similar behaviour is
found when parameters selected for $\hat{\omega}_C$ and $\chi_R$ are modified.

In consideration of an MIT symmetry an alternative option for $g_E$ is possible.
By making use of Eq.\eqref{eq:internalgI} and Eq.\eqref{eq:internalpsiBL}
with $\Xi=2$
in construction of $g_E$ we implicitly assumed a three-sheeted topology, i.e.,
black hole throats are disconnected and not 
isometric~\cite{:ref:numericalrelativity2010baumgarte}. Instead, one may
work with Misner data~\cite{:ref:methodimages1963misner}, which, based on
the technique of spherical inversion images allows for a symmetric 
identification of the throats resulting in a ``wormhole'' within what is now
a single, asymptotically flat, multiply connected manifold. For an observer
external to a horizon the consequence of this topological manipulation
is a modification to the interaction
energy~\cite{:ref:interactionwormhole1990giulini,:ref:construction1997giulini}
and hence we investigate this within the context of the gluing construction.

\begin{figure}[h]
  \begin{subfigure}
  \centering
  \includegraphics[width=.48\linewidth]{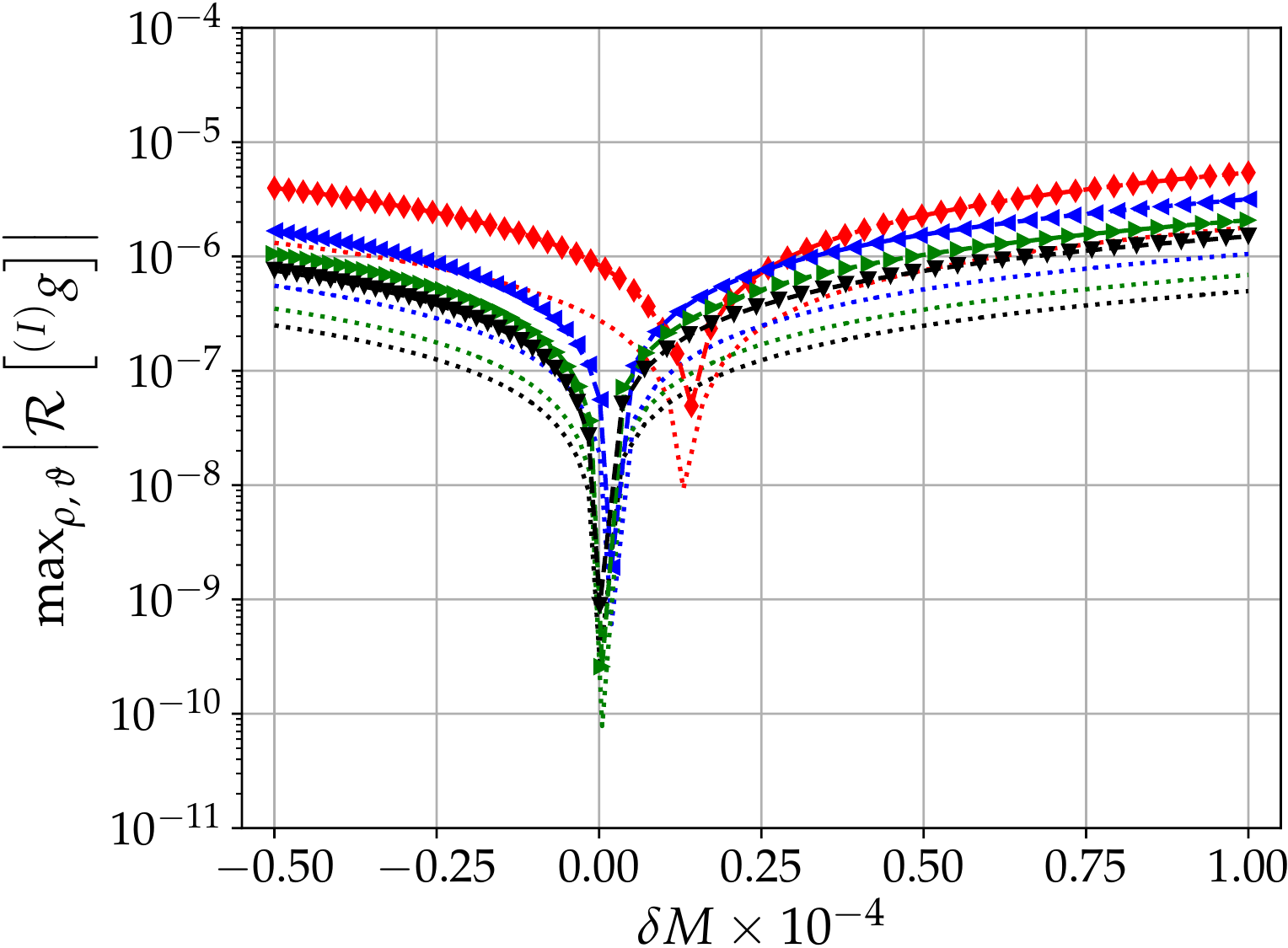}
\end{subfigure}%
\begin{subfigure}
  \centering
  \includegraphics[width=.48\linewidth]{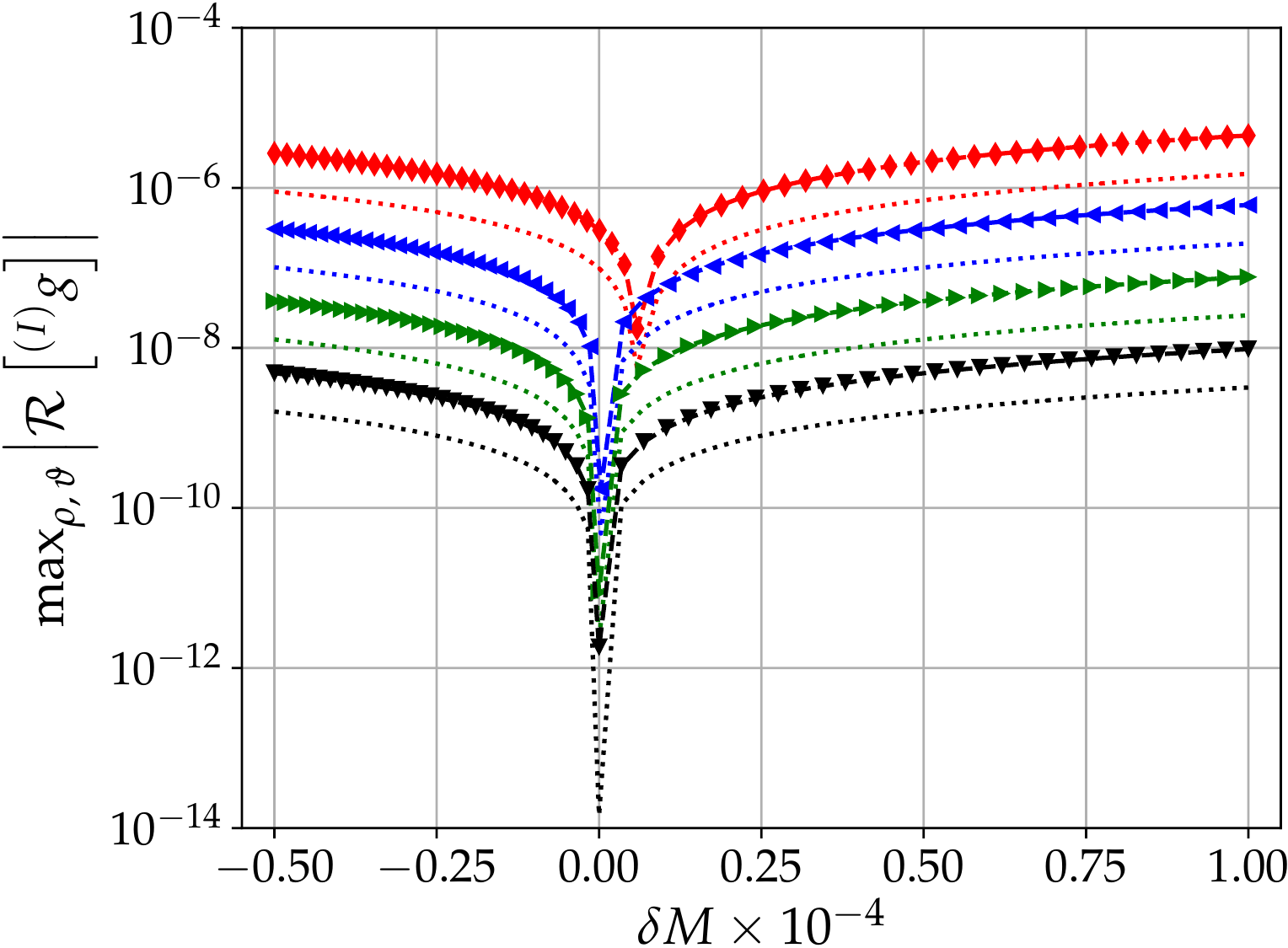}
\end{subfigure}
\caption{
Calculation of Fig.\ref{fig:axiGlue} repeated with $g_E$ constructed based
on Misner data (see text for details); all salient parameters as there
unless otherwise stated.
(\textbf{Left}) Internal value $\rmin$ fixed at $25$.
Set $\Omega_\rho=\Omega_\rho(\mu)=[25,\,30+15\mu]$.
Denoted in
(red ``{\footnotesize$\blacklozenge$}''): 
$\Omega_\rho(1)$, $\overline{r}=2.6\times10^{-5}$;
(blue ``{\footnotesize$\blacktriangleleft$}''): 
$\Omega_\rho(2)$, $\overline{r}=5.2\times10^{-6}$;
(green ``{\footnotesize$\blacktriangleright$}''): 
$\Omega_\rho(3)$, $\overline{r}=1.7\times10^{-6}$;
(black ``{\footnotesize$\blacktriangledown$}''): 
$\Omega_\rho(4)$, $\overline{r}=6.8\times10^{-7}$.
(\textbf{Right}) Scaling applied to both end-points in the radial extent
of $\Omega$. Set $\Omega_\rho=\Omega_\rho(\mu):=2^{\mu-1}[25,\,50]$.
Denoted in
(red ``{\footnotesize$\blacklozenge$}''): 
$\Omega_\rho(0)$, $\overline{r}=1.4\times10^{-5}$;
(blue ``{\footnotesize$\blacktriangleleft$}''): 
$\Omega_\rho(1)$, $\overline{r}=4.9\times10^{-7}$;
(green ``{\footnotesize$\blacktriangleright$}''): 
$\Omega_\rho(2)$, $\overline{r}=1.6\times10^{-8}$;
(black ``{\footnotesize$\blacktriangledown$}''): 
$\Omega_\rho(3)$, $\overline{r}=5.3\times10^{-10}$.
In both subfigures it is clear that $2m$ is again approached from
above with $M_{\mathrm{ADM}} \rightarrow 2m$ as $\Omega_\rho$ is enlarged;
efficiency is improved in contrast to Fig.\ref{fig:axiGlue}.
}
\label{fig:axiGlue2}
\end{figure}

For concreteness, in cylindrical coordinates $(r,\varphi,z)$ the Euclidean 
metric takes the form:
\begin{equation}
    \delta_{\mathrm{Euc}} =
 \d{r} \otimes \d{r}
    + r^2 \d{\varphi}\otimes \d{\varphi}
 + \d{z}\otimes\d{z}.
\end{equation}
Misner data representing two equal-mass black holes 
aligned with $z$ and symmetrically situated about the
origin is provided 
by~\cite{:ref:methodimages1963misner,:ref:numericalrelativity2010baumgarte}:
\begin{equation}\label{eq:MisnerConf}
    \psi_M = 1 + \sum_{n=1}^\infty \frac{1}{\sinh(n\mu)}
    \left(
    \frac{1}{\sqrt{r^2 + (z+z_n)^2}} +
    \frac{1}{\sqrt{r^2 + (z-z_n)^2}}
    \right),
\end{equation}
where $z_n:=\coth(n\mu)$ and $\mu$ is a free parameter which may be identified
with the total mass:
\begin{equation}\label{eq:MADMwh}
    \mu_{\mathrm{ADM}} = 4 \sum_{n=1}^\infty \frac{1}{\sinh(n \mu)}.
\end{equation}
In fact any representative in this family of data is completely characterised 
by selection of $\mu$ on account of the proper length $L$ of
a geodesic loop through the wormhole
being~\cite{:ref:numericalrelativity2010baumgarte}:
\begin{equation}
    L = 2\left(
    1 + 2 \mu \sum_{n=1}^\infty \frac{n}{\sinh(n\mu)}
  \right).
\end{equation}
To provide a direct comparison with the previous setup we solve
Eq.\eqref{eq:MADMwh} for $\mu$ when $\mu_{\mathrm{ADM}}=4$ using 
standard numerical techniques
to find $\mu=1.14960525757536$. This in turn fixes $\psi_M$ of 
Eq.\eqref{eq:MisnerConf} which, upon mapping to spherical coordinates allows us 
to take  $g_E = \psi_M^4 \delta_{\mathrm{Euc}}$. The results of this numerical 
calculation are shown in Fig.\ref{fig:axiGlue2}.
While it is the case that the new internal data reduces the
required $\delta M$ 
and thus may be thought of as being more ``efficient''
we again find that the $M_{\mathrm{ADM}}$ parameterising the
external Schwarzschild representative must be tuned to exceed the mass
of $g_E$, that is, the metric on $\Omega$ tends to introduce additional energy
to the gluing construction.

For both Brill-Lindquist and Misner data we numerically determined optimising 
masses (see Fig.\ref{fig:axiGlue} and Fig.\ref{fig:axiGlue2} respectively)
that allowed for gluing to exterior Schwarzschild to proceed at a variety of 
parameters. Recall that the kernel of
$L_{\overline{g}_\Omega}^*[\cdot]_{ij}$ is only approximate
and our TSVD procedure always discards
the two smallest $\sigma_i$ associated with 
$A_{mknl}$ of Eq.\eqref{eq:weakAxiExpa} constructed based on 
$\overline{g}_\Omega$. It is thus important to inspect the full singular value 
spectrum of $A_{mknl}$ directly.
Doing so (with values $\sigma_i$ ordered in descending magnitude) reveals
distinct, discrete jumps in magnitude for the smallest two values however these 
are not particularly pronounced and as 
$\rmin$ or the extent of $[\rmin,\,\rmax]$ is reduced a gradual decay is instead 
found. This is not unexpected for it is the case that
${\mathcal{R}}[\overline{g}_\Omega]\rightarrow0$ only 
when $\rho\rightarrow\infty$, i.e., we are only working with an approximate 
kernel for $L_{\overline{g}_\Omega}^*[\cdot]_{ij}$. 
This may be responsible for the larger values of
$\max_{\rho,\,\vartheta}\left|\mathcal{R}\left[{}^{(I)}g \right] \right|$
observed during use of smaller gluing regions.
On account of this, a potential alternative approach to treat the
kernel numerically may be to make use of the controlled filtering offered by a
Tikhonov regularisation 
scheme~\cite{:ref:regularized2005yagle,:ref:optimaltikhonov2001oleary,%
  :ref:deblurring2006hansen2006},
which we shall consider elsewhere.

\section{Discussion and conclusion}\label{sec:disc}

In this work we have demonstrated a new numerical technique directly inspired by
and based on the exterior asymptotic gluing (EAG) construction result of
Corvino~\cite{scalarcurvature2000corvino} that does not rely on a conformal
Lichnerowicz-York decomposition of the constraints nor the Brill-wave ansatz
approach of
\cite{corvino2005Giulini,:ref:numericalconstruction2016doulis,%
  :ref:numerical2018pook_kolb}.
Our technique enabled fashioning of new solutions to the Einstein constraints in
vacuum at a moment-in-time (MIT) symmetry based on a choice of internal
Brill-Lindquist (BL) or Misner data glued over a transition region to a
Schwarzschild exterior $g_S$. It appears that quite general asymptotically
Euclidean, internal data may be glued in this sense. Unfortunately, for all
calculations we performed, $M_{\mathrm{ADM}}$ of the interior set appeared as a
lower bound in the sense that to construct composite initial data $(\Sigma,\,g)$
the parameter $M=M_{\mathrm{ADM}}+\delta M$ entering $g_S$ and enabling the
gluing to proceed satisfied $\delta M \geq 0$. Thus a reduction of
$M_{\mathrm{ADM}}$ based on BL internal data as claimed
by~\cite{corvino2005Giulini} to be possible could not be found. This conclusion
agrees with the general indications provided by the numerical results
of~\cite{:ref:numericalconstruction2016doulis,:ref:numerical2018pook_kolb}.

It would be of considerable interest to extend our numerical technique to
incorporate a generalisation of Corvino's result to EAG on Kerr as
in~\cite{asymptotics2006Corvino}. For EAG on Kerr the proof technique remains
similar albeit the MIT symmetry condition is relaxed. In particular, this means
that the full constraint system must be considered inasmuch as the momentum
constraint is no longer trivially satisfied due to the appearance of extrinsic
curvature $K_{ij}$. From the point of view of numerical technique it should be
feasible to employ a similar approach as in the EAG Schwarzschild scenario
demonstrated here. However, clearly the system is considerably more involved.
Potentially, while the technique of truncated singular value decomposition may
still be feasible in treatment of the approximate kernel appearing in the
adjoint linearisation of the full constraints a more geometric approach based on
the Killing initial data interpretation (briefly described in
\S\ref{sec:nonlinloc}) may be required.

An upshot of the increase in intricacy is a reduction in the rigidity of the
possible composite $(\Sigma,\,g_{ij},\,K_{ij})$ forming initial data sets. For
instance an analogous investigation to that made in this work could be based on
internal Bowen-York initial data~\cite{:ref:timeasymmetric1980bowen} and a
similar question as to whether spurious gravitational wave content may be
reduced could be asked. As we have not made use of conformal techniques (other
than for the sake of convenience in specifying data to glue) this question is
not obstructed by the results of
\cite{:ref:nonexistence2000garat,:ref:nonexistence2004kroon,%
  :ref:asymptotic2004kroon}
and may be worthwhile exploring further in this new setting.

Finally, composite data sets based on EAG would be of great interest to evolve
numerically in order to better understand their dynamical properties. A
potential path towards this end has been proposed in
\cite{:ref:numericalconstruction2016doulis} where the property of an exact
Schwarzschild exterior is exploited to allow for a hyperboloidal evolution
scheme to proceed. We leave such investigations open to future work.

\section*{Acknowledgements}
The authors are grateful for a University of Otago PhD scholarship to BD and a
University of Otago Research Grant to JF.

\section{Appendix}\label{sec:appendix}
Suppose $\overline{g}\neq\delta_{\mathrm{Euc}}$ then the
$L_{\overline{g}}^*[\cdot]_{ij}$ of Eq.\eqref{eq:LstformalAdjoint} can be
seen to have trivial kernel provided that $\overline{\mathcal{R}}$ is
non-constant with the following formal 
calculation based 
on~\cite{:ref:deformations1975fischer,scalarcurvature2000corvino}.
Assume $f\in \ker(L_{\overline{g}}^*)$ then by contraction of 
Eq.\eqref{eq:LstformalAdjoint}:
\begin{equation}\label{eq:lapfRsc}
    \begin{aligned}
    0=\left(L_{\overline{g}}^*[f]\right)_{ij} \Longleftrightarrow\,&
    \overline{\nabla}_{(i}\overline{\nabla}_{j)} f =
    \overline{\mathrm{Ric}}_{ij} f + \overline{g}_{ij} \overline{\nabla}^2[f],\\
    \Longrightarrow\,& \overline{\nabla}^2[f] = 
    -\frac{1}{2} \overline{\mathcal{R}} f.
    \end{aligned}
\end{equation}
Whereas taking the divergence yields:
\begin{equation}\label{eq:divLstr}
    \begin{aligned}
    \overline{\nabla}{}^j[\overline{\nabla}{}_i[\overline{\nabla}{}_j[f]]]
    &=\overline{\nabla}^j[\overline{\mathrm{Ric}}{}_{ij}]f
    + \overline{\mathrm{Ric}}{}_{ij} \overline{\nabla}^j[f]
    + \overline{\nabla}{}_i[\overline{\nabla}{}^k[\overline{\nabla}{}_k[f]]],\\
    &= \overline{g}^{kj}
    \left(\overline{R}{}_{kij}{}^l \overline{\nabla}_l[f]
    + \overline{\nabla}{}_i[\overline{\nabla}{}_k[\overline{\nabla}{}_j[f]]]
    \right) = \overline{\mathrm{Ric}}{}_{ij}\overline{\nabla}{}^j[f]
    + \overline{\nabla}{}_i[\overline{\nabla}{}^k[\overline{\nabla}{}_k[f]]],\\
    \Longrightarrow 0&=\overline{\nabla}{}^j[\overline{\mathrm{Ric}}{}_{ij}] f.
    \end{aligned}
\end{equation}
To rewrite this in terms of the scalar curvature we make use of the
Bianchi identity~\cite{:ref:generalrelativity1984wald}:
\begin{equation}
    \overline{\nabla}{}_{[i} \overline{R}{}_{jk]l}{}^m=0,
\end{equation}
which once contracted yields:
\begin{equation}\label{eq:BianchiContr}
    \overline{\nabla}{}_i\left[\overline{R}{}_{jkl}{}^i \right]
    + \overline{\nabla}{}_j \left[\overline{\mathrm{Ric}}{}_{kl} \right]
    - \overline{\nabla}{}_k \left[\overline{\mathrm{Ric}}{}_{jl} \right]
    = 0,
\end{equation}
and once further:
\begin{equation}\label{eq:BianchiTwiceContr}
    \overline{\nabla}{}_i[\overline{\mathcal{R}}] =
    \frac{1}{2} \overline{\nabla}{}^j[\overline{\mathrm{Ric}}{}_{ij}].
\end{equation}
Thus Eq.\eqref{eq:divLstr} and Eq.\eqref{eq:BianchiTwiceContr} show:
\begin{equation}
    0 = \overline{\nabla}{}_i[\overline{\mathcal{R}}] f.
\end{equation}
That is, at points where $f$ does not vanish the gradient of 
$\overline{\mathcal{R}}$ must vanish. Consider now the behaviour of $f$ along 
an affinely parametrised geodesic $\gamma(s)$ with tangent vector $t^i$.
Then directional derivatives $\overline{D}$ of $f$ along $\gamma$ are:
\begin{equation}
    \begin{aligned}
    \overline{\mathrm{D}}_s\left[f(\gamma(s))\right]&= f'(s) = 
    t^i\overline{\nabla}{}_i[f],\\
    \overline{\mathrm{D}}^2_s\left[f(\gamma(s))\right]&= f''(s) = 
    t^j \overline{\nabla}{}_j[
    t^i\overline{\nabla}{}_i[f]]
    = \underbrace{t^j\left(\overline{\nabla}{}_j[t{}^i] \right)}_{=0}
    \overline{\nabla}{}_i[f] +
    t^j t^i \overline{\nabla}{}_j[\overline{\nabla}{}_i[f]];
    \end{aligned}
\end{equation}
where we made use of the geodesic 
equation~\cite{:ref:arelativiststoolkit2004poisson}. With
\cref{eq:LstformalAdjoint,eq:lapfRsc} 
we find an ODE for the behaviour of $f$ along $\gamma$:
\begin{equation}\label{eq:lstSurfODE}
    f''(s) = t^i t^j\left(
    \overline{\mathrm{Ric}}{}_{ij} 
    -\frac{1}{2} \overline{g}_{ij} \overline{\mathcal{R}}
    \right) f(s).
\end{equation}
Now when $f$ and $\overline{\nabla}{}_i[f]$ vanish at $x_0:=\gamma(0)$
then $f(\gamma(0))=0=f'(\gamma(0))$ and hence by Eq.\eqref{eq:lstSurfODE}
$f(\gamma(s))=0$. It follows that $f$ must vanish in an entire 
neighbourhood of $x_0$. Due to the elliptic condition
of Eq.\eqref{eq:lapfRsc}
Aronszajn's unique continuation 
theorem~\cite{:ref:uniquecontinuation1961aronszajn} implies that $f$ must vanish
everywhere which would result in a trivial $\ker(L_{\overline{g}}^*)$. 

Suppose instead $f(x_0)=0$ and $\overline{\nabla}_i[f](x_0)\neq 0$. Then
$x_0$ is a regular value and the zero-set of $f$ is an embedded
submanifold $\mathscr{S}$ of co-dimension $1$, i.e., an embedded surface in
$\Omega$~\cite{:ref:introsmoothmanifolds2003lee}. Finally, this implies that
$\overline{\nabla}_i[\overline{\mathcal{R}}]=0$ on $\Omega\setminus\mathscr{S}$
and by continuity $\overline{\mathcal{R}}$ is constant on all $\Omega$.

Therefore when $\overline{\mathcal{R}}$ is not constant 
$\ker(L_{\overline{g}}^*)$ is trivial and consequently $L_{\overline{g}}^*$ 
must be injective. 


\begin{footnotesize}
  \bibliographystyle{acm}               
  \renewcommand{\bibname}{References}   

\end{footnotesize}

\end{document}